\documentclass[11pt,letterpaper]{article}
\usepackage{jheppub}

\usepackage{graphicx}
\usepackage{bbm}
\usepackage{amsmath}
\usepackage{amssymb}
\usepackage{mathtools}
\usepackage{amsthm}
\usepackage{mathrsfs}
\usepackage{marvosym}
\usepackage{dsfont}
\usepackage[labelformat=simple]{subcaption}
\usepackage{xcolor}
\usepackage{braket}
\usepackage{cleveref}
\usepackage{comment}
\usepackage{float}
\usepackage{slashed}
\usepackage{fancybox}
\usepackage[skins,theorems]{tcolorbox}
\tcbset{highlight math style={enhanced,
		colframe=black,colback=white,arc=0pt,boxrule=1pt}}
\usepackage{overpic}

\allowdisplaybreaks

\definecolor{dark-gray}{gray}{0.20}
\definecolor{gray}{gray}{0.30}
\definecolor{light-gray}{gray}{0.80}
\definecolor{dark-red}{rgb}{0.7,0,0}
\definecolor{dark-green}{rgb}{0.1,0.4,0}
\definecolor{dark-blue}{rgb}{0.3,0.3,0.7}
\definecolor{light-blue}{rgb}{0.8,0.8,1}
\definecolor{swamp}{RGB}{240, 199, 197}

\newcommand{\be}{\begin{equation}}
	\newcommand{\ee}{\end{equation}}

\def\be{\begin{equation}}
	\def\ee{\end{equation}}
\def\bea{\begin{eqnarray}}
	\def\eea{\end{eqnarray}}

\newcommand{\beq}{\begin{equation}}  \newcommand{\eeq}{\end{equation}}
\newcommand{\bal}{\begin{aligned}}   \newcommand{\eal}{\end{aligned}}
\def\beqa{\begin{eqnarray}}
	\def\eeqa{\end{eqnarray}}

\newenvironment{eqn}{\begin{equation}\begin{aligned}}{\end{aligned}\end{equation}}
\newenvironment{eqn*}{\begin{equation*}\begin{aligned}}{\end{aligned}\end{equation*}\noindent}

\DeclareMathOperator{\csch}{csch}

\newcommand{\Tr}{\mathrm{Tr}}

\numberwithin{equation}{section}

\def\simleq{\; \raise0.3ex\hbox{$<$\kern-0.75em
		\raise-1.1ex\hbox{$\sim$}}\; }
\def\simgeq{\; \raise0.3ex\hbox{$>$\kern-0.75em
		\raise-1.1ex\hbox{$\sim$}}\; }

\numberwithin{equation}{section}

\hypersetup{
	colorlinks=true,
	linkcolor=dark-blue,
	citecolor=dark-red,
	urlcolor=dark-green,
	linktoc=page
}

\theoremstyle{remark}

\newtheoremstyle{named}{}{}{\itshape}{}{\bfseries}{.}{.5em}{#3}
\theoremstyle{named}

\title{\centering Exact Path Integral Methods\\ in Supersymmetric $\text{AdS}_2\times \mathbf{S}^2$ Backgrounds}

\author{Alberto Castellano$^{1,2}$,} 
\author{Carmine Montella$^{3}$,}
\author{Matteo Zatti$^{3}$}
\affiliation{$^1$Enrico Fermi Institute \& Leinweber Institute for Theoretical Physics,\\
University of Chicago, Chicago, IL 60637, USA}
\affiliation{$^2$Kavli Institute for Cosmological Physics,\\
University of Chicago, Chicago, IL
60637, USA}
\affiliation{$^3$Max-Planck-Institut f\"ur Physik (Werner-Heisenberg-Institut),\\
Boltzmannstrasse 8, 85748 Garching bei M\"unchen, Germany}

\emailAdd{acastellano@uchicago.edu, montella@mpp.mpg.de, zatti@mpp.mpg.de}

\abstract{We determine the exact functional determinants of charged, massive spin-0 and spin-$\frac12$ particles in $\text{AdS}_2\times \mathbf{S}^2$ backgrounds threaded by constant electric and magnetic fields. This is achieved using Schwinger proper-time formalism, which allows us to derive the full non-perturbative effective action in the 1-loop and constant background field approximations. We then specialize the computation to supersymmetric settings and we obtain the effective action for a 4d $\mathcal{N}=2$ BPS massive hypermultiplet in a supersymmetric $\text{AdS}_2\times \mathbf{S}^2$ spacetime. This setup can be seen to be equivalent to the near-horizon geometry of a BPS black hole which solves the attractor equations of 4d $\mathcal{N}=2$ supergravity. Our results provide a necessary intermediate step for the evaluation of the quantum-corrected black hole partition function. We also comment on the relation with the celebrated Gopakumar-Vafa integral representation.}

\begin{document}
	\makeatletter
	\let\old@fpheader\@fpheader
	\renewcommand{\@fpheader}{\vspace*{-0.1cm} \hfill EFI-26-02\\ \vspace* {-0.1cm} \hfill MPP-2026-36}
	\makeatother
	
	\maketitle
	\setcounter{page}{1}
	\pagenumbering{roman}

	\hypersetup{
		pdftitle={A cool title},
		pdfauthor={.. Alberto Castellano},
		pdfsubject={}
	}

\newcommand{\remove}[1]{\textcolor{red}{\sout{#1}}}

\newpage
\pagenumbering{arabic} 

\section{Introduction and Summary}
\label{s:intro}

Quantum Field Theory (QFT) is one of the most successful frameworks currently available for formulating and explaining most of the fundamental physics known to date. On one hand, it provides a unified language for describing elementary particles and their interactions, yielding theoretical predictions that have been confirmed to remarkable precision in numerous high‑energy experiments, such as those performed at particle colliders. At the same time, QFT has broad applicability beyond particle physics, from condensed matter systems to cosmology, and it even plays a central role in some modern approaches to quantum gravity \cite{Deligne:1999qp,Hori:2003ic}.

\smallskip

One particularly attractive feature of quantum field theories is that they provide a clean and compelling picture of how to relate the dynamics of the theory at widely separated length-scales. Thus, if we are only interested in the physics occurring up to certain energies, it is often possible---and sometimes even convenient---to determine an \emph{effective} description that focuses on the relevant low-energy degrees of freedom. This is encoded into the Wilsonian effective action, where only light field excitations remain dynamical and heavy fluctuations are being integrated out \cite{Polchinski:1983gv,Schwartz:2014sze}. The effects due to the latter fields are, however, of utmost importance, since they not only correct the quantum observables that can be measured and contrasted with the leading-order, two-derivative theory predictions, but are often entirely responsible for explaining certain physical phenomena. A celebrated example is given by the Euler-Heisenberg Lagrangian \cite{Euler:1935zz,Heisenberg:1936nmg,Weisskopf:1936hya}, which is customarily presented as a series expansion in the electromagnetic field strength (and its dual), and describes the non-linear corrections to the Maxwell action. These arise, in turn, as a consequence of integrating out the electron field, thereby accounting for the effective photon self-interactions.

\medskip

Interestingly, the above discussion extends beyond the familiar perturbative regime, where quantum effects can be computed through systematic expansions provided the couplings in the Lagrangian are sufficiently small. A well-known example is again provided by the Euler–Heisenberg theory, which can be recasted using Schwinger proper-time formalism \cite{Fock:1937dy,Schwinger:1951nm} as an exact, all-orders result. Furthermore, in the presence of a constant electric field, this formulation yields an imaginary contribution to the effective action that cannot be seen at any order in perturbation theory, signaling an instability. Another class of examples wherein non-perturbative physics has played a major role arises in supersymmetric non-Abelian gauge theories \cite{Wess:1992cp}. Indeed, in certain cases, when a sufficient amount of supersymmetry is present, their low energy dynamics can be exactly determined \cite{Seiberg:1994rs,Seiberg:1994aj,Seiberg:1994pq}. Remarkably, this quantum structure admits different (dual) local descriptions, which are patched together in such a way that perturbative and non-perturbative degrees of freedom are suitably exchanged.

\smallskip

\noindent A variety of techniques have been introduced to study non-perturbative effects in QFT, including the functional renormalization group \cite{Dupuis:2020fhh}, lattice computations \cite{Montvay:1994cy}, or holographic methods \cite{Kim:2012ey}. In the present work, we focus on the background field and Schwinger proper-time approaches, which are particularly well-suited to cases with constant (or slowly varying) external fields (see e.g., \cite{Schubert:2001he} for a review). These range from uniform $U(1)$ gauge profiles and vacuum expectation values for scalars to non-trivial spacetime geometries. The Schwinger formalism therefore allows us to obtain closed-form expressions for 1-loop path integrals in a plethora of backgrounds, by evaluating functional traces or determinants. 

\medskip

A particularly interesting class of backgrounds arises upon taking the near-horizon limit of asymptotically flat, extremal black hole solutions. In four dimensions, the near-horizon region of a static, extremal, charged black hole is described by some $\mathrm{AdS}_2\times \mathbf{S}^2$ geometry threaded by constant electric and magnetic fields. Hence, understanding 1-loop determinants of charged, massive particles in such spacetimes can provide crucial information about (non-)perturbative quantum effects, vacuum stability and pair production in this type of gravitational settings. Moreover, embedding these backgrounds into string theory allows one to test whether certain protected couplings \cite{Antoniadis:1995zn,Bershadsky:1993cx}---e.g., those encoded by the Gopakumar-Vafa formula \cite{Gopakumar:1998ii,Gopakumar:1998jq,Dedushenko:2014nya}, which counts BPS states in M-theory---correctly reproduce the underlying quantum-corrected black hole partition functions \cite{LopesCardoso:1998tkj,LopesCardoso:1999cv,LopesCardoso:1999fsj,LopesCardoso:1999xn,Ooguri:2004zv, Gaiotto:2006ns,Dabholkar:2010uh, Sen:2012kpz, Murthy:2015yfa, Castellano:2025ljk}.

\smallskip 

\noindent Despite extensive work on exact path integrals in flat space and related setups \cite{Comtet:1984mm,Comtet:1986ki, Pioline:2005pf,Anninos:2019oka,Anninos:2020hfj,Sun:2020ame,Grewal:2021bsu}, a full analytic evaluation of functional determinants for massive spin fields in $\mathrm{AdS}_2\times \mathbf{S}^2$ spacetimes with constant electric and magnetic backgrounds has yet not been presented in the literature. The primary goal of this paper is to fill this gap.

\medskip

Here, we derive the 1-loop effective action for charged, massive spin-$0$ and spin-$\frac12$ fields in $\mathrm{AdS}_2\times \mathbf{S}^2$ spacetimes with constant electric and magnetic flux. For simplicity, we first consider minimally coupled scalar and spinor particles interacting with fixed gravitational and gauge backgrounds. Using spectral decomposition and proper-time methods, we reduce the four-dimensional functional problem to a product of two two-dimensional Landau systems---one on $\mathbf{S}^2$ with a constant magnetic field and one on $\mathrm{AdS}_2$ with a constant electric field. The spectrum on the sphere is purely discrete, while that of $\mathrm{AdS}_2$---or rather, of its Euclidean counterpart---contains both discrete and continuous modes \cite{Wu:1976ge,Dunne:1991cs,Carinena:2011zz,Carinena:2012es,Hong:2005wp,Kordyukov_2019,Kordyukov_2022,Bolte:1990qs,Kostant:1969zz,Comtet:1984mm,Comtet:1986ki,Grosche:1988um,kim:2004rp,Kim:2003qp}. By applying a Hubbard-Stratonovich transformation to each functional trace \cite{Stratonovich1957OnAM,Hubbard:1959ub,Sun:2020ame,Grewal:2021bsu}, we obtain a double integral representation of the full 1-loop determinant. Furthermore, when considering BPS particles propagating in supersymmetric backgrounds, this representation gets simplified to a compact Schwinger integral that closely resembles the form of the original Gopakumar-Vafa generating function. This constitutes the first main result of this paper.

\medskip

\noindent We then specialize to supersymmetric setups and consider the effect of 4d hypermultiplets.
The type of backgrounds we focus on arise naturally from embedding $\mathrm{AdS}_2\times \mathbf{S}^2$ into a generic four-dimensional $\mathcal{N}=2$ supergravity with vector and hypermultiplets. These geometries describe near-horizon regions of BPS black holes carrying electric and magnetic charges. In compactifications of Type IIA string theory on Calabi–Yau threefolds, the black hole systems are realized as (wrapped) D-brane bound states, and supersymmetry restricts the gauge field, when evaluated at the horizon, to lie entirely along the graviphoton direction. In this maximally supersymmetric solution, the electromagnetic interaction of charged probe particles is thus governed by a single constant $U(1)$ field with some effective dyonic couplings.

\smallskip

On the other hand, a massive hypermultiplet in such a theory consists of two complex scalars fields and one Dirac fermion. The latter, moreover, couples non-minimally to the graviphoton, thereby inducing some amount of kinetic mixing that modifies its naïve wave operator. Using techniques developed in \cite{Banerjee:2010qc,Sen:2012kpz,Keeler:2014bra,David:2023btq,Banerjee:2011jp}, we show that this interaction introduces additional zero modes in the functional trace along the sphere, altering the index structure of the fermions. Computing the corresponding 1-loop determinant for a BPS hypermultiplet in $\mathrm{AdS}_2\times \mathbf{S}^2$ yields our second main result.

\medskip

The rest of the paper is organized as follows. In Section \ref{s:reviewAdS2S2}, we review in detail the solution to the spectral Landau problem on the two-dimensional sphere and Anti-de Sitter space, endowed with constant and everywhere orthogonal magnetic and electric fields, respectively. For the latter case, we first consider the analogous Euclidean version of the problem---involving magnetic fields in the hyperbolic plane---and subsequently perform an analytic continuation. In Section \ref{s:integrationAdS2xS2}, we outline a step-by-step derivation of the non-perturbatively exact 1-loop effective action for massive, charged particles with spin $\leq \frac12$ in certain AdS$_2 \times \mathbf{S}^2$ backgrounds. In particular, we consider such 4d spacetimes where the two characteristic radii coincide, motivated by black hole considerations (cf. Section \ref{sss:nearhorizonBH}). The main strategy consists in combining the results summarized in Section \ref{s:reviewAdS2S2} with various analytic techniques to express the effective action in a suitable Schwinger-like representation. For this, we also exploit the separability of the functional traces entering the relevant partition functions. With an eye on future string theory applications, we further restrict ourselves to those cases where the couplings and masses of the fields satisfy certain quadratic constraint, which is imposed by supersymmetry \cite{Billo:1999ip, Simons:2004nm, Castellano:2025yur, Castellano:2025rvn}. This allows us to determine the closed-form expression of the non-perturbative corrections to the 4d $\mathcal{N}=2$ effective action induced by minimally coupled BPS hypermultiplets, in the constant background field approximation. In addition, we discuss in Section \ref{sss:NonPerturbative} the non-perturbative ambiguities associated with the newly obtained formulae. In Section \ref{ss:Instability&Schwinger}, we comment on the implications of our results for the stability of the background and, in turn, on the possibility of triggering black hole decay via Schwinger pair production. To leverage these results for an actual calculation with extended supersymmetry, in Section \ref{s:SusyLoopDeterminant} we consider the effect of non-minimal couplings, which modify the fermionic determinants in a subtle but controllable way. We finally draw our conclusions in Section \ref{s:conclusions}.

\medskip

Several technical results have been relegated to the appendices. Appendix \ref{ap:densityAdS2&polygamma} provides further details on the spectral density in AdS$_2$ for both bosonic and fermionic fields. In Appendix \ref{ap:DetailsonAdS2xS2Traces}, we discuss some mathematical aspects of the integration-out procedure employed in this paper. We also review the Landau problem in $\mathbb{R}^{1,3}$ and analyze the flat-spacetime limits of our 1-loop determinants, together with their matching with the corresponding Minkowski counterpart. Finally, in Appendix \ref{ap:DetailsFermionOps} we study the diagonalization of two simple examples of fermionic kinetic operators, corresponding to massless particles and minimally coupled BPS hypermultiplets. We further explain how the chirality of the Dirac zero modes is correlated with the presence of a magnetic field on the 2-sphere.

\section{AdS$_2\times \mathbf{S}^2$ Geometry and the Spectral Problem}\label{s:reviewAdS2S2}

In this section, we provide a self-contained review of the general solution to the spectral (Landau) problem on AdS$_2$ and $\mathbf{S}^2$ for massive, charged spin-0 and spin-$\frac12$ particles, assuming minimal couplings to the gravitational and gauge background fields. For future reference, we also present the closed analytic form of the corresponding heat kernels. Readers not interested in the details can safely skip to the end of this section, where we summarize the main results. 

\subsection{The spectral problem in $\mathbf{S}^2$}\label{ss:LandauAdS2H2}

Let us first study the (non-relativistic) quantum mechanics of a charged particle living on the surface of a 2-sphere with a perpendicular and \emph{constant} magnetic field strength, which can be physically regarded as a magnetic monopole located at the center of an $\mathbf{S}^2 \hookrightarrow \mathbb{R}^3$ \cite{Dunne:1991cs, Carinena:2011zz, Carinena:2012es, Hong:2005wp, Kordyukov_2019, Kordyukov_2022, Bolte:1990qs}. Using the familiar spherical polar coordinate system, the metric tensor reads
\begin{eqn}\label{eq:metricsphere}
    ds^2= R^2 \left(d\theta^2 + \sin^2\theta\, d\phi^2\right)\, ,
\end{eqn}
where $R$ denotes the radius of the sphere. On top of this, we consider a homogeneous magnetic field $\boldsymbol{B}$ that is everywhere orthogonal to the surface. Hence, we take the field strength to be a constant factor times the volume form $\omega_{\mathbf{S}^2}$, namely 
\begin{eqn}\label{eq:spherefieldstrength}
    F = B\, \omega_{\mathbf{S}^2} = g\, \sin \theta\,d\theta \wedge d\phi\,, \qquad \omega_{\mathbf{S}^2}= R^2 \sin \theta\,d\theta \wedge d\phi\, ,
\end{eqn}
yielding a total magnetic flux equal to $4\pi B R^2=4\pi g$. The quantity $g$ can be interpreted as the charge of an underlying magnetic monopole, which moreover satisfies \cite{Kostant:1969zz,Bolte:1990qs}\footnote{In this and upcoming sections we take the minimal electric charge of probe particles to be equal to one.}
\begin{eqn}
    2g \in \mathbb{Z}\, .
\end{eqn}
Correspondingly, the 1-form connection, $A$, associated to the curvature 2-form \eqref{eq:spherefieldstrength} is
\begin{eqn}\label{eq:gaugechoicesphere}
    A^N_{\phi}=g(1-\cos\theta)\, ,\qquad A^S_{\phi} =-g(1+\cos\theta)\, ,
\end{eqn}
where we have explicitly separated between two coordinate patches that cover the whole $\mathbf{S}^2$, except for the south pole $(A^N)$ or the north pole $(A^S)$. This distinction is necessary due to the non-triviality of the principal $U(1)$ gauge bundle in the presence of the magnetic monopole. The sign of the latter charge determines the relative orientation of the magnetic field $\boldsymbol{B}$ and the two-dimensional surface---whether it is outgoing ($g>0$) or ingoing ($g<0$). Classically, changing this sign, i.e., sending $g \to -g$, causes the particle to reverse the direction of its precession around the sphere. This is equivalent to flipping its (conserved) angular momentum sending $\boldsymbol{J}\to -\boldsymbol{J}$, and it preserves the overall energy---analogously to what happens for the spectral problem on $\mathbb{R}^2$, that only depends on the quantity $\boldsymbol{L}^2=\boldsymbol{J}^2-g^2$ \cite{Castellano:2025yur}. Quantum-mechanically, a similar story holds, with the energy of the Landau levels depending just on the absolute value of the field strength (cf. eq.~\eqref{eq:algrabraicHamiltonianS2}). Therefore, in what follows we will assume that $B, g>0,$ while keeping in mind that in the most general case the exact same results hold for the density as well as the energy spectrum, both of which depend on $|B|, |g|$.

\subsubsection*{The spin-$0$ case}\label{sss:bosonS2}

The Hamiltonian for a charged, spin-less particle in the presence of such a background reads (using the position space representation)
\begin{eqn} \label{eq:2dHamiltonian}
    H = \frac12 (-i \nabla - A)^2\,.
\end{eqn}
Due to the symmetry properties exhibited by the system, it is more convenient to study this problem in terms of a different coordinate patch given by the stereographic projections $(z, \bar{z})$. Using this chart, the poles of the sphere are mapped either to the origin or to the point at infinity within the complex projective space $ \mathbb{CP}^1$. Which one is which ultimately depends on the particular patch we wish to cover. In the following, we will choose the northern projection, thus including the north pole as the origin of the complex plane $z$.\footnote{\label{fnote:morthvssouth}Note that we can actually do this without any loss of generality. Indeed, by defining a new complex coordinate $w=\sqrt{2B}\, R e^{-i\phi} \cot \left(\frac{\theta}{2}\right)= 2g/z$ (together with its complex conjugate $\bar{w}$), one can show that both the metric and Hamiltonian adopt the same form as those shown in eqs.~\eqref{eq:S2metriczcoords} and \eqref{hamiltonianS2zz} with $z \leftrightarrow w$, hence leading to the exact same solutions.} In terms of the $(\theta, \phi)$ angles, the aforementioned stereographic projection is given by
\begin{eqn}\label{eq:projcoordssphere}
    z = \sqrt{2g}\, e^{i\phi} \tan\left(\frac{\theta}{2}\right)\,, \qquad \bar{z} = \sqrt{2g}\, e^{-i\phi} \tan\left(\frac{\theta}{2}\right)\,.
\end{eqn}
Using these coordinates, the line element \eqref{eq:metricsphere} and the gauge connection \eqref{eq:gaugechoicesphere} take the form 
\begin{equation} \label{eq:S2metriczcoords}
    ds^2 =\frac{2}{B\left(1+ \frac{|z|^2}{2g}\right)^2}\, dz d\bar{z}\,, \qquad
    A =\frac{-i}{2\left(1+ \frac{|z|^2}{2g}\right)}(\bar{z}dz -z d\bar{z})\, .
\end{equation}
From this, one can easily obtain the new expression for the Hamilton operator
\begin{eqn}\label{hamiltonianS2zz}
    H_{\mathbf{S}^2} = B \left( -\left( 1 + \frac{|z|^2}{2g} \right)^2 \partial \bar{\partial} - \frac{1}{2}  \left( 1 + \frac{|z|^2}{2g} \right) (z\partial - \bar{z} \bar{\partial}) + \frac{1}{4} |z|^2 \right)\, .
\end{eqn}
In order to solve the spectral problem, we will use a simple algebraic approach. The main strategy here consists in defining a basis of Lie algebra $\mathfrak{su}(2)$ operators, whose associated quadratic Casimir controls directly the Hamiltonian of the quantum theory. Consequently, the Hilbert space can then be easily constructed using the representation theory of $SU(2)$. To show this, we first introduce (ladder) operators $J_{\pm}$ and $J_0$ as follows \cite{Dunne:1991cs}
\begin{eqn}\label{OperatorSU2}
\begin{aligned}
    J_+ &= - \frac{1}{\sqrt{2g}}z^2 \partial - \sqrt{2g}\, \bar{\partial} + \sqrt{\frac{g}{2}}\, z\, ,\quad J_- = \sqrt{2g}\, \partial + \frac{1}{\sqrt{2g}} \bar{z}^2 \bar{\partial} + \sqrt{\frac{g}{2}}\, \bar{z}\, ,\quad    J_0 = L_0 - g\,,
\end{aligned}
\end{eqn}
with the angular momentum $L_0$ about the vertical axis $x^3 = R \cos{\theta}$ defined as
\begin{eqn}\label{eq:SU2angularmomentum}
    L_0 = z \partial - \bar{z} \bar{\partial}\,, \qquad L_0^\dagger=L_0\,,
\end{eqn}
which clearly commutes with \eqref{hamiltonianS2zz}. These operators satisfy the algebra
\begin{eqn}\label{eq:SU2algebra}
    [J_+ , J_-] = 2J_0\, ,\qquad [J_0 , J_\pm ] = \pm J_\pm\, ,
\end{eqn}
and from them we can construct a quadratic Casimir element in the usual way, namely
\begin{eqn}
    C_{SU(2)} = \frac{1}{2} (J_+ J_- + J_- J_+) + J_0^2 = J_+ J_- + J_0(J_0-1)\, .
\end{eqn}
The Hamiltonian can then be rewritten in terms of the basis \eqref{OperatorSU2} simply as
\begin{eqn}\label{eq:algrabraicHamiltonianS2}
    H_{\mathbf{S}^2} = \frac{B}{2g}\, \left(C_{SU(2)} - g^2\right)\, .
\end{eqn}
From here, we conclude that the problem of finding simultaneous eigenstates of angular momentum $L_0 = J_0 + g$ and energy $H_{\mathbf{S}^2}$ is tantamount to finding similar eigenstates of $J_0$ and $C_{SU(2)}$. Furthermore, the $\mathfrak{su}(2)$ algebra allows us to deduce that the latter set can be equivalently labeled by a pair of integers $(j, m)$ defined such that 
\begin{eqn} \label{eq:systemSU2}
    C_{SU(2)} & \ket{j, m} = j (j +1)\ket{j, m}\, ,\\
    J_0 & \ket{j, m} = m \ket{j, m}\, ,
\end{eqn}
where $m=\ell-g$ ranges from $-j$ to $j$. Hence, upon noticing that lowest-weight states obtained by solving the equation $J_- \ket{j, m_{\rm min}}=0$ with integer---ensuring single-valuedness of the wavefunction---orbital angular momentum $\ell$, have the form\footnote{This can be easily shown by setting up an ansatz for the wavefunction of the particle with non-positive integer eigenvalue $\ell_{\min}=-n$ for $L_0$, namely $\psi_{j,\, m_{\rm min}}= f(|z|^2)\, \bar{z}^n$, yielding a differential equation for $f(x=|z|^2)$ 
\begin{eqn}
     (2g+x) \frac{d f}{dx}+ (n+g)f(x)=0\, ,
\end{eqn}
whose solution is precisely $f(x)= a\, (1+ \frac{|z|^2}{2g})^{-n-g}$, with $a$ some constant. Note that these are also regular in the southern $\mathbb{CP}^1$ patch (cf. footnote \ref{fnote:morthvssouth}).}
\begin{eqn}\label{eq:algebraiceigenfunctionsS2spin0}
    \psi_{j,\, m_{\rm min}} (z, \bar{z}) \propto \left(1 + \frac{|z|^2}{2g} \right)^{-g-n} \bar{z}^n\, , \qquad \forall n \in \mathbb{Z}_{\geq0}\, ,
\end{eqn}
we find that $j=n+g$, thereby obtaining that the eigenvalues of $H_{\mathbf{S}^2} $ are \cite{Wu:1976ge, Dunne:1991cs, Carinena:2011zz, Carinena:2012es, Hong:2005wp, Kordyukov_2019, Kordyukov_2022, Bolte:1990qs}
\begin{eqn}\label{spectrumS2}
    E_n = \frac{g}{R^2} \left( n + \frac{1}{2} + \frac{n(n+1)}{2g} \right)\, ,\qquad \text{with}\quad n \in \mathbb{Z}_{\geq 0}, \quad \ell = -n, -n+1, \dots, n + 2g\, ,
\end{eqn}
and each level has degeneracy
\begin{eqn}\label{degeneracyS2}
    d_n = 2j+1= 2\left( g + n + \frac{1}{2} \right)\,.
\end{eqn}
For more detailed references on this, see \cite{Bal:1996jm, etheses11236, Dunne:1991cs,libine2021}. With these results at hand, one can already determine the form of the heat kernel trace associated with the quantum-mechanical operator \eqref{eq:2dHamiltonian}. As explained in Section \ref{s:integrationAdS2xS2} below, the latter is required to obtain the relevant 1-loop determinants in this work, and can be computed as follows \cite{Vassilevich:2003xt}
\begin{equation}\label{eq:heatkernelS2_scalar}
    \mathcal{K}^{(0)}_{\mathbf{S}^2}(\tau) :=\Tr \left[e^{ -\tau H_{\mathbf{S}^2}} \right] = \sum_{n\geq 0} d_n\, e^{-\tau E_n} = 2\, \sum_{n \geq 0} \left( n + g + \frac{1}{2} \right) e^{ -\frac{\tau}{2R^2} \left( g + n(n+1+2g)\right)}\,.
\end{equation}

\subsubsection*{The spin-$\frac12$ case}\label{sss:fermionS2}

Let us now explain how the previous analysis is modified when considering spin-$\frac12$ particles. We thus start from the 2d action describing the dynamics of a charged fermion moving within some (possibly curved) Euclidean space
\begin{eqnarray} \label{eq:fermionicAction}
S = \int d^2x \sqrt{\det g}\, \bar{\Psi}\left[ (\slashed{\nabla} -i \slashed{A})+m\right] \Psi\,,
\end{eqnarray}
where $\det g$ is the determinant of the background metric, and the covariant derivative reads
\begin{eqnarray}
\nabla_i\Psi = \left( \partial_i+\frac14 \omega_{ab\, i} \sigma^{ab}\right)\Psi\,,\qquad \text{with}\quad \sigma^{ab}=\frac12\, [\gamma^a, \gamma^b]\,,
\end{eqnarray}
with $\omega_{ab}$ the connection 1-form. The Dirac matrices are defined such that 
\begin{eqnarray}
\gamma^a= e^a_{\ i} \gamma^i\,, \qquad \lbrace \gamma^a, \gamma^b \rbrace=2\delta^{ab}\, ,
\end{eqnarray}
where $e^a_{\ i}$ ($e_a^{\ i}$) is the (inverse) 2d vielbein and we take $\gamma^1=\sigma_x$ and $\gamma^2=\sigma_y$, i.e., a subset of the familiar Pauli matrices, whereas $\sigma_z=-i\sigma^{12}$. The Dirac operator is then
\begin{equation}\label{eq:gendefDiracop}
    \slashed{D}=-i (\slashed{\nabla}-i \slashed{A}) \,.
\end{equation}
To align with the conventions adopted in this chapter, it is convenient to switch to the stereographic coordinates introduced in \eqref{eq:projcoordssphere}. Using these, squaring $\slashed{D}$ yields
\begin{equation}
\begin{aligned}
\slashed{D}^2_{\mathbf{S}^2} =&\, 2B \left[ -\left( 1 + \frac{|z|^2}{2g} \right)^2 \partial \bar{\partial} - \frac{1}{2}  \left( 1 + \frac{|z|^2}{2g} \right) (z\partial - \bar{z} \bar{\partial}) + \frac{1}{4} |z|^2  \right]\\
& + \frac{1}{R^2} \left[ g\, \frac{\left( 1 + |z|^2/2g \right)^2}{8 |z|^2} + \frac14 - g\sigma_z \frac{1-|z|^4/4g^2}{2|z|^2} (z\partial - \bar{z} \bar{\partial}) + \frac{g}{2}\, \left(1 - |z|^2/2g \right) \sigma_z\right] - B \sigma_z\,,
\end{aligned}
\end{equation}
Note that the operator thus defined is diagonal, so that one may separately consider modes of positive and negative chirality---defined with respect to $\gamma^3=-i \gamma^1 \gamma^2=\sigma_z$---when solving for its eigenfunctions. Similarly to the spin-0 case, one can also find a set of $SU(2)$ operators\footnote{We remark that \eqref{eq:OperatorSU2fermi} already contains the spin contribution to the total angular angular momentum, and in fact when taking $g\to 0$ one recovers nothing but $\boldsymbol{L} = \boldsymbol{J}+\boldsymbol{S}$ in the Cartan-Weyl basis \cite{Abrikosov:2002jr}.}
\begin{eqn}\label{eq:OperatorSU2fermi}
\begin{aligned}
    J_+ &= - \frac{1}{\sqrt{2g}}z^2 \partial - \sqrt{2g}\, \bar{\partial} + \sqrt{\frac{g}{2}}\, z - \sqrt{\frac{g}{2}}\, \frac{1+ |z|^2/2g}{\bar{z}} \frac{\sigma_z}{2}\, ,\\
    J_- &= \sqrt{2g}\, \partial + \frac{1}{\sqrt{2g}} \bar{z}^2 \bar{\partial} + \sqrt{\frac{g}{2}}\, \bar{z} - \sqrt{\frac{g}{2}}\, \frac{1+ |z|^2/2g}{z} \frac{\sigma_z}{2}\, ,\\
    J_0 &= z \partial - \bar{z} \bar{\partial} - g\, ,
    \end{aligned}
\end{eqn}
satisfying the algebra \eqref{eq:SU2algebra}, and whose corresponding quadratic Casimir shall be written as
\begin{eqn}\label{eq:casimirfermiS2}
\begin{aligned}
    C_{SU(2)} = R^2\slashed{D}^2_{\mathbf{S}^2}+g^2-\frac14\, .
    \end{aligned}
\end{eqn}
As a consequence, the spectrum can be easily obtained by diagonalizing simultaneously $J_0$ and $ C_{SU(2)}$ in terms of a pair of integers $(j, m)$ defined via \eqref{eq:systemSU2}, where the kets should be now understood as bispinors. Furthermore, following the same strategy as for the spin-0 particles, one may restrict the allowed values for $j$ by solving for the lowest-weight states, i.e., those that satisfy $J_- \ket{j, m_{\rm min}}=0$. The latter take the form\footnote{To show this, one proposes an ansatz for the positive (negative) chirality wavefunction $\psi^+$ ($\psi^-$) having half-integer values $\frac12-n$ when acted on with $L_0$, namely $\psi^{\pm}_{j, m}= f^\pm(|z|^2)\, \bar{z}^{n-\frac12}$. This yields the following ODE
\begin{eqn}
     \left(1+ \frac{x}{2g}\right) \frac{d f^\pm}{dx}+ \left(\frac{n+g-\frac12}{2g} \mp \frac{1+x/2g}{4x} \right)f^\pm(x)=0\, ,\qquad \text{where} \quad x=|z|^2\, ,
\end{eqn}
which is solved by $f(x)= a\, \left(1+ \frac{|z|^2}{2g}\right)^{-(n+g-\frac12)}\, |z|^{\pm \frac12}$, with $a$ some constant. Notice that for $n=0$, only one of the above solutions is regular (and hence normalizable) in both $\mathbb{CP}_N^1$ and $\mathbb{CP}_S^1$.}
\begin{eqn}\label{eq:algebraiceigenfunctionsS2spin1/2}
    \psi^{\pm}_{j,\, m_{\rm min}} (z, \bar{z}) \propto \left(1 + \frac{|z|^2}{2g} \right)^{-g-n+\frac12}\, |z|^{\pm \frac12}\, \bar{z}^{n-\frac12}\, , \qquad \forall n \in \mathbb{Z}_{\geq0}\, ,
\end{eqn}
implying that $j=n+g- \frac12$. Hence, the eigenvalues for $\slashed{D}^2$ on $\mathbf{S}^2$ are (cf. \eqref{eq:casimirfermiS2})
\begin{eqn}\label{eq:eigenvaluesDiracS2}
    E_n^2= \frac{1}{R^2} n(n+2g)\, , \qquad n \in \mathbb{Z}_{\ge0} \,,
\end{eqn}
and can be divided into $2g = \frac{1}{2\pi} \int_{\mathbf{S}^2} F$ \emph{unpaired} zero modes (as per Atiyah-Singer \cite{Atiyah:1963zz}), together with a tower of \emph{paired} states with positive (negative) $\lambda$ corresponding to positive (negative) chirality. Thus, the energy levels have degeneracy 
\begin{equation}
    d_n = (2-\delta_{0, n}) (2j+1)= 4 n + (4 - 2  \delta_{n,0})  \, g   \,,
\end{equation}
such that computing the heat kernel trace for a Hamiltonian equal to \eqref{eq:gendefDiracop} gives 
\begin{eqn}\label{eq:heatkernelS2_fermion}
    \mathcal{K}^{(1/2)}_{\mathbf{S}^2}(\tau):=\Tr \left[e^{-\tau \slashed{D}^2_{\mathbf{S}^2}} \right] = \sum_{n \geq 0} d_n e^{-\tau E_n^2} = 2 \left(2\sum_{n \geq 1} \left( n + g\right) e^{ -\frac{\tau}{R^2} n\left( n + 2g \right)} + g\right)\,.
\end{eqn}

\subsection{The spectral problem in AdS$_2$} \label{ss:spectralAdS2}

In what follows, we review the \emph{electric} Landau problem in 2d Anti-de Sitter (AdS) space, building on a series of works within the physics and mathematics literature \cite{Pioline:2005pf, Anninos:2019oka, Carinena:2012es, Comtet:1984mm, Comtet:1986ki, Carinena:2011zz}. The AdS$_2$ metric written in conformal coordinates is \cite{Castellano:2025yur}
\begin{eqn}
ds^2 = \frac{R^2}{\rho^2} \left(-dt^2 + d\rho^2\right)\, ,
\end{eqn}
and we assume to have an electric field strength $\boldsymbol{E}$ proportional to the AdS$_2$ volume 2-form:
\begin{eqn}\label{eq:electricfieldAdS2}
F =  E\, \omega_{\text{AdS}_2}= e\, \frac{1}{\rho^2} dt \wedge d\rho\, , \qquad \omega_{\text{AdS}_2}= \left(\frac{R}{\rho}\right)^2 dt \wedge d\rho\, .
\end{eqn}
We can choose to express the 1-form connection $A$ in a convenient gauge such that $A_\rho = 0$. This implies that the gauge field $A=A_\mu dx^\mu$ may be written as
\begin{eqn}
A = e\, \frac{dt}{\rho}\,.
\end{eqn}
Instead of finding the associated eigenenergies, we will first solve the analogous spectral problem on the hyperbolic plane $\mathbb{H}^2$ with a constant magnetic field turned on. Subsequently, we translate the results to Anti-de Sitter with constant electric field strength by means of an appropriate analytic continuation of the resulting heat kernel (cf. Section \ref{ss:Traces} for details).

\subsubsection{Energy spectrum on $\mathbb{H}^2$}\label{sss:spectrumH2}

The Landau problem on the hyperbolic plane has been extensively studied \cite{Comtet:1984mm, Comtet:1986ki,Grosche:1988um, kim:2004rp, Kim:2003qp, Carinena:2011zz, Carinena:2012es, Dunne:1991cs, Bolte:1990qs}. Using the familiar upper-half coordinates $(\tau_1, \tau_2)$, one can write the line element as
\begin{eqn}\label{eq:metrichypplane}
    ds^2= R^2\, \frac{d\tau_1^2 + d\tau_2^2}{\tau_2^2}= R^2\, \frac{d\tau d\bar{\tau}}{(\text{Im}\, \tau)^2}\, , \qquad \tau= \tau_1+i\tau_2\,,
\end{eqn}
where $R$ denotes the characteristic length-scale of $\mathbb{H}^2$. Analogously to the case of the sphere, we consider a perpendicular and constant field strength
\begin{equation}\label{eq:H2fieldstrength}
     F = B\, \omega_{\mathbb{H}^2} = g\, \frac{d\tau_1 \wedge d\tau_2}{\tau_2^2}\,, \qquad \omega_{\mathbb{H}^2}= R^2\, \frac{d\tau_1 \wedge d\tau_2}{\tau_2^2}\, ,
\end{equation}
with $g$ denoting the dimensionless charge sourcing the magnetic field throughout the plane. We henceforth take $g>0$ without any loss of generality, keeping in mind that the energy spectrum is unchanged---both at the classical and quantum levels---upon flipping $g\to-g$. Despite the similarities with the $\mathbf{S}^2$ example above, there are a few crucial differences worth mentioning at this point. For instance, due to the non-compactness of the hyperbolic space, the $U(1)$ principal gauge bundle must be necessarily trivial \cite{gilligan2012lectures, Bolte:1990qs}. This implies, in turn, that the Dirac quantization condition does not apply anymore and the gauge charges shall take any positive real value. Consequently, one can define the corresponding 1-form connection $A$ associated to the curvature 2-form \eqref{eq:H2fieldstrength} in a global fashion. The latter reads
\begin{eqn}\label{eq:gaugechoiceH2}
    A=g\, \frac{d\tau_1}{\tau_2}\, , \qquad \text{with}\quad dA=F\,.
\end{eqn}

\subsubsection*{The spin-$0$ case}\label{sss:bosonH2}

For ease of comparison, we again use stereographic-like coordinates. Therefore, we define a new complex variable $z$ such that
\begin{eqn}\label{eq:geodesicpolarcoords}
    \tau=i\, \frac{1-i \frac{z}{\sqrt{2g}}}{1+i\frac{z}{\sqrt{2g}}}\, ,
\end{eqn}
which projects the upper-half plane to a disk of finite size, i.e., $|z|^2\leq 2g$. This transforms the metric \eqref{eq:metrichypplane} and the gauge connection \eqref{eq:gaugechoiceH2} into
\begin{equation}\label{eq:H2metriczcoords}
    ds^2  =\frac{2}{B\left(1- \frac{|z|^2}{2g}\right)^2}\, dz d\bar{z}\,, \qquad A  =\frac{-i}{2\left(1- \frac{|z|^2}{2g}\right)}(\bar{z}dz -z d\bar{z})\,,
\end{equation}
and thus the Hamilton operator, which is still given by \eqref{eq:2dHamiltonian}, can be written compactly as \cite{Bolte:1990qs, Comtet:1984mm, Comtet:1986ki, kim:2004rp, Kim:2003qp, Carinena:2011zz, Carinena:2012es, Dunne:1991cs}
\begin{eqn}\label{eq:HamiltonianH2}
    H_{\mathbb{H}^2} = B \left( -\left( 1 - \frac{|z|^2}{2g} \right)^2 \partial \bar{\partial} - \frac{1}{2} \left( 1 - \frac{|z|^2}{2g} \right) (z\partial - \bar{z} \bar{\partial}) + \frac{1}{4} |z|^2 \right)\, .
\end{eqn}
As before, we define ladder operators $K_\pm$ and a (hyperbolic) angular momentum $K_0$ \cite{Dunne:1991cs, Comtet:1984mm} 
\begin{eqn}\label{eq:OperatorSU11boson}
    K_+ = - \frac{1}{\sqrt{2g}}z^2 \partial + \sqrt{2g} \bar{\partial} - \sqrt{\frac{g}{2}} z\,,\quad K_- = -\sqrt{2g} \partial + \frac{1}{\sqrt{2g}} \bar{z}^2 \bar{\partial} - \sqrt{\frac{g}{2}} \bar{z}\,,\quad K_0 = L_0 + g\,,
\end{eqn}
with the difference that the relevant Lie algebra is now $\mathfrak{su}(1,1)$, whilst the operator $L_0$ is still given by \eqref{eq:SU2angularmomentum}, namely 
\begin{eqn}
    L_0 = z \partial - \bar{z} \bar{\partial}\, .
\end{eqn}
Indeed, it is easy to verify that $\{ K_\pm, K_0\}$ generate the algebra of $SU(1,1) \cong SL(2, \mathbb{R})$
\begin{eqn}\label{eq:SU11algebra}
    [K_+ , K_-] = -2K_0\, ,\qquad  [K_0 , K_\pm ] = \pm K_\pm\, .
\end{eqn}
Note that the above commutation relations differ from those of $SU(2)$ by a minus sign in $[K_+,K_-]$, cf. eq.~\eqref{eq:SU2algebra}. In terms of these, the Casimir element has the form \cite{libine2021, Bal:1996jm, etheses11236}
\begin{eqn}
    C_{SU(1,1)} = \frac{1}{2} (K_+ K_- + K_- K_+) - K_0^2 =K_+K_--K_0(K_0-1)\, ,
\end{eqn}
whereas the Hamiltonian \eqref{eq:HamiltonianH2} reads
\begin{eqn}\label{eq:HamiltonianH2algebraic}
    H_{\mathbb{H}^2} = \frac{B}{2g}\, \left(C_{SU(1,1)} + g^2\right)\, .
\end{eqn}
Using this formalism, it becomes easier to infer all the relevant properties of the energy eigenstates directly from the $\mathfrak{su}(1,1)$ algebra \cite{Bargmann:1946me, Comtet:1986ki}. Furthermore, since the symmetry group is non-compact, it features a continuous as well as a discrete part in the spectrum. The first piece corresponds to the so-called continuous principal series of $SU(1,1)$, while the second one is rather a discrete set of states similar to the one we have for the sphere \cite{Bal:1996jm, etheses11236,libine2021}. In fact, the latter set can be labeled by a pair of integers $(j, m)$ defined as 
\begin{eqn}\label{eq:diagonalizationK0C2}
    C_{SU(1,1)} & \ket{j, m} = -j (j +1)\ket{j, m}\, ,\\
    K_0 & \ket{j, m} = m \ket{j, m}\, ,
\end{eqn}
where $m=\ell+g \in [-j, \infty)$. Hence, upon using that lowest-weight states corresponding to the solutions of the equation $K_- \ket{j, m_{\rm min}}=0$ are given by functions of the form\footnote{\label{fnote:upperboundgboson}The upper bound on $n$ arises from demanding normalizability of $\psi_{j,\, m_{\rm min}}(z, \bar{z})$ close to the conformal boundary $r:=|z|^2/2g = 1$, see discussion below \eqref{eq:geodesicpolarcoords}. Indeed, the norm of the lowest-weight states is essentially determined by the following integral
\begin{eqn}
    ||\psi_{j,\, m_{\rm min}}||^2 \propto \int_0^1 dr\, r^{2n+1} (1-r^2)^{2g-2n-2}\, =\, \frac{\Gamma(n+2) \Gamma(2g-2n-1)}{2(n+1) \Gamma(2g-n)}\, ,
\end{eqn}
which is finite iff $n<g-\frac12$. Note that the discrete states are therefore present only when $g>\frac12$.}
\begin{eqn}
    \psi_{j,\, m_{\rm min}} (z, \bar{z}) \propto \left(1 - \frac{|z|^2}{2g} \right)^{g-n} \bar{z}^n\, , \qquad \text{with} \quad 0 \leq n < g -\frac12\, ,
\end{eqn}
we find that $j=n-g$, hence accounting for the \emph{discrete} energies \cite{Bolte:1990qs, Comtet:1984mm, Comtet:1986ki, Carinena:2011zz, Carinena:2012es, Dunne:1991cs}
\begin{eqn}\label{eq:discreteSpectrumH2}
    E_n = \frac{g}{R^2} \left( n + \frac{1}{2} - \frac{n(n+1)}{2g} \right)\, ,\qquad \text{with}\ \ 0 \leq n < g -\frac12\,, \quad \ell \geq -n\,.
\end{eqn}
Notice that the fact that there is no highest $K_0$-weight state may be argued from various viewpoints. For instance, one can show that the condition  $K_+ \ket{j, m_{\rm max}}=0$ yields no normalizable solution on the disc with positive energy. Relatedly, using  \eqref{eq:SU11algebra} one finds
\begin{eqn}
    K_+ \ket{j, m} = \sqrt{(m+j+1)(m-j)}\ket{j, m}\, ,
\end{eqn}
which is annihilated when $m=j, -j-1$. Hence, since both are smaller than $m_{\rm min}=-j$, these states cannot be obtained from repeatedly acting on $\ket{j, -j}$ with the raising operator.

On the other hand, the additional type of unitary irreducible representations of $SU(1,1)$ feature a \emph{continuous} set of eigenvalues for the quadratic Casimir \cite{Bargmann:1946me,Lindbland1970, Comtet:1986ki}\footnote{Note that the discrete states \eqref{eq:discreteSpectrumH2} are all such that $j<-\frac12$.}
\begin{eqn}\label{eq:jppalseries}
    j = -\frac12 + i \lambda\, , \qquad \text{with}\ \ \lambda \in \mathbb{R}_{\geq0}\, .
\end{eqn}
and $m-g\in \mathbb{Z}$. Hence, since $\overline{j\,} = -(j+1)$, we can express their associated energies as follows
\begin{eqn}\label{continSpectrum}
    E_\lambda = \frac{1}{2 R^2} \left( \frac{1}{4} + \lambda^2 + g^2 \right)\, .
\end{eqn}
The appearance of a continuous part in the spectrum is the main difference between the hyperbolic model and the spherical Landau system. Classically, these states correspond to unbounded trajectories of the particle motion in the underlying 2d space \cite{Comtet:1984mm,Comtet:1986ki, Castellano:2025yur}. It is noteworthy that in the flat space limit the continuous spectrum disappears since $E_\lambda \to \infty$, with the discrete series behaving as $E_n \to B(n+\frac12)$ (see Appendix \ref{ss:flatspacelimit} for details on this). We also note that $E_\lambda >E_n>0$ for \emph{all} energy eigenfunctions in the spectrum.

Both sets of states---i.e., those in \eqref{eq:discreteSpectrumH2} and \eqref{continSpectrum}---are infinitely degenerate with respect to the quantum number $p_{\tau_1}$, which takes values in $\mathbb{R}^+$ ($\mathbb{R}$) for discrete (continuous) energies. The infinite degeneracy can be regularized by introducing an infra-red cutoff $V_{\mathbb{H}^2}$, thereby leading to a finite density per unit area \cite{Comtet:1984mm,Comtet:1986ki}
\begin{eqn}
\rho_n = \frac{V_{\mathbb{H}^2}}{2\pi R^2} \left(g - n - \frac{1}{2} \right)\, ,
\end{eqn}
for the discrete modes, as well as \cite{Comtet:1984mm}
\begin{eqn}\label{eq:densityspin0H2}
\rho_c(\lambda) = \frac{V_{\mathbb{H}^2}}{2\pi^2R^2}\, \lambda\,\text{Im} \left[ \psi\left(\frac12 +i\lambda-g\right)+
\psi\left(\frac12+i\lambda+g\right) \right]\, ,
\end{eqn}
for the principal series, where $\psi(x) = d \log \Gamma(x) /dx$ is the Euler $\psi$-function.

With this information, we are now ready to determine the trace of the heat kernel operator constructed from the scalar Hamiltonian \eqref{eq:HamiltonianH2}. Using the above spectrum, one arrives at 
\begin{eqn}\label{eq:heatkernelAdS2_scalar}
\mathcal{K}^{(0)}_{\mathbb{H}^2}(\tau) := \Tr\, \left[e^{- \tau  H_{\mathbb{H}^2}}\right] = \sum_{n=0}^{\lfloor g-\frac12\rfloor} \rho_n\  e^{-\tau E_n} + \int_0^{\infty} d\lambda \, \rho_c(\lambda)\ e^{-\tau E_\lambda}\, ,
\end{eqn}
where we have separated the contributions due to the finite number of discrete Landau levels and the continuum set of $\delta$-normalizable states \cite{Comtet:1984mm, Comtet:1986ki,Pioline:2005pf,Grosche:2004jy,Carinena:2011zz,Carinena:2012es,Anninos:2019oka}. Notice that, from parity considerations, one may rewrite the piece due to the continuous states as an integral along the full real axis $\lambda\in\mathbb{R}$, as follows \cite{Comtet:1984mm, Pioline:2005pf}
\begin{eqn}\label{eq:heatkernelH2spin0continuous}
\mathcal{K}^{(0)}_{\mathbb{H}^2}(\tau) \supset -\frac{iV_{\mathbb{H}^2}}{(2\pi R)^2} \int_{-\infty}^{\infty} d\lambda\, \lambda\, \left[ \psi \left(\frac12 +i \lambda-g \right) + \psi\left(\frac12+i \lambda+g\right) \right] e^{-\tau E_\lambda}\, .
\end{eqn}
Next, we observe that the integrand has simple poles whenever the argument of $\psi(z)$ takes a non-positive integer value, namely when\footnote{Notice that, when using \eqref{eq:densityspin0H2} instead, the set of poles gets enlarged to $\lambda = i \left( n + \frac12 \pm g \right)$, with $n\in \mathbb{Z}$.}
\begin{eqn}\label{nun}
\lambda = i \left( n + \frac12 \pm g \right)\, ,\qquad \text{with}\quad n\in \mathbb{Z}_{\geq0}\,.
\end{eqn}
%
\begin{figure}[t!]
	\begin{center}
		\includegraphics[scale=0.33]{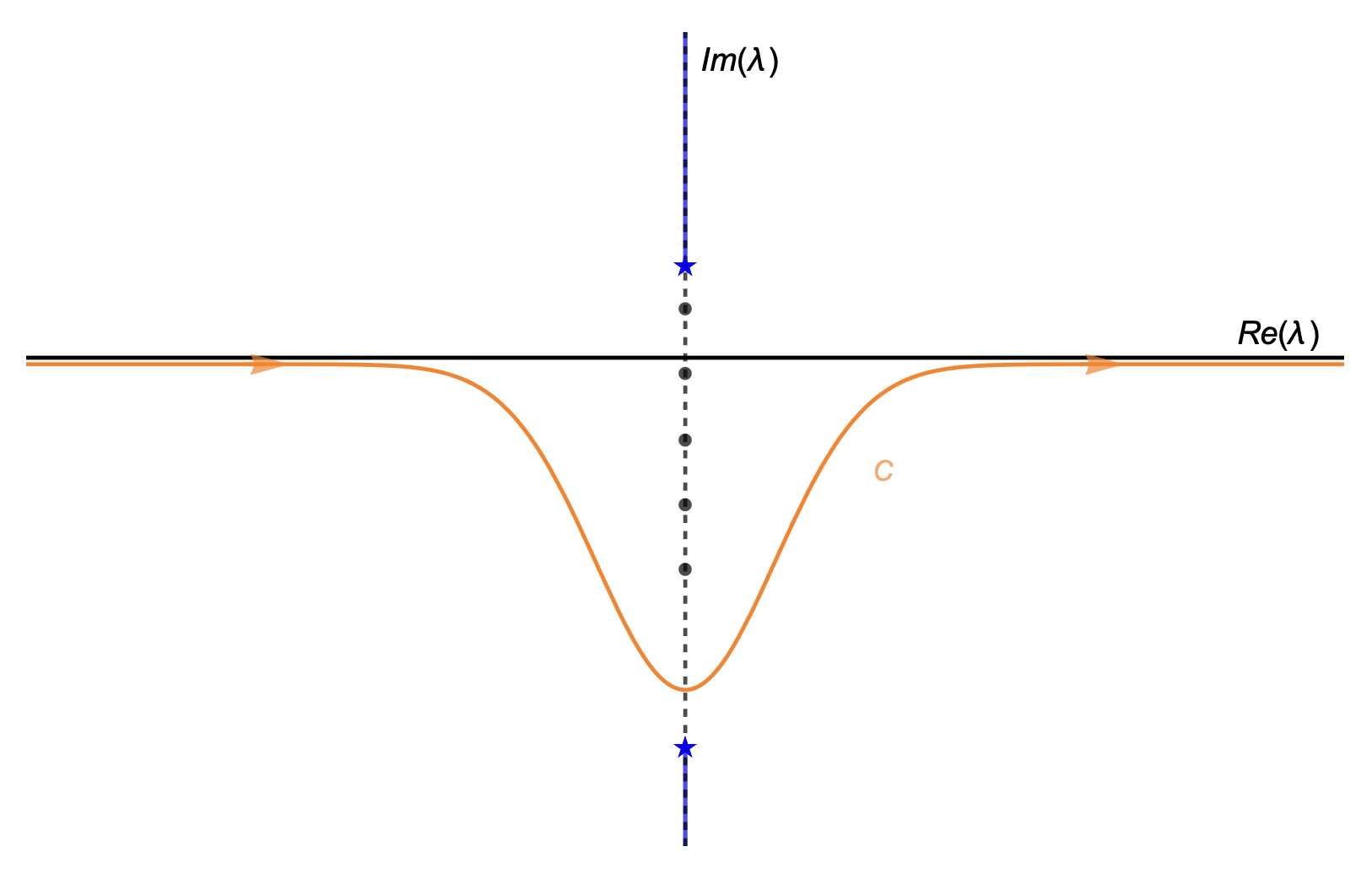}
		\caption{\small Complex $\lambda$-plane for the variable parametrizing the continuous $\mathbb{H}^2$ principal series \eqref{eq:jppalseries}. The dots denote the (simple) poles of the integrand in the heat kernel $\mathcal{K}^{(0)}_{\mathbb{H}^2}$ (cf. eq.~\eqref{eq:heatkernelH2spin0continuous}), whereas the crosses mark the branch points of the 1-loop amplitude \eqref{eq:logZphi}, which occur when $(mR)^2+E_\lambda <0$. The contour $\mathcal{C}$ is chosen so as to enclose the poles responsible for the discrete series \eqref{eq:discreteSpectrumH2}.}
		\label{fig:singusH2Kernel}
	\end{center}
\end{figure}
%
For sufficiently small fields---i.e., when $g<1/2$---all singularities lie in the upper-half $\lambda$-plane. However, as we increase $g$ past $1/2$, some of the poles associated to $\psi \left(\frac12 +i \lambda-g \right)$ cross into the lower-half plane (see Figure \ref{fig:singusH2Kernel}). Furthermore, the residue of the integrand at those points precisely equals the contribution of the discrete states with quantized energy $E_n$.\footnote{This can be checked by taking into account that the residues of the function $\psi \left(\frac12 +i \lambda-g \right)$ at the poles $\lambda=i(n+\frac12 -g)$ for $0\leq n < g - \frac12$ are all equal to $i$, which trivially follows from the property $\gamma(x)=\gamma(x+1)-\frac{1}{x}$.} Therefore, the full spectrum may be combined into a single expression by deforming the integration contour towards the lower-half $\lambda$-plane so that it goes below the poles at $\lambda = i (n + \frac12 - g)$ for any $n\geq 0$. The resulting heat kernel $\mathcal{K}^{(0)}_{\mathbb{H}^2}(\tau)$ may thus be written more compactly as \cite{Comtet:1984mm,Pioline:2005pf}
\begin{eqn}\label{eq:heatkernelH2complete}
\mathcal{K}^{(0)}_{\mathbb{H}^2}(\tau) =-\frac{iV_{\mathbb{H}^2}}{(2\pi R)^2}  
\int_{\mathcal{C}} 
d\lambda\, \lambda\, \left[ \psi\left( \frac12 +i \lambda-g\right) + \psi\left(\frac12 +i \lambda+g \right) \right] \ e^{-\tau E_\lambda}\, ,
\end{eqn}
where $\mathcal{C}=- i (g-\frac12+\epsilon) + \lambda$, with $\lambda \in\mathbb{R}$ and $\epsilon \to 0^+$.

\subsubsection*{The spin-$\frac12$ case}\label{sss:fermionH2}

Similarly to what we did for the sphere, let us consider here the analogous quantum problem involving spin-$\frac12$ fermions constrained to live in $\mathbb{H}^2$. Using the metric and the gauge connection \eqref{eq:H2metriczcoords} in stereographic coordinates, we find the Dirac operator squared to be given by
\begin{equation}\label{eq:DiracOpSquaredH2}
\begin{aligned}
\slashed{D}^2_{\mathbb{H}^2} =&\, 2B \left[ -\left( 1 - \frac{|z|^2}{2g} \right)^2 \partial \bar{\partial} - \frac{1}{2}  \left( 1 - \frac{|z|^2}{2g} \right) (z\partial - \bar{z} \bar{\partial}) + \frac{1}{4} |z|^2  \right]\\
& + \frac{1}{R^2} \left[ g\, \frac{\left( 1 - |z|^2/2g \right)^2}{8 |z|^2} - \frac14 - g\sigma_z \frac{1-|z|^4/4g^2}{2|z|^2} (z\partial - \bar{z} \bar{\partial}) +\frac{g}{2}\, \left( 1 + |z|^2/2g \right) \sigma_z\right]-B\sigma_z\, .
\end{aligned}
\end{equation}
Here, one can also construct a set of $SU(1,1)$ operators (cf. eq.~\eqref{eq:OperatorSU11boson}) 
\begin{eqn}\label{eq:OperatorSU11fermi}
\begin{aligned}
    K_+ &= - \frac{1}{\sqrt{2g}}z^2 \partial + \sqrt{2g}\, \bar{\partial} - \sqrt{\frac{g}{2}}\, z + \sqrt{\frac{g}{2}}\, \frac{1 - |z|^2/2g}{\bar{z}} \frac{\sigma_z}{2}\, ,\\
    K_- &= -\sqrt{2g}\, \partial + \frac{1}{\sqrt{2g}} \bar{z}^2 \bar{\partial} - \sqrt{\frac{g}{2}}\, \bar{z} + \sqrt{\frac{g}{2}}\, \frac{1 - |z|^2/2g}{z} \frac{\sigma_z}{2}\, ,\\
    K_0 &= z \partial - \bar{z} \bar{\partial} + g\, ,
    \end{aligned}
\end{eqn}
satisfying the algebra \eqref{eq:SU11algebra}, and whose associated quadratic Casimir is related to the square of the Dirac operator via
\begin{eqn}\label{eq:casimirfermiAdS2}
\begin{aligned}
    C_{SU(1,1)} = R^2\slashed{D}^2_{\mathbb{H}^2}-g^2+\frac14\, .
    \end{aligned}
\end{eqn}
Therefore, the energy spectrum may be obtained by diagonalizing simultaneously $K_0$ and $C_{SU(1,1)}$ (cf. eq.~\eqref{eq:diagonalizationK0C2}), taking also into account that the corresponding wavefunctions are bispinors. The discrete principal series is such that $j=n-g-\frac12$ for $0\leq n < g$, whereas $m\in [-j, \infty)$ \cite{Comtet:1984mm}. This can be deduced upon considering the equation $K_-\ket{j, m_{\rm min}}=0$, whose solutions read
\begin{eqn}\label{eq:algebraiceigenfunctionsAdS2}
    \psi^{\pm}_{j,\, m_{\rm min}} (z, \bar{z}) \propto \left(1 - \frac{|z|^2}{2g} \right)^{g-n+\frac12}\, |z|^{\pm \frac12}\, \bar{z}^{n-\frac12}\, , \qquad \text{with} \quad 0 \leq n < g\, ,
\end{eqn}
where the restriction on the quantum number $n$ comes from demanding normalizability of the wavefunction close to the boundary (see footnote \ref{fnote:upperboundgboson}). This implies, in turn, that the discrete eigenvalues of the Dirac operator squared on $\mathbb{H}^2$ are
\begin{eqn}\label{eq:eigenvaluesdiscreteDiracAdS2}
    E_n^2= \frac{n(2g-n)}{R^2}\, ,\qquad \text{with}\ \ 0 \leq n < g\,, \quad \ell \geq -n+ \frac12\,.
\end{eqn}
Introducing the infra-red cutoff $V_{\mathbb{H}^2}$, these states have a regularized density \cite{Comtet:1984mm}
\begin{equation}
    \rho_n = \frac{V_{\mathbb{H}^2}}{2 \pi R^2} \left(2(g-n) - \delta_{n,0} \, g\right) \,,
\end{equation}
which can be divided into \emph{unpaired} zero modes with spectral density $g/2\pi R^2$ (as per Atiyah-Singer), together with a tower of \emph{paired} states with spectral density given by $(g-n)/2\pi R^2$. Similarly, the continuous principal series is characterized by exhibiting a Casimir eigenvalue controlled by \cite{Comtet:1984mm}
\begin{eqn}
    j = -\frac12 + i \lambda\, , \qquad \text{with}\ \ \lambda \in \mathbb{R}_{\geq0}\, ,
\end{eqn}
thereby yielding the following energy spectrum 
\begin{eqn}\label{eq:eigenvaluescontinuousDiracAdS2}
    E_\lambda^2 = \frac{1}{R^2} \left( \lambda^2 + g^2 \right)\, ,
\end{eqn}
and whose associated density is \cite{Comtet:1984mm}
\begin{equation}\label{eq:continuousDensityFermiH2}
    \rho_c (\lambda) =   \frac{V_{\mathbb{H}^2}}{2\pi^2 R^2}\, \lambda\,\text{Im} \left[ \psi\left(i\lambda-g\right)+
    \psi\left(i\lambda+g\right) + \psi\left(i\lambda-g+1\right)+ \psi\left(i\lambda+g+1\right)\right]\,.
\end{equation}
Notice that in the flat space limit, as expected, the continuous states disappear since $E_\lambda^2 \to \infty$, whilst the discrete series behave as $E_n^2 \to 2Bn$ (see Appendix \ref{ss:flatspacelimit} for details).  

Accordingly, one may compute the trace of the heat kernel associated with \eqref{eq:DiracOpSquaredH2} to be
\begin{equation} \label{eq:heattracespin12H2}
    \mathcal{K}^{(1/2)}_{\mathbb{H}^2}(\tau) :=  \Tr\, \left[e^{ -\tau \slashed{D}^2_{\mathbb{H}^2}}\right] = \sum_{n=0}^{\lfloor g\rfloor} \rho_n\  e^{-\tau E_n^2} + \int_0^{\infty} d\lambda \, \rho_c(\lambda)\ e^{-\tau E_\lambda^2}\,.
\end{equation}
As also happened in the scalar case, upon expanding explicitly the imaginary part in \eqref{eq:continuousDensityFermiH2}, we can rewrite the integral piece of the trace above as follows
\begin{eqn}\label{eq:heatkernelH2spin12continuous}
\mathcal{K}^{(1/2)}_{\mathbb{H}^2}(\tau) \supset -\frac{iV_{\mathbb{H}^2}}{(2\pi R)^2} \int_{-\infty}^{\infty} d\lambda\, \lambda e^{-\tau E_\lambda} \left[ \psi\left(i\lambda-g\right)+
\psi\left(i\lambda+g\right) + \psi\left(i\lambda-g+1\right)+ \psi\left(i\lambda+g+1\right) \right]\, .
\end{eqn}
The singularity structure of the resulting integrand is shown in Figure \ref{3a} below, where one finds two infinite towers of simple poles of the form
\begin{eqn}\label{eq:polesspin12H2}
    \lambda_{\pm, 1}=i (n \pm g)\,,\quad \lambda_{\pm, 2} =i(n+1 \pm g)\,,\qquad \text{with}\quad  n \in \mathbb{Z}_{\geq 0}\, .
\end{eqn}
Note that some of these poles will lie below the real $\lambda$-axis. Therefore, by deforming the contour of integration in \eqref{eq:heatkernelH2spin12continuous} towards the lower-half (complex) $\lambda$-plane, one picks up some residues that reproduce the discrete piece appearing in \eqref{eq:heattracespin12H2}. This allows us to write
\begin{eqn}\label{eq:heatkernelH2completeFermions}
\mathcal{K}^{(1/2)}_{\mathbb{H}^2}(\tau) =-\frac{iV_{\mathbb{H}^2}}{(2\pi R)^2}  
\int_{\mathcal{C}} 
d\lambda\, \lambda e^{-\tau E_\lambda} \left[ \psi\left(i\lambda-g\right)+
    \psi\left(i\lambda+g\right) + \psi\left(i\lambda-g+1\right)+ \psi\left(i\lambda+g+1\right) \right]\,,
\end{eqn}
where $\mathcal{C}=- i (g+\epsilon) + \lambda$, with $\lambda \in\mathbb{R}$ and $\epsilon \to 0^+$. 

\subsubsection{Analytic continuation to AdS$_2$}\label{sss:analyticcontAdS2}

\subsubsection*{The spin-0 case}\label{sss:bosonAdS2}

\begin{figure}[t!] \centering
\subcaptionbox{\label{2a}}{
	\includegraphics[width=0.4\linewidth, height = 6cm]{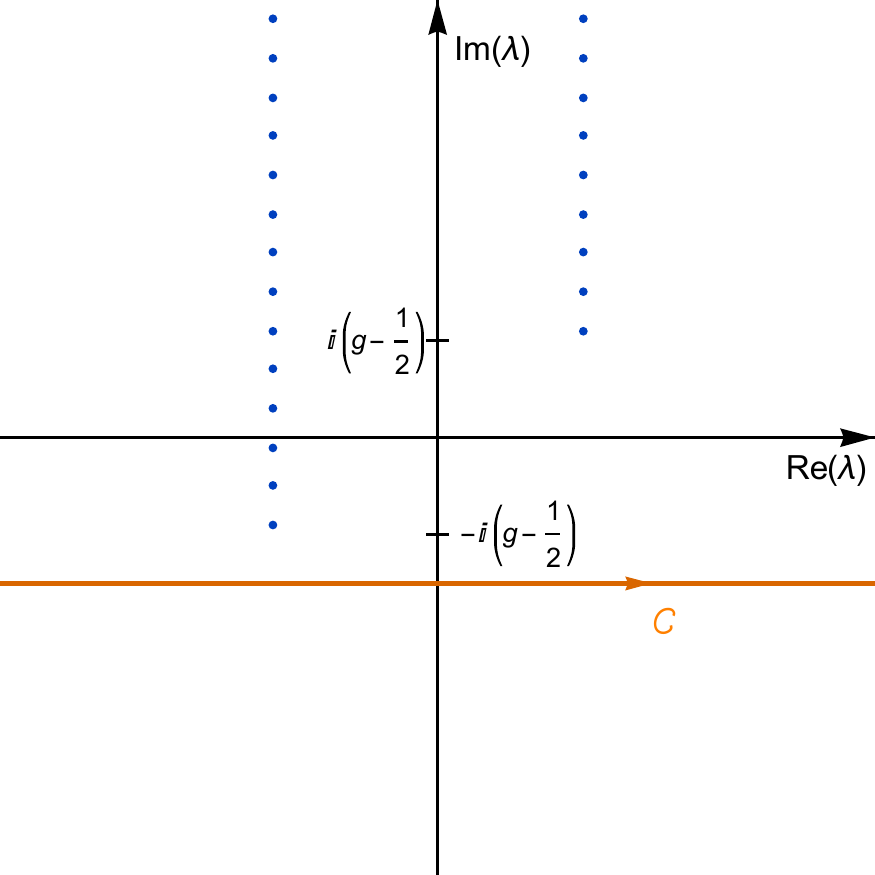}}
       \hspace{0.3cm}
\subcaptionbox{\label{2b}}{\includegraphics[width=0.4\linewidth, height = 6cm]{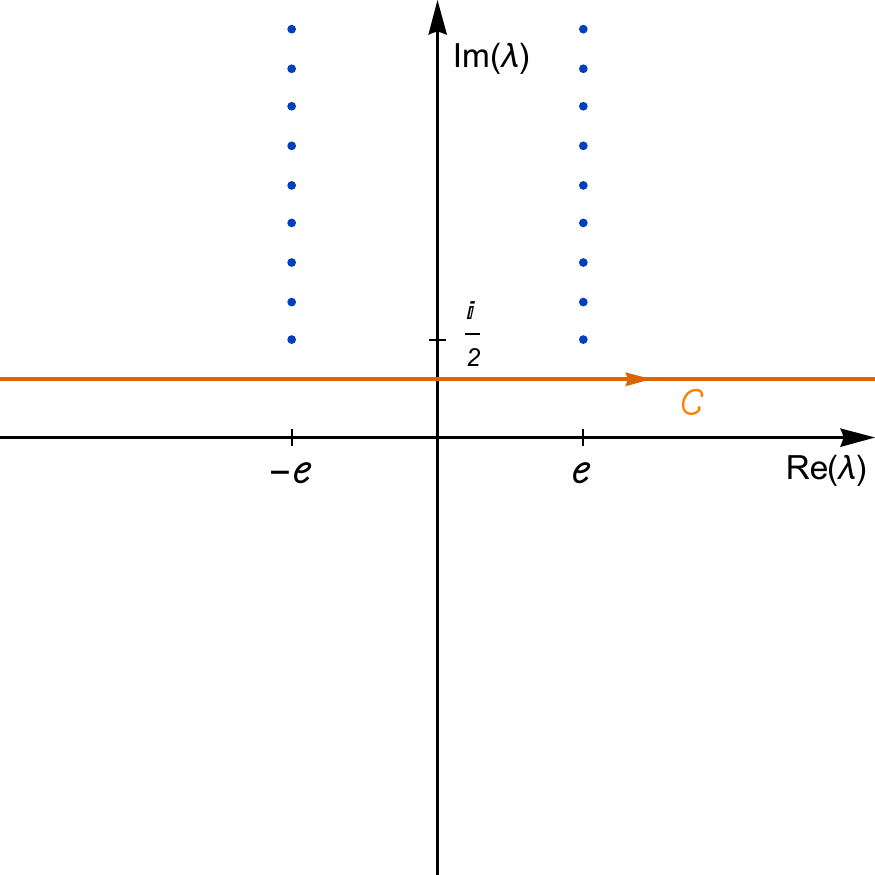}}
\caption{\small Singularity structure of the density of (continuous) energy states for a charged spin-0 particle in $\mathbb{H}^2$ (left) and AdS$_2$ (right), when expressed as in eqs.~\eqref{eq:heatkernelH2spin0continuous} and \eqref{eq:heatkernelAdS2spin0complete}. The $\mathbb{H}^2$ poles have been slightly separated from the imaginary axis for clarity. The red lines indicate the contour of integration $\mathcal{C}=- i (g-\frac12+\epsilon) + \mathbb{R}$ (left) and its continuation $\mathcal{C}'= \frac{i}{2}-i\epsilon + \mathbb{R}$ (right) via \eqref{eq:H2->AdS2continuation}.} \label{fig:Spin0PolesH2&AdS2} 
\end{figure}

Let us finally return to the original Lorentzian problem posed at the beginning of Section \ref{ss:spectralAdS2}. The kinetic operator for a charged spin-0 field in AdS$_{2}$ reads (ignoring the mass term)
\begin{eqn}\label{eq:helec}
\mathcal{D}_{\text{AdS}_2}^2= -\frac{\rho^2}{R^2} \left[ \partial_\rho^2 - (\partial_t - i e/\rho)^2 \right]\, .
\end{eqn}
To compute its associated heat kernel $\mathcal{K}^{(0)}_{\rm AdS}(\tau)$, we perform an analytic continuation of the magnetic one for the hyperbolic plane already obtained in \eqref{eq:heatkernelH2complete}, following closely ref. \cite{Pioline:2005pf}. One first notices that the worldline Hamiltonian $H_{\text{AdS}_2} = \mathcal{D}_{\text{AdS}_2}^2/2$ is transformed into the corresponding Landau analogue in $\mathbb{H}^2$ (cf. \eqref{eq:HamiltonianH2})
\begin{eqn}
\label{maglap}
H_{\mathbb{H}^2}=\frac12\mathcal{D}^2_{\mathbb{H}^2}  =  \frac12(-i \nabla - A)^2 =  - \frac{\tau_2^2}{2R^2 } \left[ \partial_{\tau_2}^2 + (\partial_{\tau_1} - i g/\tau_2)^2 \right]\,,
\end{eqn}
upon identifying
\begin{eqn}\label{eq:H2->AdS2continuation}
\tau_1 = i t\, ,\qquad g = -i e\,, \qquad H \stackrel{!}{=} -\frac12 m^2\, .
\end{eqn}
However, by doing so, we must be extremely careful to keep track of the appropriate changes occurring in both the singularity structure of the integrand and the integration contour. Indeed, one readily observes that the relevant poles now appear at (see Figure \ref{2b})
\begin{eqn}\label{eq:polesSpin0AdS2}
    \lambda_\pm = i\left(n+\frac12\right) \pm e\,,\qquad n \in \mathbb{Z}_{\geq 0}\, ,
\end{eqn}
thus comprising two symmetric towers lying entirely within the upper-half (complex) $\lambda$-plane, with `energies' given by
\begin{eqn}
    2R^2 E_{\lambda_\pm}= \pm 2i e \left(n+\frac12\right)- n(n+1)\, .
\end{eqn}
Note that the fact that these are no longer real-valued is not a peculiarity of the present setup. For instance, the magnetic Landau levels in $\mathbb{R}^2$ also exhibit purely imaginary energies once we Wick rotate to $\mathbb{R}^{1,1}$ and send $B \to -iE$, as in eq.~\eqref{eq:H2->AdS2continuation} above (see Appendix \ref{ss:flatspacelimit}). 

Similarly, the integration contour defined in eq.~\eqref{eq:heatkernelH2complete}, which was introduced so as to incorporate the effect of the discrete spectrum, gets deformed into a straight line lying slightly below the poles \eqref{eq:polesSpin0AdS2} and within the upper-half plane, as shown in Figure \ref{fig:Spin0PolesH2&AdS2}. We can hence freely deform the latter towards the real axis, and obtain
\begin{eqn}
\label{eq:heatkernelAdS2spin0complete}
\Tr\, \left[e^{- \tau H_{\text{AdS}_2}}\right] = -\frac{iV_{\text{AdS}}}{(2\pi R)^2}  
\int_{\mathbb{R}} 
d\lambda\, \lambda\, \left[ \psi\left( \frac12 +i (\lambda-e) \right) + \psi\left(\frac12 +i (\lambda+e) \right) \right] \ e^{-\tau E_\lambda}\,, 
\end{eqn}
where 
\begin{eqn}
    2R^2 E_\lambda=\lambda^2+\frac{1}{4}-e^2\, . 
\end{eqn}
Reversing now the argument that took us from eq.~\eqref{eq:heatkernelAdS2_scalar} to \eqref{eq:heatkernelH2complete}, this may be equivalently rewritten as a line integral over the continuous spectrum associated with the Klein-Gordon operator \eqref{eq:helec}
\begin{eqn}\label{KE2}
\mathcal{K}^{(0)}_{\rm AdS}(\tau) =\Tr\, \left[e^{-\tau H_{\text{AdS}_2}}\right] = \int_0^{\infty} 
d\lambda\, \rho_B(\lambda)\, e^{-\tau E_\lambda} \,,
\end{eqn}
with the density\footnote{We stress that this density (and the analogous spin-$\frac12$ one appearing in \eqref{eq:tracefermiAdS2ppalvalue}) is non-negative $\forall\, \lambda, g \in \mathbb{R}$.} of states being captured by (see Appendix \ref{ap:densityAdS2&polygamma} for a careful derivation)
\begin{eqn}\label{eq:densityspin0AdS2}
\rho_B(\lambda) = \frac{V_{\text{AdS}}}{2\pi R^2}\, 
\frac{\lambda\, \sinh(2 \pi \lambda)}{ \cosh(2 \pi \lambda) + \cosh(2 \pi e)}\,.
\end{eqn}

\subsection*{The spin-$\frac12$ case}\label{sss:fermionAdS2}

\begin{figure}[t!] \centering
\subcaptionbox{\label{3a}}{
	\includegraphics[width=0.4\linewidth, height = 6cm]{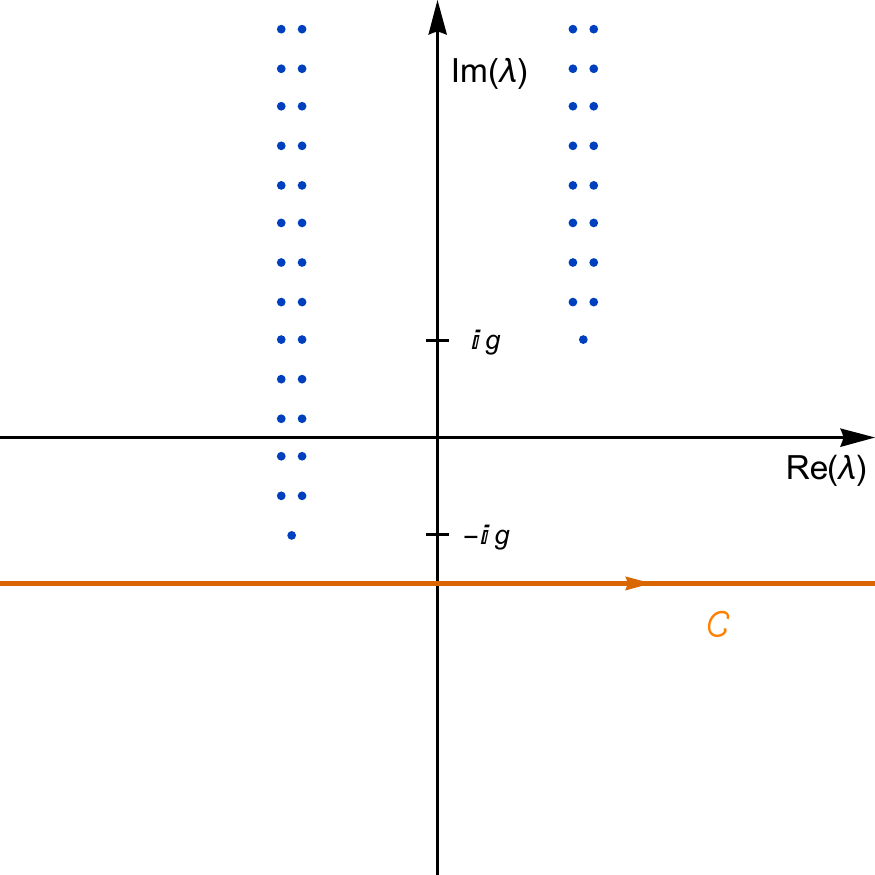}}
       \hspace{0.3cm}
\subcaptionbox{\label{3b}}{\includegraphics[width=0.4\linewidth, height = 6cm]{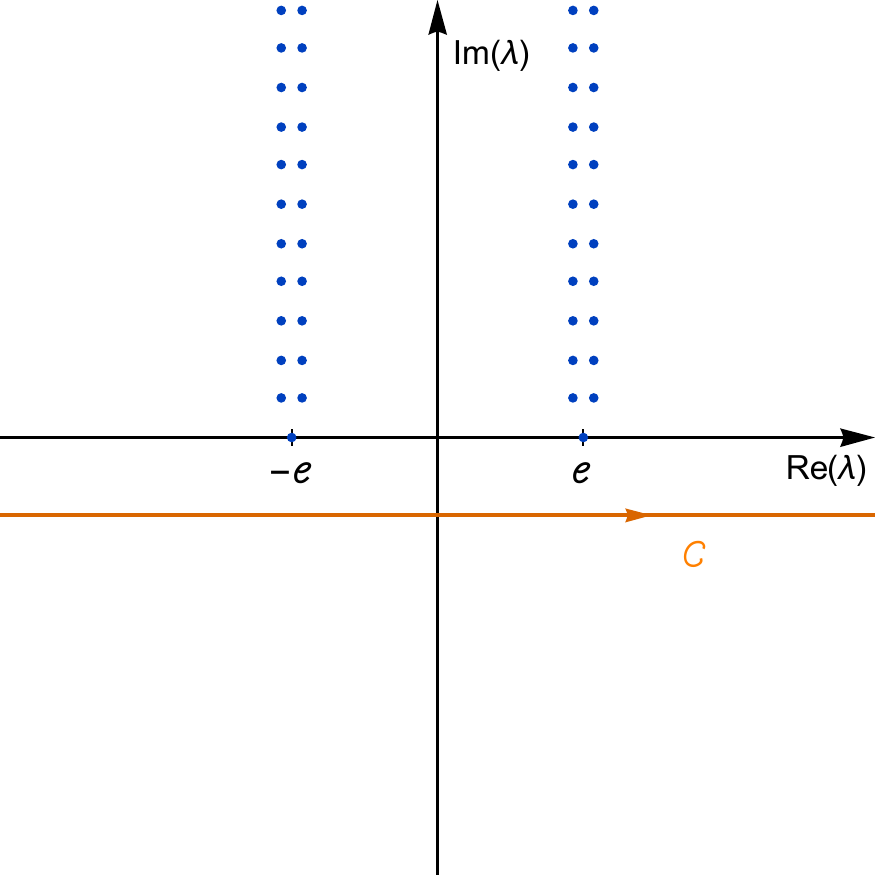}}
\caption{\small Singularity structure of the density of (continuous) energy states for a charged spin-$\frac12$ particle in $\mathbb{H}^2$ (left) and AdS$_2$ (right), when expressed as in eqs.~\eqref{eq:heatkernelH2spin12continuous} and \eqref{eq:heatkernelAdS2spin12complete}. The $\mathbb{H}^2$ poles have been slightly separated from the imaginary axis for clarity. The red lines indicate the contour of integration $\mathcal{C}=- i (g+\epsilon) + \mathbb{R}$ (left) as well as its analytic continuation $\mathcal{C}'= -i\epsilon + \mathbb{R}$ (right) via \eqref{eq:H2->AdS2continuation}.} \label{fig:Spin12PolesH2&AdS2} 
\end{figure} 

In order to perform the analytic continuation to AdS$_2$ for a spin-$\frac{1}{2}$ field, we proceed analogously to the scalar case. After transforming the background according to \eqref{eq:H2->AdS2continuation}, the poles of the integrand in \eqref{eq:heatkernelH2completeFermions} get reshuffled, yielding two symmetric towers parametrized by
\begin{eqn}
    \lambda_{\pm, 1}=i n \pm e\,,\quad \lambda_{\pm, 2} =i(n+1) \pm e\,,\qquad \text{with}\quad n \in \mathbb{Z}_{\geq 0}\, ,
\end{eqn}
among which one finds two zero modes located at $\lambda=\pm e$, cf. Figure \ref{3b}. Notice that the integration path now lies slightly below the real axis, such that by deforming it towards the latter one ends up with two contributions, namely the principal value of the integral along $\mathbb{R}$ as well as (half the) residues of the poles at $\lambda=\pm e$. Crucially, these two cancel against each other, since they are given, respectively, by
\begin{eqn}\label{eq:residueszeromodes}
     \pm i\frac{V_{\text{AdS}}}{4\pi R^2}\,e\, . 
\end{eqn}
Consequently, the correct analytic continuation of the full heat kernel trace in $\mathbb{H}^2$ for the spin-$\frac12$ particle reads
\begin{eqn}
\label{eq:heatkernelAdS2spin12complete}
\begin{aligned}
\Tr\, \left[e^{- \tau \slashed{D}_{\rm{AdS}_2}^2}\right] =-\frac{iV_{\text{AdS}}}{(2\pi R)^2}\,  
\text{P.V.}\, \bigg\{&\int_{-\infty}^{\infty} 
d\lambda\, \lambda\, \big[ \psi\left(i(\lambda-e)\right)+
\psi\left(i(\lambda+e)\right)\\
& + \psi\left(i(\lambda-e)+1\right)+ \psi\left(i(\lambda+e)+1\right) \big] \ e^{-\tau E_\lambda}\bigg\}\, ,
\end{aligned}
\end{eqn}
which, thanks to parity, together with the identity \eqref{eq:densityspin12H2&AdS2}, can be written as
\begin{eqn}\label{eq:tracefermiAdS2ppalvalue}
\begin{aligned}
\mathcal{K}^{(1/2)}_{\rm AdS}(\tau) =\Tr\, \left[e^{- \tau \slashed{D}_{\rm{AdS}_2}^2}\right] =  
\text{P.V.}\, \bigg\{&\int_{0}^{\infty} 
d\lambda\, \rho_F(\lambda)\,  e^{-\tau E_\lambda^2}\bigg\}\,,
\end{aligned}
\end{eqn}
with
\begin{eqn} \label{eq:densityspin12AdS2}
    \rho_F(\lambda) &= \frac{V_{\text{AdS}}}{\pi R^2} \frac{\lambda\,\sinh(2\pi \lambda)}{\cosh\left(2\pi \lambda\right) - \cosh\left(2\pi e\right)} \,,
\end{eqn}
the corresponding density of eigenstates, whose energies are given by
\begin{eqn} \label{eq:spectrumspin12AdS2}
    R^2 E_\lambda &= \lambda^2 - e^2\,.
\end{eqn}

\subsection{Summary and conventions} \label{ss:SummaryConventions}

Before closing, let us summarize the main results that will be used in later parts of this work.
In this section, we considered the following two-dimensional operators
\begin{equation}\label{eq:opsS2&AdS2}
    \mathcal{D}^2 = -(\nabla - i A)^2 \,, \qquad \qquad \slashed{D}^2= - (\slashed{\nabla}-i \slashed{A})^2 \,,
\end{equation}
and we determined their spectrum in $\mathbf{S}^2$ and $\text{AdS}_2$. In particular, $\mathcal{D}^2 + m^2$ gives the kinetic operator for a spin-$0$ particle minimally coupled to gravity and a $U(1)$ gauge field, whereas $\slashed{D}^2 + m^2$ corresponds to (the square of the) kinetic operator for a similar spin-$\frac{1}{2}$ field. 

In our conventions, the $\mathbf{S}^2$ background is characterized by
\begin{align}\label{eq:MetricMaxwellSphere}
     ds^2 = R^2_{\mathbf{S}} \left(d\theta^2 + \sin^2\theta\, d\phi^2\right) \,,\qquad F = B\, \omega_{\mathbf{S}^2} = g\, \sin \theta\,d\theta \wedge d\phi\,, 
\end{align}
where $R_{\mathbf{S}}$ denotes the radius of the sphere, and $B = g/R_{\mathbf{S}}^2$ is the constant magnetic field, with $g \in \mathbb{Z}/2$ the quantized magnetic charge. The heat kernel trace associated with $\mathcal{D}^2$, $\slashed{D}^2,$ on the sphere can be computed using the eigenvalues $E_n$ and the degeneracies $d_n$
\begin{subequations}
\begin{align}
    & \underline{\text{spin-}0}: \qquad  E_n =  \frac{2 g}{R_{\mathbf{S}}^2} \left[ n + \frac{1}{2} + \frac{n(n+1)}{2g} \right]\,, \qquad  d_n = 2\left( g + n + \frac{1}{2} \right)\,, \label{summary:bosonS2}\\[2mm]
    & \underline{\text{spin-}1/2}: \qquad  E_n^2 =  \frac{1}{R^2_{\mathbf{S}}} n (n+2g)  \,, \qquad d_n =  4 n + (4 - 2  \delta_{n,0})\, g\,,\label{summary:fermionS2}
\end{align}
\end{subequations}
with  $n \in \mathbb{Z}_{\ge0}$, and reads 
\begin{subequations}\label{eq:heatkernelS2_scalar&fermion}
\begin{align}
    & \mathcal{K}^{(0)}_{\mathbf{S}^2}(\tau)  = \sum_{n} d_n\, e^{-\tau E_n} = 2\, \sum_{n \geq 0} \left( 2n + 2g + 1 \right)\, e^{ -\frac{\tau}{R^2} \left( g + n(n+1+2g)\right)}\,,\\[2mm]
 & \mathcal{K}^{(1/2)}_{\mathbf{S}^2}(\tau)= \sum_{n} d_n\, e^{-\tau E^2_n} = 4\sum_{n} \left( n + g\right) e^{ -\frac{\tau}{R^2} n\left( n + 2g \right)} + 2g\,.
\end{align}
\end{subequations}
Similarly, the $\text{AdS}_2$ background is described by 
\begin{align}\label{eq:MetricMaxwellAdS}
ds^2 = \frac{R_{\rm{A}}^2}{\rho^2} \left(-dt^2 + d\rho^2\right)\,, \qquad F =  E\, \omega_{\text{AdS}_2}= e\, \frac{1}{\rho^2} dt \wedge d\rho\, ,
\end{align}
where $R_{\rm{A}}$ is the AdS radius, and $E = e/R_{\rm{A}}^2$ the constant electric field. The heat kernel traces for the operators in \eqref{eq:opsS2&AdS2} can be computed via the eigenvalues $E_\lambda$ and energy densities $\rho_\lambda$\footnote{Remember that the density is defined up to the subtlety discussed around \eqref{eq:heatkernelAdS2spin12complete}. The heat kernel for fermions requires taking a principal value.}
\begin{subequations}
\begin{align}
    & \underline{\text{spin-}0}: \qquad  E_\lambda =  \frac{1}{R_{\rm{A}}^2} \left( \lambda^2 + \frac{1}{4} - e^2 \right)\,, \qquad  \rho_B  = \frac{V_{\text{AdS}}}{2\pi R_{\rm{A}}^2} \frac{\lambda\,\sinh(2\pi \lambda)}{\cosh(2\pi \lambda) + \cosh(2\pi e)}\,,\label{summary:bosonAdS2} \\[2mm]
    & \underline{\text{spin-}1/2}: \qquad  E_\lambda^2 =  \frac{1}{R_{\rm{A}}^2} (\lambda^2 - e^2) \,, \qquad \rho_F = \frac{V_{\text{AdS}}}{\pi R_{\rm{A}}^2}  \frac{\lambda\,\sinh(2\pi \lambda)}{\cosh\left(2\pi \lambda\right) - \cosh\left(2\pi e\right)}\,,\label{summary:fermionAdS2}
\end{align}
\end{subequations}
with $\lambda \in \mathbb{R}_{>0}$, as follows
\begin{eqn}\label{eq:heatkernelAdS2_scalar&fermion}
\mathcal{K}^{(0)}_{\rm AdS}(\tau) = \int_0^{\infty} 
d\lambda\, \rho_B(\lambda)\, e^{-\tau E_\lambda}\, ,\qquad \mathcal{K}^{(1/2)}_{\rm AdS}(\tau) = \int_0^{\infty} d\lambda\, \rho_F(\lambda)\ e^{-\tau E_\lambda^2} \,.
\end{eqn}

\section{Integrating Out Charged Massive Particles in AdS$_2 \times \mathbf{S}^2$}\label{s:integrationAdS2xS2}

The aim of this section is to determine the 1-loop partition function for minimally coupled spin-0 and spin-$\frac12$ particles propagating within certain AdS$_2 \times \mathbf{S}^2$ spacetimes. This is the content of Section \ref{ss:ExactComputation}. Before doing so, we first provide a brief review of the aforementioned type of backgrounds (Section \ref{sss:nearhorizonBH}), as well as introduce the necessary techniques to perform such quantum computation (Section \ref{sss:1loopbasics}). Subsequently, in Section \ref{ss:Susic1LoopMinimal}, we specialize the calculation to the case of a BPS-like spectrum in a fully supersymmetric 4d $\mathcal{N}=2$ solution. A more complete analysis, including non-minimal couplings, will be deferred to Section \ref{s:SusyLoopDeterminant}.

\subsection{Preliminary considerations} \label{ss:Traces}

We begin by reviewing the structure of the near-horizon region of asymptotically flat charged, static, and extremal black holes in four dimensions. The spacetime geometry is AdS$_2 \times \mathbf{S}^2$, with the same radius of curvature for the two factors $R_{\rm{A}} = R_{\mathbf{S}}$. In particular, we describe an embedding of AdS$_2 \times \mathbf{S}^2$ in 4d $\mathcal{N} = 2$ supergravity with vector- and hyper-multiplets, and its realization as the near-horizon geometry of a BPS black hole. We then review the derivation of the general expression for the effective action arising upon integrating out minimally coupled, massive fields in 4d. Finally, we show how the 1-loop partition function of AdS$_2 \times \mathbf{S}^2$ can be computed in terms of the heat kernel operators of the sphere \eqref{eq:heatkernelS2_scalar&fermion} and Anti-de Sitter \eqref{eq:heatkernelAdS2_scalar&fermion}.

\subsubsection{The near-horizon (extremal) black hole geometry}\label{sss:nearhorizonBH} 

The Reissner-Nordström spacetime is the unique static, spherically symmetric solution to the Einstein-Maxwell equations with non-trivial sources \cite{nordstr,reissner}. The metric and the Maxwell field strength are given, respectively, by
\begin{equation}\label{eq:MetricGaugeFieldRN}
ds^{2}=-\frac{\Delta}{r^{2}}dt^{2}+\frac{r^{2}}{\Delta}dr^{2}+r^{2}d\Omega_{2}^{2}\,,  \qquad F = - \frac{Q_e}{r^2} dt \wedge dr + Q_m \sin\theta \, d\theta \wedge d\phi \,,
\end{equation}
where $\Delta = (r - r_{-})(r - r_{+})$, $d\Omega_{2}^{2}$ is the line element on the unit 2-sphere and $Q_{e,m}$ are the electric and magnetic charges sourced at the origin. This solution exhibits two horizons, namely the Cauchy horizon ($r=r_{-}$) and the black hole horizon ($r=r_{+}$), which are uniquely determined by the mass $M$ and the physical charges $Q_{e,m}$ of the black hole as follows
\begin{equation} r_{\pm} = M \pm \sqrt{M^{2} - Q^{2}}\, , \qquad \text{with}\quad Q^2 = Q_e^2 + Q_m^2 \,.
\end{equation}
The (Hawking) temperature of the solution is a function of the inner and outer horizon radii
\begin{equation} \label{eq:TS} 
T_{H} = \frac{r_{+} - r_{-}}{4\pi r_{+}^{2}}\,.
\end{equation}
In the \emph{extremal} case, where $M^{2} = Q^{2}$, the two horizons coincide and the temperature vanishes. Correspondingly, the line element shown in \eqref{eq:MetricGaugeFieldRN} gets simplified to
\begin{equation} 
ds^{2} = -f(r)^2 dt^{2} + f(r)^{-2} dr^{2} + r^{2} d\Omega_{2}^{2}\, , \qquad \text{with}\quad f(r)=1-\frac{r_h}{r}\,, 
\end{equation}
and $r_h = Q$. In order to study more closely the near-horizon geometry, we can define a new radial coordinate $y = r - r_{h} > 0$. Taking the near-horizon limit, parametrized by $y/r_h \ll 1$, yields the following Lorentzian metric
\begin{equation}\label{eq:BertRobmetric}
ds^{2} = -\frac{y^{2}}{Q^{2}} dt^{2} + \frac{Q^{2}}{y^{2}} dy^{2} + Q^{2} d\Omega_{2}^{2}\,, 
\end{equation} 
which corresponds to a Bertotti-Robinson spacetime of mass $M_{\rm BR}^2=Q^2$ \cite{Gibbons:1982ih,Gibbons:1987ps,Garfinkle:1990qj} and topology given by AdS$_{2} \times \mathbf{S}^{2}$. Here, the AdS$_2$ line element is written in Poincaré coordinates,\footnote{See Section 4.2.1 of \cite{Castellano:2025yur} for a discussion of the different coordinate systems and/or patches that are relevant.} and the connection with the coordinates used in Section \ref{ss:SummaryConventions} becomes manifest upon defining yet another variable $\rho=Q^2/y$, thus providing a new metric tensor and Maxwell field strength 
\begin{equation}\label{eq:conformalcoords}
ds^{2} = \frac{Q^{2}}{\rho^2} \left(- dt^{2} + d\rho^2 +\rho^2 d\Omega_{2}^{2} \right)\,, \qquad F = \frac{Q_e}{\rho^2} dt \wedge d\rho + Q_m \sin\theta \, d\theta \wedge d\phi \,.
\end{equation} 
Comparing with eqs.~\eqref{eq:MetricMaxwellSphere} and \eqref{eq:MetricMaxwellAdS}, we have $R_{\rm A} = R_{\mathbf{S}^2} = Q$ and $Q_e = e$, $Q_m = g$.

\medskip

\noindent Let us now review the supersymmetric embedding of the AdS$_2 \times \mathbf{S}^2$ backgrounds above. We are interested in four-dimensional $\mathcal{N}=2$ supergravity theories with vector- and hyper-multiplets. For concreteness, we focus on realizations via Calabi-Yau compactification of Type IIA string theory \cite{Bodner:1990zm, Ceresole:1995ca}. The low-energy theory comprises one gravity multiplet, $n_V$ Abelian vector multiplets, and $n_H$ massless, neutral hypermultiplets. Restricting ourselves to the bosonic sector, the latter includes the standard metric, as well as various scalar and $U(1)$ gauge fields. In what follows, we will concentrate just on the gravity and vector multiplets, since the black hole solutions we care about only depend on those. At two derivatives, the kinetic functions associated with the scalar and gauge bosons may appear rather complicated; however, they are fully specified by a holomorphic function referred to as the prepotential $\mathcal{F}(X^A)$. The $X^A$, with  $A = 0, \ldots, n_V,$ are homogeneous (projective) coordinates on a special K\"ahler manifold and they are related to the lowest spin components $z^i$ of the vector multiplets via $z^a = X^a/X^0$, with $a = 1,\ldots, n_V$. In total, we have $n_V + 1$ field strengths labeled as $F^A$, $n_V$ of which come from the vector multiplets and one from the gravity multiplet. In this setup, one can easily build BPS black hole geometries solving the attractor equations \cite{Ferrara:1995ih,Strominger:1996kf,Ferrara:1996dd,Ferrara:1996um,Ferrara:1997yr}. The near-horizon region has the universal form AdS$_{2} \times \mathbf{S}^2$, that in Poincaré coordinates reads 
\begin{equation}\label{eq:Poincaremetric}
 ds^2=\frac{R^2}{\rho^{2}}\left( -dt^2+d\rho^2\right)+ R^2d\Omega_2^2\, ,\qquad \text{with}\quad R^2\,=|Z_{\rm BH}|^2\, .
\end{equation}
Here, $Z_{\rm BH}$ denotes the central charge of the black hole evaluated at the attractor point
\begin{equation}\label{eq:BHCentralCharge}
    Z_{\rm BH}= e^{K/2} \left( p^A{}'\mathcal{F}_A-q_A{}' X^A\right) \,,
\end{equation}
where $\mathcal{F}_A = \partial\mathcal{F}/\partial X^A $, $K  = - \log \left( i\bar{X}^A \mathcal{F}_{A}-i X^A \bar{\mathcal{F}}_{A} \right)$ is the K\"ahler potential of the theory (from which the field space metric may be obtained via $G_{a \bar{b}}= \partial^2 K/\partial z^a \partial z^b$), and $(q_A{}', p^A{}')$ correspond to the quantized charges of the black hole. The latter are defined by
\begin{equation}\label{eq:BHcharges}
 p^A{}'=\frac{1}{4\pi}\int_{\mathbf{S}^2} F^A\, ,\qquad  q_A'=\frac{1}{4\pi}\int_{\mathbf{S}^2} G_A\, ,
\end{equation}
where $G_A$ are the magnetic duals of $F^A$. These satisfy the linear constraint $G_A^-=\bar{\mathcal{N}}_{AB}F^{B,\, -}$, with $\mathcal{N}_{AB}$ having the form \cite{Ceresole:1995ca} 
\begin{equation}
    \mathcal{N}_{AB} = \overline{\mathcal{F}}_{AB} + 2i\, \frac{(\text{Im}\, \mathcal{F})_{AC} X^C (\text{Im}\, \mathcal{F})_{BD} X^D}{X^C (\text{Im}\, \mathcal{F})_{CD} X^D}\,.
\end{equation}
More precisely, the $U(1)$ field strengths at the attractor locus are given by
\begin{equation}\label{eq:backgroundfieldsBH}
 R^2\,F^A= p^A{}' \omega_{\mathbf{S}^2} - 2\text{Re}\, CX^A \omega_{\rm{AdS}_2}\, ,\qquad R^2\, G_A= q_A' \omega_{\mathbf{S}^2} - 2\text{Re}\, C\mathcal{F}_A \omega_{\rm{AdS}_2}\, , 
\end{equation}
where
\begin{equation}\label{eq:attvalues}
 CX^A= \text{Re}\, CX^A +\frac{i}{2}p^A{}'\, ,\qquad C\mathcal{F}_A= \text{Re}\, C\mathcal{F}_A +\frac{i}{2}q_A'\, ,\qquad C=e^{K/2} \bar{Z}_{\rm BH}\,. 
\end{equation}
One can check that the background \eqref{eq:backgroundfieldsBH} lies completely along the graviphoton direction, as required by the attractor equations, see \cite{Castellano:2025yur} for details on this. This implies that a probe BPS particle with charges $(q_A, p^A)$ couples to a single effective constant gauge field aligned with the graviphoton. Therefore, the bosonic part of the 1d worldline action \cite{Billo:1999ip,Simons:2004nm} reads\footnote{We have changed slightly the convention compared to that of \cite{Castellano:2025yur,Castellano:2025rvn} when writing down the action \eqref{eq:BPSwordlineactionAdS2xS2}. In particular, the relative factor of 2 between the kinetic and gauge terms in $S_{wl}$ now appears within the latter.}
\begin{equation}\label{eq:BPSwordlineactionAdS2xS2}
 S_{wl} = -|Z_{\rm p}|  \int_\gamma d\sigma R \,\sqrt{\rho^{-2}\left(\dot{t}^2-\dot{\rho}^2\right) -\dot{\theta}^2-\sin^2\theta \dot{\phi}^2} + \frac12\int_\Sigma p^AG_A-q_A F^A\, ,
\end{equation}
where $\dot{x}^{\mu} := dx^\mu/d\sigma$, and $\sigma$ is any convenient parameter along the worldline trajectory, which we denote by $\gamma$. Similarly, $\Sigma$ corresponds to any 2d surface that ends on the worldline, whereas $m = |Z_{\rm p}|$ is the mass of the BPS particle in Planck units, with
\begin{equation}
    Z_{\rm p} = e^{K/2} \left( p^A{}\mathcal{F}_A-q_A{} X^A\right)\,,
\end{equation}
its central charge (cf. eq.~\eqref{eq:BHCentralCharge}). Notice that the last term of \eqref{eq:BPSwordlineactionAdS2xS2} can be written as 
\begin{equation}\label{eq:DyonicInteraction}
    \frac{1}{2}\int_\Sigma p^AG_A-q_A F^A = -q_e \int_\gamma \frac{dt}{\rho}-q_m \int_\gamma \cos \theta d\phi \,,
\end{equation}
with 
\begin{equation}\label{eq:chargeDefSugra}
    q_e= \text{Re}\, ( \bar{Z}_{\rm BH} Z_{\rm p} )\, ,\qquad q_m= \text{Im} \,  (\bar{Z}_{\rm BH} Z_{\rm p}) \,.
\end{equation}
and it satisfies the identity \cite{Castellano:2025yur}
\begin{equation}\label{eq:ChargeMassIdentity}
    m^2 R^2 = |Z_{\rm p} \bar{Z}_{\rm BH}|^2  = q_e^2 + q_m^2 \,.
\end{equation}
Comparing with Section \ref{ss:SummaryConventions}, we can conclude that the interaction between a BPS particle and a BPS black hole near its horizon can be equivalently described in terms of a massive probe particle with unit electric charge moving in a AdS$_2\times \mathbf{S}^2$ spacetime with $R_{\rm A} = R_{\mathbf{S^2}} = R$, threaded by effective electric and magnetic fields $E R^2 = - q_e$, $B R^2 = q_m$, and such that
\begin{equation}\label{eq:QuadraticConstraint}
m^2 = R^2 \left(E^2 + B^2\right) \,.
\end{equation}

\subsubsection{Basics of functional determinants}\label{sss:1loopbasics} 

The other important piece of information that we will need for our purposes herein concerns the exact evaluation of functional (path-)integrals and effective actions, in the presence of non-trivial background fields. To that end, we consider the 4d Lorentzian quadratic action
\begin{equation}\label{eq:QuadraticScalarAction}
    S[\varphi,\phi] = S[\varphi] + \int d^4x \sqrt{-\det g} \, \phi^\dagger \left[ -\mathcal{D}^2 - m^2 \right] \phi \,,
\end{equation}
where $\phi$ is a complex massive scalar, $\varphi$ is used to collectively denote all the remaining fields, and $\mathcal{D}^2 = -(\nabla - i A)^2$. Integrating out the former, we shall define the effective action $\Gamma[\varphi]$:
\begin{equation}
    e^{i \Gamma[\varphi]} := \int D\phi \, e^{i S[\varphi,\phi] }\,.
\end{equation}
Since $\phi$ appears quadratically in \eqref{eq:QuadraticScalarAction}, we can perform this integration via Gaussian methods 
\begin{equation}
    \int D\phi^\dagger D\phi\, \exp\left[ i \int d^4x \sqrt{-\det g}\, \phi^\dagger \left[ -\mathcal{D}^2 - m^2 \right] \phi\right] = \frac{\mathcal{N}}{\det[ -\mathcal{D}^2 - m^2 + i \varepsilon]} \,,
\end{equation}
where $\mathcal{N}$ denotes some (divergent) normalization constant and the $+i\varepsilon$ shift, with $\varepsilon >0$, enforces the standard Feynman prescription, ensuring convergence of the Gaussian integral. Using standard textbook manipulations (see e.g., \cite{Schwartz:2014sze}), one can moreover write
\begin{equation} \label{eq:definitionDeltaGamma}
     \Delta\Gamma [\varphi] := \Gamma[\varphi] - S[\varphi] = - i \log \mathcal{N} + i\, \Tr \log[ -\mathcal{D}^2 - m^2 + i \varepsilon] \,.
\end{equation}
Next, inserting the Lorentzian version of Schwinger's proper-time parameterization\footnote{The $i\varepsilon$ prescription selects the Feynman (time-ordered) propagators among the possible Green functions, thereby implementing causal boundary conditions in the diagrammatic expansion.}
\begin{eqnarray}\label{eq:lorentzianSwchwinger}
\frac{i}{{\cal O} + i \varepsilon} = \int^{\infty}_0 ds\, e^{is \left({\cal O}+i\varepsilon\right)} \,,
\end{eqnarray}
for a positive operator $\mathcal{O}$, we easily obtain the following identity
\begin{equation} \label{eq:identityDerivativeMsquare}
    \frac{\partial}{\partial m^2 }\,  i \, \Tr \log[ -\mathcal{D}^2 - m^2 + i \varepsilon] = - \int_0^\infty ds  \, e^{-i s m^2} \Tr \left[ e^{- i s \mathcal{D}^2} \right] e^{- s\varepsilon} \,,
\end{equation}
such that, upon integrating \eqref{eq:identityDerivativeMsquare} with respect to $m^2$, we can express \eqref{eq:definitionDeltaGamma} as
\begin{equation}  \label{eq:stepIntermediate}
     \Delta\Gamma[\varphi] = - i \int_{\epsilon_{\rm uv}}^\infty \frac{ds}{s}  \, e^{-i s m^2} \Tr \left[ e^{- i s \mathcal{D}^2} \right] e^{- s\varepsilon} \,.
\end{equation}
Here, $\epsilon_{\rm uv}$ is a UV cutoff that absorbs both the integration constant and the $\log {\cal N}$ factor, packaging them together within the divergent contribution associated to the limit $\epsilon_{\rm uv} \to 0$. By changing coordinates $s\rightarrow -i\tau$ and rotating the integration contour back onto the positive real axis, we recover the standard proper-time representation of the 1-loop partition function
\begin{equation} \label{eq:PartitionFunction}
      \log \mathcal{Z}_\phi := - i \Delta\Gamma[\varphi] = - \int_{\epsilon_{\rm uv}}^\infty \frac{d\tau}{\tau}  \, e^{-\tau m^2} \Tr \left[ e^{- \tau  \mathcal{D}^2} \right] e^{i \tau \varepsilon}     
\end{equation}
Notice that the regulator $\varepsilon$ is not anymore necessary, and hence can be discarded. Moreover, we can trade the cutoff $\epsilon_{\rm uv}$ appearing in the integration contour with the insertion of a smooth damping operator $\exp[-\epsilon^2/4\tau]$. Doing so, we find
\begin{equation} \label{eq:logZphi}
      \log \mathcal{Z}_\phi = - \int_{0}^\infty \frac{d\tau}{\tau}  \, e^{-\frac{\epsilon^2}{4\tau}} \, e^{-\tau m^2} \Tr \left[ e^{- \tau  \mathcal{D}^2} \right]= - \int_{0}^\infty \frac{d\tau}{\tau}  \, e^{-\frac{\epsilon^2}{4\tau}} \, e^{-\tau m^2} \mathcal{K}^{(0)}(\tau)\,,
\end{equation}
where $\mathcal{K}^{(0)}$ denotes the heat kernel trace associated with kinetic operator $\mathcal{D}^2$ \cite{Vassilevich:2003xt}. Specializing $\mathcal{Z}_\phi$ to the case of interest here, namely AdS$_2 \times \mathbf{S}^2$ threaded by constant $U(1)$ gauge fields
\begin{equation} \label{eq:background}
     ds^2 = ds^2_{\rm AdS} + ds^2_{\mathbf{S}^2}\, \qquad F =  E\, \omega_{\text{AdS}_2} + B\, \omega_{\mathbf{S}^2} \,,
\end{equation}
it is straightforward to verify that the four-dimensional kinetic operator $\mathcal{D}^2$ can be expressed in terms of two commuting operators acting on the corresponding two-dimensional subspaces 
\begin{equation}
    \mathcal{D}^2 = \mathcal{D}^2_{{\rm AdS}} + \mathcal{D}^2_{\mathbf{S}^2}  \,, \qquad [\mathcal{D}^2_{{\rm AdS}} ,\mathcal{D}^2_{\mathbf{S}^2}] = 0 \,.
\end{equation}
The trace of \eqref{eq:logZphi} can then be further decomposed as a product of traces, such that upon expressing it in terms of the heat kernel operator introduced in Section \ref{ss:SummaryConventions}, we get
\begin{equation} 
      \log \mathcal{Z}_\phi = - \int_{0}^\infty \frac{d\tau}{\tau}  \, e^{-\frac{\epsilon^2}{4\tau}} \, e^{-\tau m^2} \mathcal{K}^{(0)}_{\rm AdS}(\tau) \, \mathcal{K}^{(0)}_{\mathbf{S}^2}(\tau)  \,.
\end{equation}

\medskip

Let us now consider the case of a massive Dirac spinor minimally coupled to a given gravitational and gauge backgrounds. We start from the Lorentzian action\footnote{\label{fnote:conventionDirac}Our convention is to use the mostly plus signature, with the Dirac matrices satisfying $\{ \gamma^\mu, \gamma^\nu\}=g^{\mu \nu}$, yielding anti-hermitian $\gamma^0$ and hermitian $\gamma^i$. The fermionic kinetic operator reads $\slashed{\nabla} - i \slashed{A} - m $, and $\bar{\psi} =  i \psi^\dagger \gamma^0 $.}
\begin{equation}\label{eq:FermionAction}
    S[\varphi,\Psi] = S[\varphi] + \int d^4 x \sqrt{-\det g}\, \bar{\Psi} (i \slashed{D} - m) \Psi  \,,
\end{equation}
where $\Psi$ is a Dirac spinor, $\varphi$ collectively denotes all other dynamical fields in the theory, and $\slashed{D} = - i (\slashed{\nabla} - i \slashed{A})$. In this case, we can also integrate $\Psi$ exactly via the relation 
\begin{equation}
    \int D\bar{\Psi} D\Psi  \, \exp\left[ i \int d^4 x \sqrt{-\det g} \, \bar{\Psi} (i \slashed{D} - m) \Psi \right] = \mathcal{N}\, \det (i \slashed{D} - m + i \varepsilon)\, ,
\end{equation}
with $\mathcal{N}$ a (possibly divergent) normalization constant, and the resulting quantum corrections to the Wilsonian effective action are determined by
\begin{equation}
    \Delta\Gamma[\varphi] = -i \log \mathcal{N} - i \log \det (i \slashed{D} - m + i \varepsilon)  \,.
\end{equation}
Furthermore, using that\footnote{A simple way to prove this is provided in \cite{Dunne:2007rt}. One uses the fact that $\gamma_5^2 = 1$, and that it commutes with $-m + i \varepsilon$ but anti-commutes with $\slashed{D}$, together with the standard properties of the determinant.}
\begin{equation}
    \det (i \slashed{D} - m + i \varepsilon)  = \det (-i \slashed{D} - m + i \varepsilon) \,,
\end{equation} 
allows us to write 
\begin{equation}
    \Delta\Gamma[\varphi]  = -i \log \mathcal{N} - \frac{i}{2}\, \Tr \log ( \slashed{D}^2 + m^2 - i \varepsilon')  \,,
\end{equation}
where we dropped terms of $\mathcal{O}(\varepsilon^2)$ and we introduced a new positive regulator $\varepsilon' = 2 m \varepsilon$. Repeating the same steps we did to obtain \eqref{eq:stepIntermediate} from eq.~\eqref{eq:definitionDeltaGamma}, we eventually arrive at
\begin{equation} \label{eq:stepIntermediateFermions}
\Delta\Gamma[\varphi] = \frac{i}{2}\, \Tr  \int_{\epsilon_{\rm uv}}^\infty \frac{ds}{s}\, e^{i s (- \slashed{D}^2 - m^2 + i \varepsilon')} \,,
\end{equation}
after absorbing again both the integration constant and $\mathcal{N}$ into the UV cutoff $\epsilon_{\rm uv}$. Next, changing coordinates $s = - i\tau$ and rotating the integration contour, we end up getting
\begin{equation} \label{eq:PartitionFunctionFermions}
\log \mathcal{Z}_\Psi := - i \Delta\Gamma[\varphi] = \frac{1}{2} \int_{\epsilon_{\rm uv}}^\infty \frac{d\tau}{\tau} e^{- \tau m^2} \Tr\left[ e^{- \tau \slashed{D}^2 } \right] e^{i \tau  \varepsilon'}\,.
\end{equation}
which, upon setting $\varepsilon'=0$ and replacing the regulator $\epsilon_{\rm uv}$ with the insertion $\exp[-\epsilon^2/4\tau]$, yields
\begin{equation} \label{eq:logZpsi}
    \log \mathcal{Z}_\Psi =  \frac{1}{2}\int_{0}^\infty \frac{d\tau}{\tau}  \, e^{-\frac{\epsilon^2}{4\tau}} \, e^{-\tau m^2} \Tr \left[ e^{- \tau  \slashed{D}^2} \right]= \frac{1}{2} \int_{0}^\infty \frac{d\tau}{\tau}  \, e^{-\frac{\epsilon^2}{4\tau}} \, e^{-\tau m^2} \mathcal{K}^{(1/2)}(\tau)\,.
\end{equation}
Here, $\mathcal{K}^{(1/2)}$ denotes the heat kernel trace associated with the kinetic operator $\slashed{D}^2$. 

We can now specialize $\mathcal{Z}_\psi$ defined above to AdS$_2 \times \mathbf{S}^2$ spacetimes with constant electric and magnetic fields. A convenient choice of local-frame Dirac matrices, $\gamma^a$, compatible with our conventions (cf. footnote \ref{fnote:conventionDirac}) is
\begin{eqn}
    {\gamma}^0= -i \tau^1 \otimes \sigma^3\, , \quad {\gamma}^1= \tau^2 \otimes \sigma^3\, , \quad {\gamma}^2= \mathds{1} \otimes \sigma^1\, , \quad {\gamma}^3= \mathds{1} \otimes \sigma^2\, ,
\end{eqn}
where $\tau^i$ and $\sigma^i$ are two independent sets of Pauli matrices. It is then easy to verify that the Dirac operator specialized to the background \eqref{eq:background} splits into two anti-commuting operators
\begin{equation}
    \slashed{D} = \slashed{D}_{\rm AdS} \otimes \sigma^3 + \mathds{1} \otimes \slashed{D}_{\mathbf{S}^2} \,, \qquad \left\{\slashed{D}_{\rm AdS} \otimes \sigma^3, \mathds{1} \otimes \slashed{D}_{\mathbf{S}^2}\right\} = 0 \,,
\end{equation}
hence implying the following decomposition
\begin{equation}\label{eq:SumOfSquares}
    \slashed{D}^2 = \slashed{D}_{\rm AdS} ^2 \otimes \mathds{1}  + \mathds{1} \otimes \slashed{D}_{\mathbf{S}^2}^2 \,.
\end{equation}
Therefore, expressing the functional trace in \eqref{eq:logZphi} in terms of the heat kernel operators introduced in Section \ref{ss:SummaryConventions}, we get
\begin{equation} 
      \log \mathcal{Z}_\Psi = \frac{1}{2} \int_{0}^\infty \frac{d\tau}{\tau}  \, e^{-\frac{\epsilon^2}{4\tau}} \, e^{-\tau m^2} \mathcal{K}^{(1/2)}_{\rm AdS}(\tau) \,\mathcal{K}^{(1/2)}_{\mathbf{S}^2}(\tau)\,.
\end{equation}

We close this section by commenting briefly on a subtlety hidden in our discussion above. Focusing for simplicity on the scalar case, we note that in order to obtain eq.~\eqref{eq:PartitionFunction}, we had to deform the contour integral in \eqref{eq:stepIntermediate} within the complex $s$-plane. However, the fact that this can be done without encountering any singularities along the way is not a priori guaranteed. Indeed, in the context of Wick rotations, it is well-known that this procedure can in principle introduce non-perturbative ambiguities and potentially spoil the identification between Lorentzian and Euclidean path integrals. An alternative route that one can follow consists in starting from the very beginning with the Euclidean formulation of the theory, with action $S_E$. Then, one can extract a similar 1-loop partition function directly from $\log \mathcal{Z}_\phi = \Delta \Gamma_E$. Following the same steps that led us to \eqref{eq:stepIntermediate}, we would get that $\log \mathcal{Z}_\phi$ is equal to the right-hand-side of \eqref{eq:PartitionFunction}, but with $\mathcal{D}^2$ replaced by its Euclidean analogue $\mathcal{D}^2_E$. In order to reproduce \eqref{eq:logZphi}, one would need to perform an analytic continuation from $\mathbb{H}^2$ to AdS$_2$, which is equivalent to continuing the heat kernel trace as reviewed in Section \ref{sss:analyticcontAdS2}. Such procedure does not seem to exhibit any crucial obstruction, and thus we conclude that \eqref{eq:logZphi} must be correct. Similar considerations apply in the fermionic case for the rotation we performed to obtain eq.~\eqref{eq:PartitionFunctionFermions} from \eqref{eq:stepIntermediateFermions}. Finally, notice that any issues with the Wick rotation in the Schwinger plane would imply that $\Gamma_E = - i \Delta\Gamma$ is spoiled by non-perturbative effects. Throughout the paper, we evaluate explicitly the traces of the heat kernel operators and we find no evidence of any singularities captured by such rotations (cf. Appendix \ref{ss:domainoftauintegration}).

\subsection{Exact computation of 1-loop partition functions}
\label{ss:ExactComputation}

To calculate the relevant functional determinants in AdS$_2 \times \mathbf{S}^2$, we first analyze the simpler cases of bosonic and fermionic fields living on a two-dimensional sphere (or Anti-de Sitter) subject to a constant and everywhere orthogonal magnetic (electric) field. We rely heavily on the Hubbard-Stratonovich transformation (cf. eq.~\eqref{eq:HStrick} below), which allows us to recast the heat kernel traces as a useful integral representation. The main advantage of this approach is that the resulting dependence on Schwinger proper time factorizes in a convenient way. We exploit this factorization to extract closed-form expressions for the 1-loop partition functions.

\subsubsection{The sphere trace}\label{sss:HalfS2}

\subsubsection*{The spin-$0$ case}\label{sss:spin0S2}

To begin with, and following the notation introduced in the previous discussion, we express the regularized partition function $\mathcal{Z}_{\phi}$ for a minimally coupled complex scalar field as follows
\begin{eqn}\label{eq:bosonicS2trace}
    \log \mathcal{Z}_\phi = -\int_0^{\infty} \frac{d\tau }{\tau}\, e^{-\frac{\epsilon^2}{4\tau}}\, \Tr \left[e^{-\tau \left(\mathcal{D}^2 + m^2 \right)}\right]\,.
\end{eqn}
As shown in \eqref{eq:heatkernelS2_scalar}, we can explicitly perform this trace by computing the infinite series
\begin{equation}\label{eq:traceLaplacianS2}
    \Tr \left[e^{-\tau \mathcal{D}^2} \right] = \sum_{n\geq 0} \rho_n\, e^{-\tau E_n} = 2\, e^{ \frac{\tau}{R_{\mathbf{S}}^2} \left( g^2 + \frac{1}{4} \right)} \sum_{n \geq 0} \left( n + g + \frac{1}{2} \right) e^{ -\frac{\tau}{R_{\mathbf{S}}^2} \left( n + g + \frac{1}{2} \right)^2}\,,
\end{equation}
with $BR_{\mathbf{S}}^2=g$.\footnote{\label{fnote:absolutevalue}Recall that the density of energy eigenstates as well as the spectrum itself were computed in Section \ref{sss:bosonS2} for $g>0$. The exact expressions hold when the monopole charge has opposite sign upon substituting $g \to |g|$.} For convenience, we may write the above series in terms of an integral representation, which is achieved via the Hubbard-Stratonovich (HS) transformation \cite{Stratonovich1957OnAM,Hubbard:1959ub}. The latter is defined through the identity
\begin{equation} \label{eq:HStrick}
    e^{- \frac{a}{2}x^2} = \frac{1}{\sqrt{2 \pi a}} \int_{-\infty}^{+\infty} e^{-\frac{y^2}{2a}} \, e^{i x y} \,dy := \mathcal{F} \left[ \frac{1}{\sqrt{a}}e^{-\frac{y^2}{2a}} \right](x)\,, \qquad a \in \mathbb{R}_{+}\, ,
\end{equation}
where our convention for the Fourier transform is
\begin{equation}
    \mathcal{F}\left[f\right](y) = \frac{1}{\sqrt{2\pi}} \int_{-\infty}^{+\infty} f(x) e^{i x y} dx \,.\notag
\end{equation}
Upon doing so, one obtains \cite{Grewal:2021bsu,Anninos:2020hfj}
\begin{equation}\label{eq:HStransfbosonsS2}
    2 \sum_{n \geq 0} \left( n + g + \frac{1}{2} \right) e^{-\frac{\tau}{R_{\mathbf{S}}^2} \left( n + g + \frac{1}{2} \right)^2} = \int_{\mathbb{R}+i \delta_u} du\, \frac{R_{\mathbf{S}}}{\sqrt{4\pi \tau}}\, e^{ -\frac{R_{\mathbf{S}}^2 u^2}{4\tau} }\, f_B(u)\, ,
\end{equation}
with $f_B(u)$ being
\begin{equation}\label{eq:fB}
    f_B(u) = 2 \sum_{n \geq 0} \left( n + g + \frac{1}{2} \right) e^{ iu \left( n + g + \frac{1}{2} \right)} = \frac{d}{du} \left(\frac{e^{i g u}}{\sin\left(\frac{u}{2}\right)}  \right) \,, 
\end{equation}
and where in order to correctly perform the Fourier transform---thereby avoiding the poles of $f_B(u)$ when integrating over $u$---we deformed the integration contour within the complex $u$-plane to $\mathcal{C}=\mathbb{R} \to \mathcal{C}'=\mathbb{R}+i\delta_u$, with $\delta_u >0$.

\subsubsection*{The spin-$\frac12$ case}\label{sss:spin12S2}

The fermionic case introduces additional complexity compared to the bosonic one, primarily due to the slightly richer structure of its spectrum. Specifically, as shown in Section \ref{sss:fermionS2}, Dirac modes consist of paired excited states with definite angular momentum $\boldsymbol{J}^2=(n+g)^2-\frac14$, together with $2g$ unpaired zero modes (see \cite{Abrikosov:2002jr,Grewal:2021bsu} for the original references). Accordingly, the 1-loop partition function
\begin{eqn}\label{eq:fermionicS2trace}
    \log \mathcal{Z}_\Psi = \frac12 \int_0^{\infty} \frac{d\tau }{\tau}\, e^{-\frac{\epsilon^2}{4\tau}}\, \Tr \left[e^{-\tau \left(\slashed{D}^2 + m^2 \right)}\right]\, ,
\end{eqn}
is defined by the following trace 
\begin{eqn}\label{eq:traceD2ferm}
    \Tr \left[e^{-\tau \slashed{D}^2} \right] &= 2\, e^{ \frac{\tau}{R_{\mathbf{S}}^2} g^2} \left(2\sum_{n \geq 1} \left( n + g\right) e^{ -\frac{\tau}{R_{\mathbf{S}}^2} \left( n + g \right)^2} + g\,e^{ -\frac{\tau}{R_{\mathbf{S}}^2} g^2}\right)\,\\
&=
 2\, e^{ \frac{\tau}{R_{\mathbf{S}}^2} g^2} \left(2\sum_{n \geq 0} \left( n + g\right) e^{ -\frac{\tau}{R_{\mathbf{S}}^2} \left( n + g \right)^2} - g\,e^{ -\frac{\tau}{R_{\mathbf{S}}^2} g^2}\right)\,.
\end{eqn}
To write this expression in a more manageable way, we observe that the first part of the trace corresponds to the bosonic one with a shift $g \to g-\frac12$ (cf. eq.~\eqref{eq:traceLaplacianS2}), while the second piece accounts for the unpaired zero-mode contribution. Hence, applying the HS trick, we get
\begin{eqn}\label{eq:HStrickS2Fermion}
    2\sum_{n \geq 0} \left( n + g\right) e^{ -\frac{\tau}{R_{\mathbf{S}}^2} \left( n + g \right)^2} - g\,e^{ -\frac{\tau}{R^2} g^2} = \int_{\mathbb{R}+i\delta_u} du\, \frac{R_{\mathbf{S}}}{\sqrt{4\pi \tau}}\, e^{ -\frac{R_{\mathbf{S}}^2u^2}{4\tau}}\, f_F(u)\,,
\end{eqn}
where the closed form expression for $f_F(u)$ reads
\begin{eqn}\label{fF2}
    f_F(u) = 2 \sum_{n \geq 0} (n+g) e^{iu g} - g e^{iu g}= \frac{d}{du} \left[  \frac{e^{i g u}}{\tan\left(\frac{u}{2}\right)}\right] \,.
\end{eqn}

\subsubsection{The AdS$_2$ trace}\label{sss:HalfAdS2}

Let us now consider the case of massive spin-0 and spin-$\frac12$ particles constrained to AdS$_2$ space, with the latter being threaded by a constant and everywhere orthogonal electric field $\boldsymbol{E}$.

\subsubsection*{The spin-$0$ case}\label{sss:spin0AdS2}

\begin{figure}[t!]
	\begin{center}
		\includegraphics[scale=0.7]{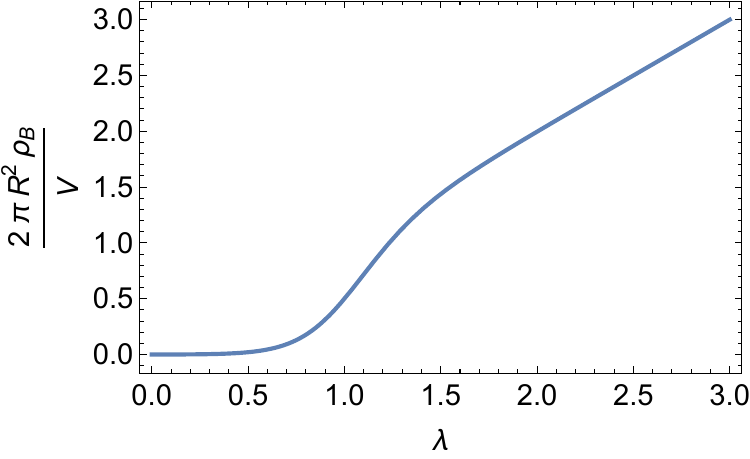}
		\caption{\small Spectral density of eigenstates for a charged spin-less particle in AdS$_2$ with $ER_{\rm{A}}^2 =1$. The variable $\lambda$ parametrizes the different (continuous) $SU(1,1)$ representations and is non-negative.} 
		\label{fig:spectraldensityspin0}
	\end{center}
\end{figure}

Proceeding as in $\mathbf{S}^2$, we start by writing the $\epsilon$-regularized partition function $\mathcal{Z}_{\phi}$:
\begin{eqn}
    \log \mathcal{Z}_\phi = -\int_0^{\infty} \frac{d\tau }{\tau}\, e^{-\frac{\epsilon^2}{4\tau}}\, \Tr \left[e^{-\tau(\mathcal{D}^2 + m^2)} \right]\,.
\end{eqn}
As shown in \cite{Comtet:1984mm, Comtet:1986ki}, and reviewed in detail in Section \ref{ss:spectralAdS2}, the heat kernel trace for the scalar operator in AdS$_2$ can be recast, using some analytic continuation from $\mathbb{H}^2$, as a line integral
\begin{eqn}\label{FirstPartFunBos}
    \Tr \left[e^{-\tau \mathcal{D}^2} \right] = \frac12\int_{-\infty}^{\infty} d\lambda\, \rho_B(\lambda)\, e^{-\frac{\tau}{R_{\rm{A}}^2}(\lambda^2 + \frac{1}{4} - e^2)}\,,
\end{eqn}
where $ER_{\rm{A}}^2=e$, and the density of (bosonic) states $\rho_B(\lambda)$ is given by (cf. eq.~\eqref{summary:bosonAdS2})
\begin{eqn}\label{eq:rhoB1}
    \rho_B(\lambda) = \frac{V_{\text{AdS}}}{2\pi R_{\rm{A}}^2} \frac{\lambda\,\sinh(2\pi \lambda)}{\cosh(2\pi \lambda) + \cosh(2\pi e)}\,, 
\end{eqn}
which is well-defined and non-negative for all $\lambda \in \mathbb{R}$ (see Figure \ref{fig:spectraldensityspin0}). Our aim is now to express \eqref{FirstPartFunBos} as a line integral over a Fourier conjugate variable, in analogy with the 2-sphere case. To do so, we first perform the following formal manipulation
%
\begin{eqn}\label{eq:FourierAdS2}
    \int_{\mathbb{R}} d\lambda \, \rho_B(\lambda) \, e^{-\frac{\tau}{R_{\rm{A}}^2} \lambda^2} = \int_{\mathbb{R}} d\lambda \, \mathcal{F}^{-1}\left[\mathcal{F}\left[\rho_B \right]\right](\lambda)\, e^{-\frac{\tau}{R_{\rm{A}}^2} \lambda^2 }\,
 = \int_{-\infty}^\infty dt  \,\mathcal{F}\left[\rho_B\right](t)  \,\mathcal{F}^{-1} \left[ e^{-\frac{\tau}{R_{\rm{A}}^2} \lambda^2} \right](t) \,.
\end{eqn}
By carrying out the aforementioned Fourier transform, we can write the 1-loop partition function of a scalar field in AdS$_2$ as (see also \cite{Anninos:2019oka, Sun:2020ame, Grewal:2021bsu})
\begin{eqn}\label{eq:PathIntegralBosonAdS2}
    \log \mathcal{Z}_\phi = - \frac{V_{\text{AdS}}}{4\pi R_{\rm{A}}^2} \int_0^{\infty} \frac{d\tau}{\tau}\, e^{-\frac{\epsilon^2}{4\tau}} e^{-\tau m^2} e^{-\frac{\tau}{R_{\rm{A}}^2}\left(\frac{1}{4} - e^2\right)}\int_{\mathbb{R}+ i\delta_t} dt\, \frac{R_{\rm{A}}}{\sqrt{4\pi \tau}}\,e^{-\frac{R_{\rm{A}}^2 t^2}{4\tau}}\, W_B(t)\,,
\end{eqn}
where, similarly to what we did in \eqref{eq:HStransfbosonsS2}, we have shifted the integration contour in order to have a well-defined Fourier transform ({see Appendix \ref{ss:TraceDetailsAdS2} for details on this point}), with
\begin{eqn}\label{eq:FourierTransfBosonDensity}
    W_B(t) = -\frac{1}{2}\, \frac{\cos(e t)}{\sinh\left(\frac{t}{2}\right)} \left( \coth\left(\frac{t}{2}\right) + 2e \tan(e t) \right) = \frac{d}{dt} \left(\frac{\cos(e t)}{\sinh\left(\frac{t}{2}\right)}\right)  \,.
\end{eqn}
%

\subsubsection*{The spin-$\frac12$ case}\label{sss:spin12AdS2}

The fermionic computation presents a similar structure to the bosonic one, with key differences arising from the distinct density of states, which now reads
\begin{eqn}
\begin{aligned}\label{eq:fermionicdensityAdS2}
    \rho_F(\lambda) =\frac{V_{\text{AdS}}}{\pi R_{\rm{A}}^2}  \frac{\lambda\,\sinh(2\pi \lambda)}{\cosh\left(2\pi \lambda\right) - \cosh\left(2\pi e\right)}\, ,
\end{aligned}
\end{eqn}
as well as the absence of the $\frac{1}{4}$ zero-point energy contribution to the continuous spectrum of $\slashed{D}^2_{\text{AdS}_2}$, i.e., $E^2_\lambda = (\lambda^2 -e^2)/R_{\rm{A}}^2$ (cf. eq.~\eqref{summary:fermionAdS2}). Notice that $\rho_F(\lambda)$ exhibits a simple pole at $\lambda=e$, which can be associated with the (analytic continuation of some) zero modes in $\mathbb{H}^2$, since $E (\lambda=e)=0$. Consequently, one finds that for $\lambda < e$ the spectral density, which must be positive definite, becomes strictly negative. However, one should recall that the actual integration contour---when doubled as in eq.~\eqref{eq:heatkernelH2completeFermions}---lies slightly below the real axis, namely
\begin{eqn}\label{FirstPartFunFerm}
    \Tr \left[e^{-\tau \slashed{D}^2} \right] = \frac12\int_{\mathbb{R}-i\epsilon} d\lambda\, \rho_F(\lambda)\, e^{-\frac{\tau}{R_{\rm{A}}^2}(\lambda^2 - e^2)}\,,
\end{eqn}
such that, upon using the Sokhotski–Plemelj theorem
\begin{figure}[t!]
	\begin{center}
		\includegraphics[scale=0.7]{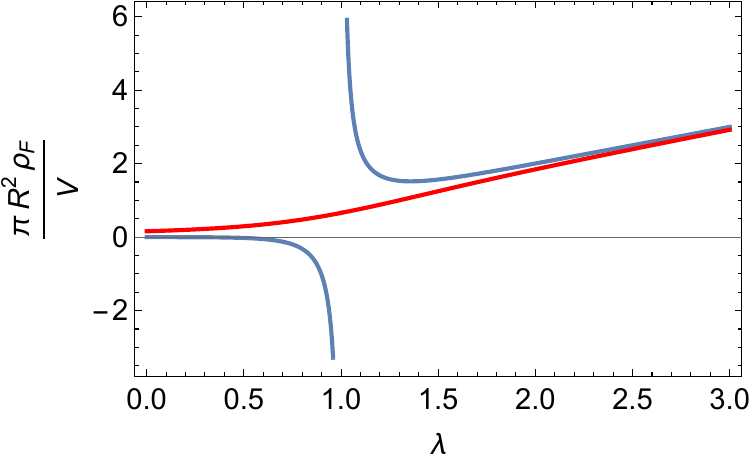}
		\caption{\small Spectral density of energy eigenstates for a charged spin-$\frac12$ particle in AdS$_2$ with $ER_{\rm{A}}^2 =1$. Naïvely, the function (blue) presents a simple pole at $\lambda=e$, thus rendering the density negative definite for $\lambda <e$. However, upon taking the principal value (red) and separating the localized zero mode contribution at $\lambda=e$ (cf. discussion around eq.~\eqref{eq:densitydistributionfermions}), one effectively restores positivity in $\rho_F(\lambda)$.} 
		\label{fig:spectraldensityspin1/2}
	\end{center}
\end{figure}
%
\begin{eqn}
   \frac{1}{x\pm i 0^+} = \text{P.V.}\, \bigg\{\frac{1}{x} \bigg\} \mp i\pi \delta(x)\,,
\end{eqn}
the spectral function \eqref{eq:fermionicdensityAdS2} should be properly understood as the following distribution 
\begin{eqn}\label{eq:densitydistributionfermions}
    \rho_F(\lambda) = \text{P.V.}\, \big\{\rho_F(\lambda) \big\} + i\pi \left[\text{Res} \left(\rho_F, e\right)\,\delta(\lambda-e) + \text{Res} \left(\rho_F, -e\right)\,\delta(\lambda+e)\right]\,.
\end{eqn}
Notice that, by taking the principal value, one effectively resolves the negative density issue, provided we interpret the integrand as a distribution (see Figure \ref{fig:spectraldensityspin1/2}). The $\delta$-functions, on the other hand, represent the discrete contribution associated to the would-be zero modes in $\mathbb{H}^2$. However, as explained in Section \ref{sss:fermionAdS2}, their effect ultimately cancels since their residues are equal in magnitude but opposite in sign (cf. eq.~\eqref{eq:residueszeromodes}). In what follows, we assume that the principal value of $\rho_F(\lambda)$ is taken, unless otherwise stated.

Therefore, applying the same methodology as for the spin-0 case, we get\footnote{One may also obtain the Fourier transform of $\rho_F(\lambda)$ by noticing that the latter is nothing but $2\rho_B(\lambda)$ with $e\rightarrow e+i/2$. However, care must be taken when performing such identification, since it yields an additional term proportional to $-i e \cos (et)$ associated with half the residues at $\lambda=\pm e$, which must be subtracted.}
\begin{eqn}\label{eq:TraceSpin12Fourier}
    \Tr \left[e^{ -\tau \left(\slashed{D}^2 + \frac{e^2}{R_{\rm{A}}^2}\right)} \right] = \frac12 \int_{-\infty}^{\infty} d\lambda \, \rho_F(\lambda) \, e^{-\frac{\tau}{R_{\rm{A}}^2} \lambda^2} = \frac12 \int_{-\infty}^{\infty} dt  \,\mathcal{F}\left[\rho_F\right](t)  \, \frac{R_{\rm{A}}}{\sqrt{4 \pi \tau}} e^{-\frac{R_{\rm{A}}^2 t^2}{4\tau}}\,,
\end{eqn}
such that performing the Fourier transform, we arrive at the fermionic 1-loop partition function
\begin{eqn}\label{eq:fermitraceAdS2}
    \log \mathcal{Z}_\Psi = \frac{V_{\text{AdS}}}{4\pi R_{\rm{A}}^2} \int_0^{\infty} \frac{d\tau }{\tau} e^{-\frac{\epsilon^2}{4\tau}} e^{-\tau m^2} e^{\frac{\tau}{R_{\rm{A}}^2} e^2}\int_{\mathbb{R}+ i \delta_t} dt\, \frac{R_{\rm{A}}}{\sqrt{4\pi \tau}}\,e^{-\frac{R_{\rm{A}}^2 t^2}{4\tau}}\, W_F(t)\,,
\end{eqn}
with \cite{Anninos:2019oka, Sun:2020ame, Grewal:2021bsu}
\begin{eqn}\label{eq:WFAdS2}
    W_F(t) = - \frac{1}{2}\, \frac{\cos(e t)}{\sinh\left(\frac{t}{2}\right)} \left( \csch\left(\frac{t}{2}\right) + 2e \tan(e t) \cosh\left(\frac{t}{2}\right) \right)\, = \frac{d}{dt} \left(\frac{\cos(e t)}{\tanh\left(\frac{t}{2}\right)}\right)\,.
\end{eqn}

\subsubsection{The full AdS$_2 \times \mathbf{S}^2$ trace}\label{ss:AdS2xS2Trace}

Finally, exploiting the (anti-)commutation properties of the corresponding (fermionic) bosonic kinetic operators (see Section \ref{sss:1loopbasics} for details), we are now in a position to perform the full trace computation for massive and charged particles propagating in four-dimensional AdS$_2\times \mathbf{S}^2$ spacetimes. In general, these fields may carry both electric and magnetic charges with respect to a given $U(1)$ field strength $F=dA$. However, as explained in Section \ref{sss:nearhorizonBH} (see, in particular, the discussion around \eqref{eq:DyonicInteraction}), the resulting minimal couplings in these types of 4d backgrounds can be conveniently described in terms of an equivalent particle with unit electric charge and some \emph{effective} electric and magnetic fields that we denote by $E R^2 = - q_e$ and $B R^2 = q_m$, borrowing the notation from $\mathcal{N}=2$ supergravity \cite{Castellano:2025yur, Castellano:2025rvn}. Therefore, the first-quantized (worldline) action will be assumed to take the form 
\begin{equation}\label{eq:effectivecharges}
 S_{wl} = -m \int_\gamma ds -q_e \int_\gamma \frac{dt}{\rho}-q_m \int_\gamma \cos \theta d\phi\, ,
\end{equation}
where $\gamma$ denotes the worldline path with proper length $ds=\sqrt{-ds^2}$ computed using \eqref{eq:conformalcoords}. In practice, this simply amounts to replacing $e$ with $-q_e$, and $g$ with $q_m$, in the 1-loop path integrals previously obtained for AdS$_2$ and $\mathbf{S}^2$, respectively. Furthermore, as an important simplifying assumption, and to make contact with the discussion presented in Section \ref{sss:nearhorizonBH}, we take the radii of the two two-dimensional factors to be equal, i.e., we set $R_{\mathbf{S}} = R_{\rm{A}} \equiv R$. This, in turn, allows us to derive closed analytic expressions that will be useful later on. 

\subsubsection*{The spin-$0$ case}\label{sss:spin0AdS2xS2}

To compute the full AdS$_2 \times \mathbf{S}^2$ functional determinant for charged, spin-0 particles, we need to combine the results from the previous section into a single Schwinger-like integral. Thus, using the fact that the relevant heat kernel factorizes into a direct product, and inserting the resulting traces obtained in sections \ref{sss:spin0S2} and \ref{sss:spin0AdS2}, we arrive at the following 1-loop partition function for a massive, charged scalar field
\begin{eqn}\label{eq:fullintegralspin0}
    &\log \mathcal{Z}_\phi = -\int_0^{\infty} \frac{d\tau }{\tau}\, e^{-\frac{\epsilon^2}{4\tau}}\, \Tr \left[e^{-\tau \left(\mathcal{D}^2 + m^2 \right)}\right]\\
    &= -\int_0^{\infty} \frac{d\tau}{\tau^2} e^{-\frac{\epsilon^2}{4\tau}-\tau \tilde{m}^2} \left[ \frac{V_{\text{AdS}}}{(4\pi R)^2} e^{-\tau \left(\frac{1}{4}-q_e^2\right)}\int_{\mathbb{R} + i\delta_t} dt\, e^{-\frac{t^2}{4\tau}} W_B(t)  \right] \left[ e^{\tau \left(q_m^2 + \frac{1}{4}\right)}  \int_{\mathbb{R} + i\delta_u} du\, e^{-\frac{u^2}{4\tau}}\, f_B(u) \right]\,,
\end{eqn}
where in the second step we rescaled Schwinger proper time such that $\tau \to \tau R^2\,,\; \epsilon \rightarrow R \epsilon$, and we defined $\tilde{m}=Rm$. To proceed, it is natural to first perform the Schwinger $\tau$-integral\footnote{\label{fnote:CorrectContour}The actual path of integration in Schwinger proper time for the integral \eqref{FirstSintegral} would be along the positive imaginary axis in a Lorentzian theory (cf. Section \ref{ss:Traces}). However, as argued in Appendix \ref{ss:domainoftauintegration}, properly taking this into account does not seem to change the final result.}
\begin{eqn}\label{FirstSintegral}
    &I_S = \frac{1}{4\pi} \int_0^{\infty} \frac{d\tau }{\tau^2}\, e^{-\frac{\epsilon^2+t^2 +u^2}{4\tau}}\, e^{-\tau \Delta^2}\, ,
\end{eqn}
with $\Delta^2 := \tilde{m}^2 - q_e^2 - q_m^2$. In supersymmetric setups, the latter quantity can be shown to be non-negative \cite{Castellano:2025yur}. Henceforth, we will assume this condition to hold, leaving the super-extremal case along with the physics associated with black hole instabilities to Section \ref{ss:Instability&Schwinger}. We can further massage this formula by redefining $\tau = \frac{1}{x}$, which yields
\begin{eqn}\label{eq:intBes}
    I_S = \frac{1}{4\pi} \int_0^{\infty}  dx\, e^{-x\,\frac{\epsilon^2+t^2 +u^2}{4}}\, e^{-\frac{\Delta^2}{x}}\,,
\end{eqn}
thus resembling the integral representation of the modified Bessel function of degree $\nu$
\begin{eqn}\label{eq:defModifiedBessel}
    K_\nu(z) = \frac{1}{2^{1+\nu}}z^{\nu} \int_0^{\infty} e^{-t - \frac{z^2}{4t}} \frac{dt}{t^{1+\nu}} = K_{-\nu}(z)\, .
\end{eqn}
To be more explicit, we change variables again to $y = \frac{x}{4} (\epsilon^2 + t^2 + u^2)$, such that \eqref{eq:intBes} becomes
\begin{eqn}\label{SIntegral}
    I_S = \frac{1}{\pi} \frac{1}{\epsilon^2 + t^2 + u^2} \int_0^{\infty}  dy\, e^{-y -\frac{\Delta^2(\epsilon^2 + t^2 + u^2)}{4y}} = \frac{\Delta}{\pi} \frac{1}{\sqrt{\epsilon^2 + t^2 + u^2}}\, K_{1}(\Delta \sqrt{\epsilon^2 + t^2 + u^2})\, .
\end{eqn}
Furthermore, whenever the scalar field is part of a BPS supermultiplet, one can prove that $\tilde{m}^2 = q_e^2 + q_m^2$ (cf. \eqref{eq:ChargeMassIdentity}). In that case, one simply computes the $\Delta \rightarrow 0$ limit of \eqref{SIntegral}, obtaining
\begin{eqn}\label{BPSLimit}
    \lim_{\Delta \rightarrow 0} I_S(\Delta) = \frac{1}{\pi} \frac{1}{\epsilon^2 + t^2 + u^2}\, .
\end{eqn}
We will consider in what follows only BPS particles, and hence the formula \eqref{BPSLimit} should be understood as resulting from a direct evaluation of the integral \eqref{FirstSintegral} with $\Delta = 0$. 

Inserting \eqref{BPSLimit} into \eqref{eq:fullintegralspin0} leads to
\begin{eqn}\label{eq:BPSlikebosonAdS2xS2}
    \log \mathcal{Z}_\phi = -\frac{V_{\text{AdS}}}{4\pi^2 R^2} \int_{\mathbb{R} + i\delta_t} dt\, W_B(t) \int_{\mathbb{R} + i\delta_u} du\, \frac{1}{u^2 + t^2 + \epsilon^2}\, f_B(u)\, .
\end{eqn}
To simplify this even more, let us focus on the second integral in \eqref{eq:BPSlikebosonAdS2xS2}, namely
\begin{eqn}\label{eq:Iuintegral}
   I_U(t) =  \int_{\mathbb{R} + i\delta_u} du\,  \frac{1}{u^2 + t^2 + \epsilon^2}\, f_B(u)\, ,
\end{eqn}
where we recall that the function $f_B(u)$ is given by (cf. eq.~\eqref{eq:fB})
\begin{equation*}
     f_B(u) = \frac{d}{du} \left(\frac{e^{iq_mu}}{\sin\left(\frac{u}{2}\right)}  \right)  \,.
\end{equation*}
Notice that the singularities of $f_B(u)$ lie along the real axis. In fact, this is the reason why we had to perform the integral over $\mathbb{R} + i\delta_u$, with $\delta_u > 0$ rather than just along the real line. The integrand also exhibits a pair of simple poles at $u_\pm = \pm i\sqrt{t^2 + \epsilon^2}$.
%
\begin{figure}[t]
    \centering
    \includegraphics[scale=0.8]{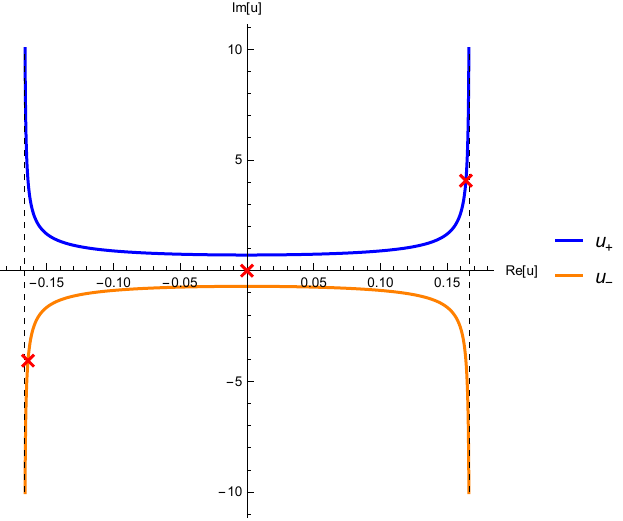}
    \caption{\small Singularity structure of the integral \eqref{eq:Iuintegral}. All poles are marked as red crosses, whereas in blue/orange we show the loci of the simple poles parametrized by $u= \pm i \sqrt{t^2 + \epsilon^2}$, respectively, with $t \in \mathbb{R}+ i\delta_t$. 
    For $t=i\delta_t$, the latter lie at $\pm i \sqrt{\epsilon^2 - \delta_t^2}$, where one should take $\epsilon > \delta_t >  0$. The dotted lines correspond to $u = \pm \delta_t + i \mathbb{R}$.}
    \label{fig:poles_u_integral}
\end{figure}
%
To evaluate \eqref{eq:Iuintegral}, we deform the contour of integration by adding an arc at infinity in the complex upper-half $u$-plane. We see that in such case $\text{Im}\, u > 0$, which implies that for large $u$ none of the terms in the integrand explodes provided $q_m > -1/2$. Importantly, we remark that this is always the case since, as commented already around eq.~\eqref{eq:traceLaplacianS2}, in all our expressions $q_m$ must be actually understood as $|q_m|$.
We can then evaluate the line integral as the residue of the simple pole at $u=+i\sqrt{t^2 + \epsilon^2}$ surrounded by a counter-clockwise contour (cf. Figure \ref{fig:poles_u_integral}), yielding\footnote{Notice that, in order to be able to pick up the residue at $u=i \sqrt{t+\epsilon^2}$ for all $t \in \mathbb{R}+i\delta_t$, we must take the regulators to satisfy the inequalities $\delta_t<\epsilon$ and $\delta_u < \sqrt{\epsilon^2-\delta_t^2}$.}
\begin{eqn}
     I_U(t)  
     = 2\pi i\, \text{Res}\left({\frac{1}{u^2 + t^2 + \epsilon^2}\, f_B(u)}\, , \, i\sqrt{t^2 + \epsilon^2}\right) = \frac{\pi}{\sqrt{t^2 + \epsilon^2}}\, f_B(i\sqrt{t^2 + \epsilon^2})\, .
\end{eqn}
Substituting this result into the full expression \eqref{eq:BPSlikebosonAdS2xS2}, we obtain
\begin{eqn} \label{eq:LogZintermediate}
     \log \mathcal{Z}_\phi = -\frac{V_{\text{AdS}}}{4\pi R^2}  \int_{\mathbb{R} + i\delta_t} \frac{dt}{\sqrt{t^2 + \epsilon^2}} \, W_B(t) f_B(i\sqrt{t^2 + \epsilon^2})\, .
\end{eqn}

Finally, in order to express \eqref{eq:LogZintermediate} in a way that resembles a proper Schwinger integral, we split this formula at the origin
\begin{eqn}
\begin{aligned}
     \log \mathcal{Z}_\phi = -\frac{V_{\text{AdS}}}{4\pi R^2}   \bigg[ &\int_{0 +i\delta_t}^{\infty + i \delta_t } \frac{dt}{\sqrt{t^2 + \epsilon^2}} \, W_B(t) f_B(i\sqrt{t^2 + \epsilon^2})\\
     &+ \int^{0 + i\delta_t}_{-\infty + i \delta_t } \frac{dt}{\sqrt{t^2 + \epsilon^2}} \, W_B(t) f_B(i\sqrt{t^2 + \epsilon^2})\bigg]\, .
\end{aligned}
\end{eqn}
Notice that $t = 0$ is the only pole of the integrand lying on the real axis, which we are in fact avoiding by integrating along $\mathbb{R}+i\delta_t$. Furthermore, since the residue of $W_B(t)$ at that point is zero, we can safely remove the regulator and deform the contour towards the real axis, thereby picking the principal value of the resulting integral (see Appendix \ref{ss:TraceDetailsAdS2} for details). 

Additionally, using the parity properties of the integrand, we find
\begin{eqn}\label{SintegralBoson0}
     \log \mathcal{Z}_\phi = -\frac{V_{\text{AdS}}}{2\pi R^2}   \int_{0^+ }^{\infty} \frac{dt}{\sqrt{t^2 + \epsilon^2}} \, W_B(t) f_B(i\sqrt{t^2 + \epsilon^2}) \, .
\end{eqn}
As the poles of both $W_B(t)$ and $f_B(i t)$
lie now on the imaginary axis---with the exception of the one at the origin, which is avoided by integrating from $0^+$, we can remove the $\epsilon$-regulator. The final result then reads
\begin{eqn} \label{SintegralBoson}
     \log \mathcal{Z}_\phi = -\frac{V_{\text{AdS}}}{2\pi R^2}   \int_{0^+ }^{\infty} \frac{dt}{t} \, W_B(t) f_B(i t) \,,
\end{eqn}
with 
\begin{equation}
    {f}_B(it) = -\frac{d}{dt} \left(\frac{e^{-q_mt}}{\sinh\left(\frac{t}{2}\right)}\right)\,, \qquad  {W}_B(t) = \frac{d}{dt}\left( \frac{\cos(q_et)}{\sinh\left(\frac{t}{2}\right)} \right)\,. 
\end{equation}

\subsubsection*{The spin-$\frac12$ case}\label{sss:spin12AdS2xS2}

For the spin-$\frac12$ fields, the computation proceeds in a similar manner to the scalar case above. Therefore, exploiting the separability of the trace  and inserting the results from sections \ref{sss:spin12S2} and \ref{sss:spin12AdS2}, we obtain
\begin{eqn}\label{eq:fullintegralspin1/2}
    \log \mathcal{Z}_\Psi &= \frac12 \int_0^{\infty} \frac{d\tau }{\tau}\, e^{-\frac{\epsilon^2}{4\tau}}\, \Tr \left[e^{-\tau \left( -\slashed{D}^2 + m^2 \right)}\right]\\
    &= \int_0^{\infty} \frac{d\tau }{\tau } e^{-\frac{\epsilon^2}{4\tau}-\tau \tilde{m}^2} \left[ \frac{V_{\text{AdS}}}{2\pi R^2} e^{\tau q_e^2}\int_{\mathbb{R} + i\delta_t} dt \frac{e^{-\frac{t^2}{4\tau}}}{\sqrt{4\pi \tau}} W_F(t)  \right]  \left[ e^{\tau q_m^2}  \int_{\mathbb{R} + i\delta_u} du \frac{e^{-\frac{u^2}{4\tau}}}{\sqrt{4\pi \tau}}\, f_F(u) \right]\, ,
\end{eqn}
where in the second line we have performed the same rescaling as in eq.~\eqref{eq:fullintegralspin0}. It is worth remarking that, unlike the bosonic trace, the fermionic computation has no extra $\frac{1}{4}$ factor in the energy spectrum and hence exhibits no zero-point contribution in AdS$_2 \times \mathbf{S}^2$ spacetimes.\footnote{Note that this happens even if the `external' and `internal' radii do not coincide, namely when $R_{\textbf{S}} \neq R_{\textbf{A}}$.} Moreover, the fermionic density contains no additional poles along the $t$ ($u$) imaginary (real) axis with respect to its bosonic analogue. Hence, up to the different density functions $W_F(t)$ and $f_F(u)$, one can follow the same steps outlined in Section \ref{sss:spin0AdS2xS2} in order to evaluate \eqref{eq:fullintegralspin1/2}.

Upon doing so, one eventually finds a simplified analytic expression for the 1-loop path integral associated with a massive, charged spin-$\frac12$ field in AdS$_2 \times \mathbf{S}^2$:
\begin{eqn}\label{SintegralFermions}
   \log \mathcal{Z}_{\Psi} = \frac{V_{\text{AdS}}}{\pi R^2} \int_{0^+ }^{\infty} \frac{dt}{t} \, W_F(t) f_F(i t) \,,
\end{eqn}
where
\begin{equation}
        {f}_F(it) = -\frac{d}{dt} \left(\frac{e^{-q_mt}}{\tanh\left(\frac{t}{2}\right)}  \right) \,, \qquad    {W}_F(t) = \frac{d}{dt}\left( \frac{\cos(q_e t)}{\tanh\left(\frac{t}{2}\right)} \right)\,. 
\end{equation}

\subsection{Supersymmetric-like determinants and non-perturbative effects}\label{ss:Susic1LoopMinimal}

As a first interesting application of our results, in what follows we restrict the computation of the 1-loop determinant to a minimally coupled supermultiplet in 4d $\mathcal{N}=2$ supergravity. In particular, we focus on the simplest such example corresponding to a BPS hypermultiplet, which is comprised by two complex scalars and one Dirac fermion with equal mass and charges. Strictly speaking, however, to ensure closure under $\mathcal{N}= 2$ supersymmetry transformations one must include some additional interactions in the Lagrangian that couple the fermionic degrees of freedom to the graviphoton background in a non-minimal way \cite{Andrianopoli:2007gt,deWit:1984rvr}. A detailed account of these terms and their effect on the functional traces is presented in Section \ref{s:SusyLoopDeterminant} below.

\subsubsection{A four-dimensional effective action}\label{sss:HyperEFTMinimal}

As previously mentioned, the hypermultiplet furnishes one of the basic matter representations of the 4d $\mathcal{N}=2$ supersymmetry algebra \cite{Weinberg:2000cr}. Concretely, its on-shell field content consists of two complex scalar fields---forming an $SU(2)_R$ doublet---and one ($R$-singlet) Dirac fermion.

\smallskip

Using this information, one may derive the four-dimensional effective action resulting from integrating out such multiplet within a given AdS$_2 \times \mathbf{S}^2$ supersymmetric background. Indeed, assuming minimal couplings and combining the expressions \eqref{SintegralBoson} and \eqref{SintegralFermions} obtained in the previous section, one finds
\begin{eqn}\label{eq:susyeffaction}
    \log{\mathcal{Z}_{\rm hm}} :=  2\log{\mathcal{Z}_\phi} + \log{\mathcal{Z}_{\Psi}} =\lim_{\epsilon \rightarrow 0^+} \frac{V_{\text{AdS}}}{\pi R^2} \int_{\epsilon}^{\infty} \frac{dt}{t} \left[W_F(t) f_F(i t)-W_B(t) f_B(i t)\right] \,,
\end{eqn}
where (restoring the appropriate absolute values\footnote{Interestingly, the formula \eqref{eq:susycomb} does not require to take the absolute value of $q_e$. The reason for this stems from the fact that it only accounts for the continuous states in AdS$_2$, whose energies are symmetric under $q_e \to -q_e$, even if we do not impose---as one should---the absolute value on $\boldsymbol{E}$.})
\begin{eqn}\label{eq:susycomb}
   W_F(t) f_F(i t) - W_B(t) f_B(i t) =\frac{e^{-|q_m| t}}{4} \frac{ \cos(q_e t)}{ \sinh^2\left(\frac{t}{2}\right)} - |q_m| q_e e^{- |q_m| t} \sin(q_e t) \,.
\end{eqn}
Remarkably, the second term in the expression above can be further simplified by performing the integral over $t$ in \eqref{eq:susyeffaction} explicitly, which yields
\begin{equation}\label{eq:newPieceFormula}
   I_\theta = -\frac{V_{\text{AdS}}}{4\pi R^2} \int_{0^+}^{\infty} \frac{dt}{t} e^{- |q_m| t} \left[ 4 |q_m| q_e  \sin(q_e t) \right]  
   = -\frac{V_{\text{AdS}}}{\pi R^2} q_e |q_m| \tan^{-1}\left(\frac{q_e}{|q_m|}\right)\, .
\end{equation}
Hence, putting everything together, we finally arrive at 
\begin{eqn}\label{eq:hyperformulanonmixing}
    \log{\mathcal{Z}_{\rm hm}} &= -\frac{V_{\text{AdS}}}{\pi R^2}\, q_e |q_m| \tan^{-1}\left(\frac{q_e}{|q_m|}\right)  +  \frac{V_{\text{AdS}}}{4\pi R^2}\int_{0^+}^{\infty} \frac{dt}{t} e^{- |q_m| t} \frac{ \cos(q_e t) }{ \sinh^2\left(\frac{t}{2}\right)} \,\\
    &= -\frac{V_{\text{AdS}}}{\pi R^2}\, \,\text{Re}\left[  q_e q_m \tan^{-1}\left(\frac{q_e}{q_m}\right)  -  \frac{1}{4}\int_{0^+}^{\infty} \frac{dt}{t}  \frac{e^{i(q_e + i |q_m|) t}}{ \sinh^2\left(\frac{t}{2}\right)} \right]\, ,
\end{eqn}
which constitutes one of the main results of this work.

\smallskip

The 1-loop determinant may be thus written in a more suggestive way as follows
\begin{equation} \label{eq:AdS2xS2effaction}
\begin{aligned}
    \Gamma_{\rm hm} [A_\mu] =\, &\frac{1}{(4\pi R^2)^2}\int_{\text{AdS}_2\times \mathbf{S}^2} d^4x \sqrt{-\det g}\, \int_{0^+}^{\infty} \frac{dt}{t} e^{- |B|R^2 t} \left[\frac{ \cos(E R^2 t) }{ \sinh^2\left(\frac{t}{2}\right)}\right]\\
    &-\int_{\text{AdS}_2\times \mathbf{S}^2}\frac{\theta_{\rm eff}}{4\pi^2}\, E\, B\, \omega_{\text{AdS}_2} \wedge \omega_{\mathbf{S}^2}\,, 
\end{aligned}
\end{equation}
where we used \eqref{eq:QuadraticConstraint} and we defined $\theta_{\rm eff}:=\tan^{-1}\left(E/B\right)$. The notation has been chosen to reflect the fact that the second term has a form very reminiscent of a topological $\theta$-term. The latter is moreover related to the phase of the functional determinant associated with the fermionic degrees of freedom, and it can be readily determined to be $\theta=\pi/2+ \text{arg}(Z \bar{Z}_{\rm BH})$ using eq.~\eqref{eq:chargeDefSugra} (see Appendix \ref{ss:freemassivehyper} for more details on this point).

\subsubsection{Comments on non-perturbative phenomena}\label{sss:NonPerturbative}

Let us comment on various salient features exhibited by the 1-loop determinant thus obtained. Since the topological $\theta$-term was already discussed in the previous section, we focus on the Schwinger integral appearing in the first line of \eqref{eq:AdS2xS2effaction}. After factoring out constant and volume prefactors, the latter takes the form
\begin{equation}\label{eq:SchingerManipulated}
    \mathcal{I} =   \text{Re} \left[\int_{0^+}^{\infty} \frac{dt}{t}\, \frac{ e^{-i \beta t} }{ \sinh^2\left(\frac{t}{2}\right)} \right] \,, \qquad \text{with}\quad \beta = ( |E| - i |B|) R^2 \,.
\end{equation}
Observe that the parameter $\beta$ is related to the black hole radius $R$ and the particle mass $m$ via $|\beta| = m R $. Interpreting $m^{-1}$ as the Compton wavelength of the particle, it becomes clear that for $|\beta| \gg 1 $, the curvature and graviphoton corrections are suppressed. We should then identify such a regime as the weak coupling limit of the system, and similarly interpret $\beta^{-1}$ as the (complexified) expansion parameter in the present background.

\smallskip

\begin{figure}[t]
        \centering
        \includegraphics[scale=0.30]{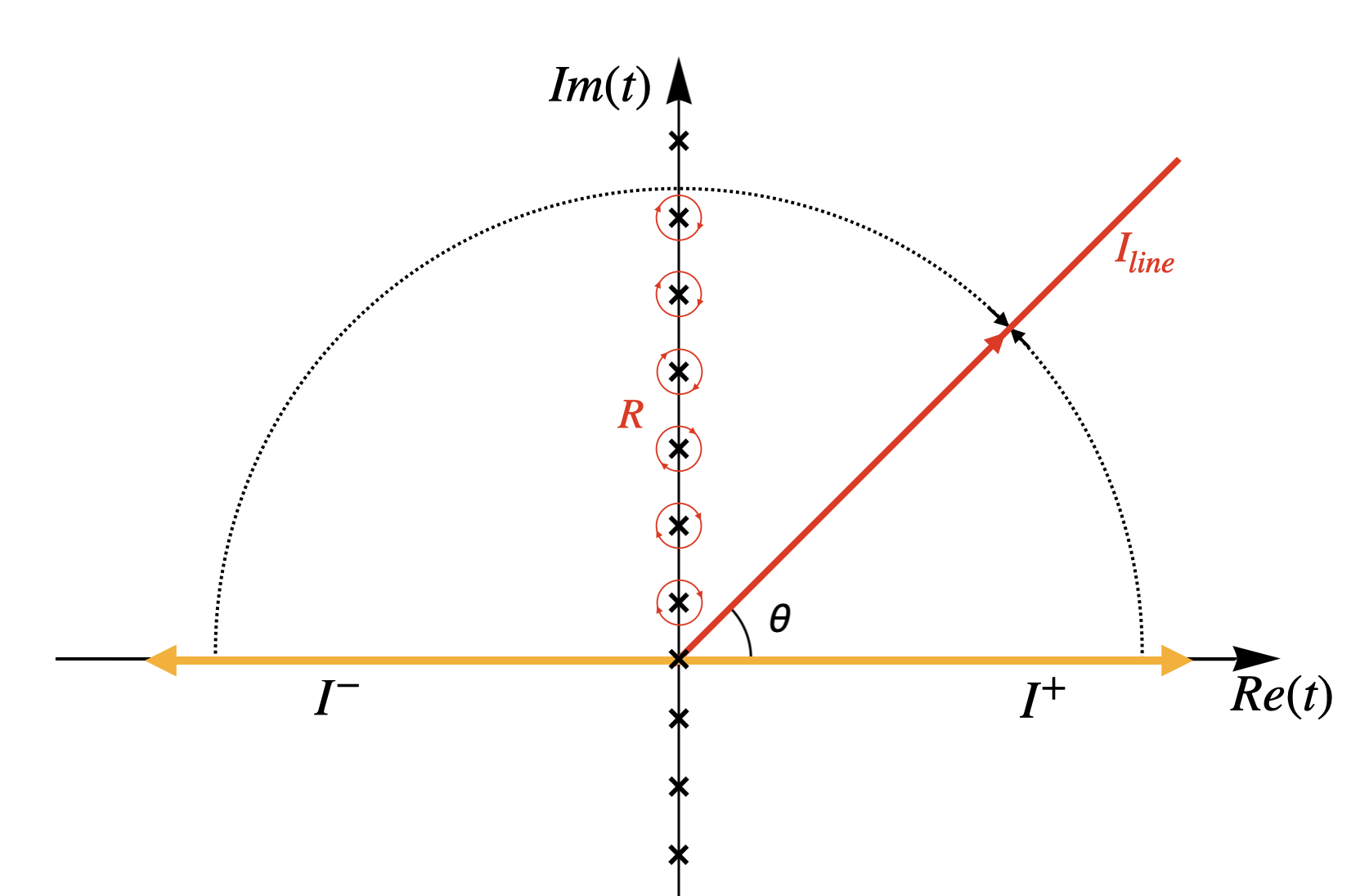}
        \caption{\small As explained in the text, one can make manifest the perturbative and non-perturbative structure of the integral \eqref{eq:SchingerManipulated} by separating the latter into two and deforming the contours towards the preferred ray at an angle $\theta=\tan^{-1} (|B|/|E|)$. One gets a line integral $\mathcal{I}_{\rm line}$ plus some residues $\mathcal{R}$.}
        \label{fig:NonPertContour2}
\end{figure}

We now demonstrate that the presence of poles in the complex Schwinger $t$-plane implies that $\mathcal{I}$ defined above can be decomposed into a line integral plus an infinite sum over residues, with the latter having a structure that is reminiscent of non-perturbative corrections in $\beta^{-1}$. To do so, we first rewrite \eqref{eq:SchingerManipulated} as 
\begin{equation}
    \mathcal{I} = \mathcal{I}^+ + \mathcal{I}^- \,, \qquad \mathcal{I}^{\pm} =  \frac{1}{2} \text{Re} \left[\int_\mathbb{R^\pm} \frac{dt}{t}\, \frac{ e^{\mp i \beta t} }{ \sinh^2\left(\frac{t}{2}\right)} \right] \,.
\end{equation}
Next, we align the two contours performing a rotation in the complex $t$-plane (see Figure \ref{fig:NonPertContour2}). Notice that for $\mathcal{I}^\pm$, adding an arc at infinity will not contribute provided $ |E| \,  \text{Im} \,t  \lessgtr  |B| \, \text{Re}\, t$. 
This allows us to continuously deform both line integrals to $\mathbb{R}_+ e^{i\theta}$, with $\theta = \tan^{-1} (|B|/|E|)$. In doing so one must pick up the residues associated with the poles of the integrand in $\mathcal{I}^-$ located along the positive imaginary axis, namely at $t = 2 \pi i k $ for $k \in \mathbb{Z}$. Therefore, absorbing the contribution from the arc at the origin into the UV cutoff, we obtain
\begin{equation}  \label{eq:SchwingerDecomposition}
    \mathcal{I} =  \mathcal{I}_{\text{line}} +  \mathcal{R} \,,
\end{equation}
with
\begin{subequations}
\begin{align} 
    \mathcal{I}_{\text{line}} & = \text{Re} \left[\int_{\mathbb{R}_+ e^{i\theta}} \frac{dt}{t}  \frac{ \cos(\beta t) }{ \sinh^2\left(\frac{t}{2}\right)} \right] \,, \\[2mm]
 \mathcal{R} & = \text{Re} \left[  \, - 2 \pi i\, \sum_{k \ge 1} \text{Res} \, \left(  \frac{ e^{i \beta t} }{ 2t \sinh^2\left(\frac{t}{2}\right)},\,  2 \pi i k\right) \, \right] \,. \label{eq:residuesR}
\end{align}
\end{subequations}
The sum of residues can be evaluated explicitly and yields
\begin{equation}\label{eq:Residues}
    \mathcal{R}= \frac{1}{\pi} \, \text{Im} \left[ \mathrm{Li}_2(e^{-2 \pi \beta }) + 2 \pi \beta \mathrm{Li}_1(e^{-2 \pi \beta })\right]\,.
\end{equation}
This formula is non-perturbative in $1/\beta$, with $|E|$ entering via the exponential factor $e^{-2\pi |E| R^2}$. Note that this is not the same behaviour one finds for the Schwinger effect in flat space, where non-perturbative pair production takes the form $\mathcal{O}\left(e^{-m^2/|E|}\right)$ (see Appendix \ref{ss:flatspacelimit} for details). However, it gives the right dependence in the presence of an AdS$_2$ geometry, were wordline (Euclidean) instantons are associated with actions exhibiting precisely this scaling \cite{Pioline:2005pf, Lin:2024jug, Castellano:2025yur}. This also confirms that the perturbative parameter previously identified is the correct one, and that the residues \eqref{eq:Residues} depend on the latter non-perturbatively.

\smallskip

A natural question to ask is what precise information is encoded in the residues \eqref{eq:Residues}, and whether additional contributions of a similar type are contained in the line integral $\mathcal{I}_{\text{line}}$. In general, the latter will provide a non-perturbative definition of the asymptotic series at weak coupling, and thus may contain non-perturbative information. However, there might be special choices of contours for which it describes purely perturbative physics whereas all the non-perturbative one is captured by the residues. Typically, this is realized when the line integral matches the Borel resummation of the EFT perturbative expansion. Unfortunately, without knowing the latter one usually cannot fix the `ambiguity' in the choice of contour within the complex Schwinger $t$-plane. In the case at hand, we do not have the explicit form of the perturbative series. Still, we can make an educated guess based on our knowledge of the correct contour for the Schwinger integral of the Gopakumar--Vafa (GV) partition function. Indeed, the system considered herein and the GV one are closely related because in both cases we integrate out supermultiplets in a maximally supersymmetric configuration of 4d $\mathcal{N}=2$ supergravity with constant graviphoton field strength. We can then proceed by making use of the same prescription as in \cite{Castellano:2025ljk}, which identifies $\mathbb{R}_+ e^{i\theta}$ as the ray along which $\mathcal{I}_\text{line}$ is purely perturbative.
Hence, in the following we will sum over D-instanton contributions corresponding to a tower of particles with the prescription given in \cite{Castellano:2025ljk}, thereby showing that the resulting line integral reproduces the structure of the GV perturbative series.

\smallskip

\noindent For concreteness, we consider an infinite tower of BPS states and assume that the particle charges scale as $q_{e,m}^{(n)} = n\,  q_{e,m}$, with  $n\in \mathbb{Z}$.\footnote{This example is physically realized as e.g., the tower of D0-branes in a D0–D2–D4–D6 background \cite{Castellano:2025ljk, Castellano:2025yur}.} As explained around eq.~\eqref{eq:effectivecharges}, this is equivalent to having unit charges and electric-magnetic fields $E^{(n)} = n\, E$, $B^{(n)} = n \, B$.  The total contribution to the 1-loop determinant is then obtained by summing over all particles
\begin{equation} \label{eq:Itot}
   \mathcal{I}_{\text{tot}} = \sum_{n\in \mathbb{Z}} \, \mathcal{I}_n \,.
\end{equation}
Restricting our attention to the line integrals and making the rotation $t \rightarrow t e^{i\theta}$, we may write
\begin{equation}
  \mathcal{I}_{\text{line},\,n} = \text{Re} \left[ \int_{\mathbb{R}_+ e^{i\theta}} \frac{dt}{t}  \frac{ \cos(|n| \beta t) }{ \sinh^2\left(\frac{t}{2}\right) }\right] =  \text{Re} \left[ \int_{\mathbb{R}_+ } \frac{dt}{t}  \frac{ \cos(|n| |\beta| t) }{ \sinh^2\left( \frac{t |\beta|}{2 \beta}\right)} \right] \,,
\end{equation}
such that using the identity 
\begin{equation}
    \sum_{n\in\mathbb{Z}} \cos(|n| |\beta| t) = \sum_{n\in \mathbb{Z}} e^{i n |\beta| t} = \sum_{k\in \mathbb{Z}} \delta\left(\frac{t |\beta|}{2\pi} - k  \right) \,,
\end{equation}
we can perform a Poisson resummation of the line integrals
\begin{equation} \label{eq:PoissonRersummation}
    \sum_{n\in\mathbb{Z}}  \mathcal{I}_{\text{line},\, n} = \text{Re} \left[ \sum_{k \ge 1}  \frac{1}{k \sinh^2(\pi \, k \, \beta^{-1})} \right] = \text{Re} \left[ - 4 \sum_{k \ge 1}  k \, \mathrm{Li}_1\left(e^{- \frac{2 \pi k}{\beta} } \right)\right] \,.
\end{equation}
Equation \eqref{eq:PoissonRersummation} shows that this sum is perturbative (yet non-analytic) around $\beta^{-1}=0$. Summing over the residues we instead find
\begin{equation}
    \sum_{n\in \mathbb{Z}} \mathcal{R}_n = \frac{2}{\pi} \sum_{k \ge 1} \, \text{Im} \left[ \mathrm{Li}_2(e^{-2 \pi k \beta }) + 2 \pi k \beta \, \mathrm{Li}_1(e^{-2 \pi k  \beta })\right] \,,
\end{equation}
which is non-perturbative in $\beta^{-1}$. Thus, $\mathcal{I}_{\text{tot}}$ contains in general both perturbative and non-perturbative contributions as
\begin{equation}
    \mathcal{I}_{\text{tot}} =  \mathcal{I}_{\text{tot}}^{(p)} +  \mathcal{I}_{\text{tot}}^{(np)}  \,, \qquad \mathcal{I}_{\text{tot}}^{(p)}  =  \sum_{n\in\mathbb{Z}}  \mathcal{I}_{\text{line},\,n}\,, \qquad \mathcal{I}_{\text{tot}}^{(np)} =    \sum_{n\in \mathbb{Z}} \mathcal{R}_n \,,
\end{equation}
and it moreover has the compact expression
\begin{equation}  \label{eq:SummedItot}
    \mathcal{I}_{\text{tot}} =   -\frac{ 2}{\pi }\text{Re}\left\{ \beta^2   \frac{d}{d\beta} \sum_{k\ge 1} \left( \mathrm{Li}_2\left(e^{-\frac{2\pi k}{\beta}}\right)  + \frac{i}{\beta}  \mathrm{Li}_2\left(e^{-2\pi \beta k }\right) \right) \right\} \,.
\end{equation}

Finally, we remark that there exist two special configurations, corresponding to the purely electric ($B = 0$) and purely magnetic ($E = 0$) backgrounds. In the former case, the residues appearing in \eqref{eq:residuesR} become real and $\mathcal{R}$ vanishes identically. Therefore, we can still divide the Schwinger integral according to \eqref{eq:SchwingerDecomposition}, but the poles do not contribute. On the other hand, in the purely magnetic scenario the decomposition $\mathcal{I} =  \mathcal{I}_{\text{line}} +  \mathcal{R}$ breaks down. The ray $\mathbb{R}_+ e^{i \pi/2}$ would lie now on the imaginary axis, which prevents us from picking up the residues by deforming the contour in \eqref{eq:SchingerManipulated}. Notice that the exact same phenomenon was observed in \cite{Castellano:2025yur} when studying the Wald entropy of supersymmetric black holes in Type IIA string theory. This connection becomes more transparent when summing over a tower of particles. In that case, \eqref{eq:SummedItot} is equal to the corrections of $\text{Im} \, \mathcal{F}$ computed with the Gopakumar-Vafa prepotential evaluated precisely at the attractor geometry \cite{Castellano:2025ljk}. Indeed, expressing \eqref{eq:SchingerManipulated} in terms of the $\mathcal{N}=2$ central charges via \eqref{eq:chargeDefSugra}---and reminding $q_e> 0 $ and $q_m > 0$, we get 
\begin{equation}
    \mathcal{I} =   \text{Re} \left[\int_{0^+}^{\infty} \frac{dt}{t}  \frac{ e^{-i Z_{\rm BH} \bar{Z}_{\rm p} t} }{ \sinh^2\left(\frac{t}{2}\right)} \right] \,.
\end{equation}

\subsection{Background (in)stability and Schwinger effect}\label{ss:Instability&Schwinger}

In this section, we briefly comment of the stability of the AdS$_2 \times \mathbf{S}^2$ backgrounds studied herein as well as the implications for black hole decay due to Schwinger pair production \cite{Schwinger:1951nm}.

\medskip

We start by reviewing the close relation between vacuum instabilities and the development of an imaginary part in the effective action, both in the Euclidean and Lorentzian pictures. In the Lorentzian quantum theory, one defines the vacuum persistence as the ratio of the vacuum amplitudes with and without the gauge background $A$ turned on, namely
\begin{equation} \label{eq:vacuumPersistence}
     \frac{\braket{0_A|0_A}} {\braket{0|0}} = \frac{\int D \varphi \, e^{i S[\varphi,A]}}{\int D \varphi \, e^{i S[\varphi]}} \,.
\end{equation}
The norm of such quantity is then interpreted as the probability that the system persists in the vacuum state $\ket{0_A}$, and it is related to the decay rate (per unit volume) $\Gamma$ via
\begin{equation}
   \left|    \frac{\braket{0_A|0_A}} {\braket{0|0}}\right|^2  = e^{-\Gamma V} \,,
\end{equation}
where $V$ is a regulator accounting for the divergent spacetime volume. Interpreting the right-hand-side of \eqref{eq:vacuumPersistence} as the definition of an effective action $\Gamma[A]$, one easily obtains
\begin{equation}
    \Gamma =   \frac{2\,  \text{Im} \,\Gamma[A]}{V}\, \sim\,  2\, \text{Im} \, \mathcal{L}_{\text{eff}} \,.
\end{equation}
The Euclidean picture is slightly more subtle. Upon performing a Wick rotation $t\to-it$, the imaginary part of the Lorentzian effective action translates into an imaginary part of its Euclidean analogue. However, the converse is not necessarily true. A simple example is provided by theories with Chern–Simons (more generally topological) terms, where an imaginary term in the Euclidean Lagrangian is mapped to a real one in Lorentzian signature. Still, in order for the Lorentzian theory to exhibit an instability, the Euclidean action must develop an imaginary part. In 1-loop exact theories $\Delta \Gamma \sim \log \det \mathcal{O}$, and such an imaginary contribution can arise from the phase of the determinant of the quadratic fluctuation operator, $\mathcal{O}$. Notice that existence of unstable directions, i.e., non-positive eigenmodes of $\mathcal{O}$, does not automatically imply an instability, since their associated phases may cancel each other (e.g., when there is an even number of negative modes). Additional phases may also arise from the regularization of the zero modes, which are typically associated with anomalies \cite{Fujikawa:1979ay}.

\smallskip

If one instead works with a Schwinger integral representation of the 1-loop effective action, the emergence of instabilities is typically seen in the form of a divergence when integrating over Schwinger-proper time. The reason being that when $\text{Tr} \log \mathcal{O}$ is complex, one needs to perform an extra analytic continuation to be able to apply the identity \eqref{eq:lorentzianSwchwinger}. We illustrate this in what follows for the determinants considered in Section \ref{ss:ExactComputation}. In particular, we consider  the integral in \eqref{SIntegral}, which we write here again for convenience
\begin{equation}
     I_S(\Delta) = \frac{1}{\pi} \frac{1}{\epsilon^2 + t^2 + u^2} \int_0^{\infty}  dy\, \, e^{-y -\frac{\Delta^2(\epsilon^2 + t^2 + u^2)}{4y}}  \,,
\end{equation}
where $\Delta^2 = \tilde{m}^2 - q_e^2 - q_m^2$. Let us focus now on the case where $\tilde{m}^2<q_e^2 + q_m^2$, corresponding to a super-extremal particle. Notice that now $\Delta \in i \mathbb{R}$. However, we cannot insert this directly into the integral because the result would be infrared divergent. To overcome this issue, we consider instead the analytic continuation of $I_S(\Delta)$ expressed in terms of the order-one modified Bessel function (cf. \eqref{SIntegral}). Hence, we perform the replacement
\begin{equation}
     I_S(i |\Delta|)  = \frac{1}{\pi} \frac{i |\Delta| }{\sqrt{\epsilon^2 + t^2 + u^2}} K_{1}(i |\Delta| \sqrt{\epsilon^2 + t^2 + u^2})\,.
\end{equation}
The above expression can be further simplified using the connection formulas \cite{watson1944bessel}
\begin{subequations}
\begin{align}
    K_\nu(z) & = - \frac{1}{2} \pi i e^{- i \nu \frac{\pi}{2}} H_{\nu}^{(2)}(z e^{- i \frac{\pi}{2}}) \,,  \qquad \theta_z \in \left[- \frac{\pi}{2}, \pi \right] \,,  \\
    H_{\nu}^{(2)}(z) & = J_\nu (z) - i Y_\nu(z) \,,
\end{align}
\end{subequations}
where $J_\nu$, $Y_\nu$ and $H_\nu^{(2)}$ denote Bessel functions of the first, second and third kind, respectively. We then find
\begin{equation}\label{analyticContK}
   I_S(i |\Delta|) = -\frac{ |\Delta|}{2\sqrt{\epsilon^2 + t^2 + u^2}} \left[Y_1(|\Delta|\sqrt{\epsilon^2 + t^2 + u^2})) + i J_1(|\Delta|\sqrt{\epsilon^2 + t^2 + u^2})) \right]  \, .
\end{equation}
As is well-known $K_\nu(z)$, as well as the other relevant Bessel functions, are real when $z \in \mathbb{R}$. Taking negative values for $\Delta^2$, we obtain an imaginary part defined by $J_1(z)$, together with a real piece from $Y_1(z)$. Notice that with \eqref{analyticContK} we cannot perform the same steps we carried out in Section \ref{ss:AdS2xS2Trace} for the BPS case $\Delta = 0$. Indeed, the Bessel functions have a branch point at $z = 0$ and they are defined on the complex plane up to a branch-cut, usually placed along the negative real axis. Therefore, eq.~\eqref{eq:Iuintegral} cannot be written as a contour integral.

\smallskip

An immediate consequence of the analysis carried out here is that super-extremal particles are required in order to trigger black hole decay via Schwinger pair production.\footnote{See also \cite{Castellano:2025yur} for a related semiclassical analysis} We also verified explicitly that the supersymmetric AdS$_2 \times \mathbf{S}^2$ background is stable in the BPS theory. We postpone a more systematic analysis of the 1-loop determinant of super-extremal particles in AdS$_2 \times \mathbf{S}^2$ to \cite{WGCandSPP}, where the precise decay condition will be presented.

\section{Integrating Out Supersymmetric Particles in AdS$_2 \times \mathbf{S}^2$}\label{s:SusyLoopDeterminant}

In Section \ref{ss:Susic1LoopMinimal}, we combined the exact functional traces derived for charged, massive scalar and spinor fields in AdS$_2 \times \mathbf{S}^2$ to obtain the 1-loop effective action induced by a BPS hypermultiplet minimally coupled to the near-horizon background of a supersymmetric black hole in four-dimensional $\mathcal{N}= 2$ supergravity. However, as already commented therein, the actual two-derivative Lagrangian describing the relevant dynamical fields contains additional interactions that couple the fermionic degrees of freedom to the graviphoton in a non-minimal fashion. This is reviewed in detail in Section \ref{ss:4dN=2HyperLagrangian} below. Accordingly, our main task in this section will be to modify the previous analysis by taking into account the effect of such Pauli-like terms. We will do so adopting two, very different strategies. First, in Section \ref{ss:ExactHyperDeterminant} we exploit the superconformal symmetries of AdS$_2\times \mathbf{S}^2$ to set up an on-shell computation that manifestly diagonalizes the bulk interactions, following the approach of \cite{Keeler:2014bra}. This provides a streamlined derivation from which one may readily see that including the kinetic mixing introduces two additional zero modes on the sphere compared to the minimally coupled case, cf. eq.~\eqref{eq:AdS2xS2effaction}. To give further evidence, we present in Section \ref{ss:TwistedDiracDiagonalization} the explicit diagonalization of the relevant fermionic operator similar to the one performed in \cite{Sen:2012kpz}, which agrees with our previous result.

\subsection{The BPS hypermultiplet action}\label{ss:4dN=2HyperLagrangian}

Hypermultiplets provide the simplest matter content in theories preserving eight supercharges. In four spacetime dimensions, each of these superfields contains four real scalars as well as one Dirac spin-$\frac12$ fermion. When coupled to 4d $\mathcal{N}=2$ supergravity, and assuming a purely graviphoton background, the BPS hypermultiplet Lagrangian takes the following form \cite{Freedman:2012zz,Lauria:2020rhc}
\begin{eqn}\label{eq:N=2HypermultipletLagrangian}
    \mathcal{L} =-\delta_{ij} \left( (D_\mu \phi^i)^\dagger D^\mu \phi^j + |Z|^2 (\phi^i)^\dagger\phi^j\right) + i \bar{\Psi} \slashed{D} \Psi +\frac{1}{4} \bar{\Psi} \gamma^{\mu \nu} W_{\mu \nu} \Psi -i \bar{\Psi} \left(Z P_+ -\bar{Z} P_-\right) \Psi\, .
\end{eqn}
Here, we have grouped the scalars into two complex fields $\phi^i$, $i=1,2$, whereas our conventions for the $\gamma$-matrices and Dirac conjugation are summarized in footnote \ref{fnote:conventionDirac}.  The relevant kinetic operators are $D_\mu =\partial_\mu -i V_\mu$ for the scalar field, and $\slashed{D}=-i\gamma^\mu(\nabla_\mu-i V_\mu)$ for the Dirac fermion, which are covariant both with respect to $U(1)$ gauge transformations of the graviphoton field as well as diffeomorphisms. Notice that the scalar and spin-$\frac12$ fields exhibit identical masses, being these controlled by the central charge $Z$ of the hypermultiplet
\begin{eqn}
    Z= e^{K/2} \left( p^A\mathcal{F}_A-q_A X^A\right) \, .
\end{eqn}
Relatedly, they share the same charge with respect to the graviphoton gauge connection, whose curvature 2-form
\begin{equation}
    dV = \frac{i}{2}\bar{Z}W^- - \frac{i}{2}Z W^+\,,\qquad \text{with}\quad W^{\pm}= \frac12 (W\mp i\star W)\, ,
\end{equation}
can be deduced from the effective field strength $F (p,q)=p^A G_A-q_AF^A$ seen by a particle with quantized charges $(p^A, q_A)$ \cite{Castellano:2025ljk, Castellano:2025rvn}. When specialized to the AdS$_2 \times \mathbf{S}^2$ near-horizon geometry of interest in here (see Section \ref{sss:nearhorizonBH} for details), the graviphoton background reads
\begin{eqn}
    W= \frac{2}{R^2}\left( \text{Re}\, Z_{\rm BH}\, \omega_{\mathbf{S}^2}-\text{Im}\, Z_{\rm BH}\, \omega_{\text{AdS}_2}\right) =\frac{2}{R}\left( \cos \varphi\,\omega_{\mathbf{S}^2}-\sin \varphi\, \omega_{\text{AdS}_2}\right)\, ,
\end{eqn}
with $Z_{\rm BH}= R\, e^{i\varphi}$ the central charge of the underlying black hole solution, and $R$ denotes the common radius of AdS$_2$ and $\mathbf{S}^2$ (cf. eq.~\eqref{eq:Poincaremetric}). The above expression is obtained upon inserting the explicit form of the anti-/self-dual components of the corresponding field strength
\begin{eqn}\label{eq:(Anti)SelfDualW}
    W^-=\frac{Z_{\rm BH}}{R^2} \left( \omega_{\mathbf{S}^2}+i \omega_{\text{AdS}_2}\right)\, ,\qquad  W^+= \overline{W}^-=\frac{\bar{Z}_{\rm BH}}{R^2} \left( \omega_{\mathbf{S}^2}-i \omega_{\text{AdS}_2}\right)\, .
\end{eqn}

From \eqref{eq:N=2HypermultipletLagrangian} we also realize that the complex scalars couple to the gauge and gravitational backgrounds in a minimal way, with their mass and gauge parameters being determined by the vacuum expectation values of the massless fields in the supergravity theory. Consequently, for those the analysis performed in sections \ref{s:reviewAdS2S2} and \ref{s:integrationAdS2xS2} readily applies. The fermionic field, on the other hand, exhibits two important differences with respect to the action shown in \eqref{eq:FermionAction}. First, we observe that the mass term is complexified and chiral, such that (negative) positive chirality modes, defined in terms of the projectors $P_\pm =\frac12 (1\pm \gamma^5)$ with $\gamma^5=-i \gamma^0 \gamma^1 \gamma^2 \gamma^3$, couple to the (anti-)holomorphic central charge.  Second, we notice that there is an additional Pauli interaction that mixes the two chiralities in the presence of a non-trivial field strength. This latter piece is ultimately the one preventing us from identifying the 1-loop hypermultiplet determinant with the result already obtained in \eqref{eq:AdS2xS2effaction}. 

\smallskip

In the following sections, we explain in detail how one can take these two effects into account and derive the actual supersymmetric trace we seek for. Before doing so, however, and in order to simplify the fermionic operator that needs to be diagonalized, we will make use of the gauge redundancy associated with the holomorphic line bundle inherent to the vector multiplet moduli. Recall that the space of vacua can be parametrized by some set $X^A$, with $A = 0, \ldots, n_V,$ of complex projective coordinates, which may be rescaled by an arbitrary holomorphic function, since the actual fields are given by the (invariant) ratios $z^a=X^A/X^0$. If we moreover restrict to $U(1)$ transformations of the form
\begin{eqn}\label{eq:Rsymmtransf}
    X^A \to e^{-i \beta} X^A\, ,\qquad \mathcal{F}_A \to e^{-i \beta} \mathcal{F}_A\, ,
\end{eqn}
we see that this actually modifies the way in which we describe the central charge of any BPS particle in the theory via the map $Z \to e^{-i \beta} Z$, as well as the graviphoton background field, since one also has that $W_{\mu \nu}^- \to e^{-i \beta} W_{\mu \nu}^-$. Nevertheless, this kind of transformations should not alter the relevant physics, given that both the attractor point and the physical couplings of the particles (mass, charges, etc.) are left unchanged under such field redefinitions. Therefore, in what follows it will be convenient to perform a rotation of this type with $\beta =\varphi-\pi/2$, which amounts to selecting a frame where $Z_{\rm BH}=iR$ and thus $W=-\frac{2}{R} \omega_{\text{AdS}_2}$ lies along AdS$_2$. Under this transformation, the fermionic piece of the Lagrangian \eqref{eq:N=2HypermultipletLagrangian} gets mapped to
\begin{eqn}\label{eq:N=2HypermultipletLagrangianRotated}
    \mathcal{L}_{\Psi} = i \bar{\Psi} \slashed{D} \Psi +\frac{i}{4} \bar{\Psi} \gamma^{\mu \nu} \gamma^5 e^{-i\varphi \gamma^5} W_{\mu \nu} \Psi +\frac{1}{R} \bar{\Psi} \left(Z \bar{Z}_{\rm BH}P_+ +\bar{Z} Z_{\rm BH} P_-\right) \Psi\, ,
\end{eqn}
whilst the effective electric and magnetic charges, $q_e=\text{Re}\, \bar{Z}_{BH} Z$ and $q_m= \text{Im}\, \bar{Z}_{BH}Z$, perceived by the BPS particle---and hence the corresponding Dirac operator---remain unchanged. Notice that, in the constant background field approximation, the transformation \eqref{eq:Rsymmtransf} is seen to be completely equivalent to an axial redefinition of the form $\Psi \to e^{-i\beta/2 \gamma^5} \Psi$, which should only modify the 1-loop determinant through the familiar chiral anomaly \cite{Adler:1969gk,Bell:1969ts,Fujikawa:1979ay}. 

\subsection{Computing the 1-loop determinant}\label{ss:ExactHyperDeterminant}

In general, the determination of the 1-loop effective action due to massive (and massless) multiplets in supersymmetric AdS$_2 \times \mathbf{S}^2$ spacetimes can become significantly challenging. This happens because of the rich matter content of such supermultiplets, as well as the non-minimal interactions among the different fields, which typically lead to complicated kinetic operators whose diagonalization is non-trivial (see Section \ref{ss:TwistedDiracDiagonalization} below).

Interestingly, there exists an alternative, much simpler route that one can follow so as to perform this kind of computations. The approach, pioneered by \cite{Keeler:2014bra}, consists in exploiting the superconformal symmetries exhibited by the background to extract the on-shell spectrum of the kinetic operator in a way that manifestly diagonalizes all the relevant interactions appearing in the supergravity theory. This is what we review next.

\medskip

Recall that the isometry group of AdS$_2 \times \mathbf{S}^2$ is given by $SU(1,1)\times SU(2)$. Moreover, these symmetries are exactly preserved in the presence of constant (and everywhere orthogonal) electric and magnetic fields. Consequently, as already explained in Section \ref{s:reviewAdS2S2}, they can be used to classify the different irreducible representations associated with physical particles propagating within these spacetimes. The latter are characterized by a pair $(h, j)$ of quantum numbers, where $h$ corresponds to the lowest-weight of the $K_0$ generator of $SU(1,1)$ (hence defining a primary state), whereas $j$ labels the appropriate $SU(2)$ representation. From this, one may construct an infinite tower of `descendants' by acting with $K_+$, thereby raising the $h$ number by one unit each time. Similarly, the allowed values of $j$ depend on the magnetic field $B=q_m/R^2$ threading the 2-sphere, and correlate with the spin of the field as follows
\begin{equation}\label{eq:SU2NumbersHyper}
\begin{aligned}
    & \underline{\text{spin-}0}: \qquad  \quad j=n+|q_m|\, ,\\
    & \underline{\text{spin-}1/2}: \qquad  j=n+|q_m|-\frac12\, ,
\end{aligned}
\end{equation}
with $n=0,1\ldots, \infty$. The on-shell condition then amounts to a certain constraint satisfied by the quadratic Casimir operators of $SU(1,1)$ and $SU(2)$ \cite{Castellano:2025rvn}, which relates, in turn, $h$ and $j$.

\smallskip

\noindent When embedded into $\mathcal{N}=2$ supergravity, the situation is improved thanks to supersymmetry. In that case, the background solution preserves 8 supercharges, and the symmetry group enhances to $SU(1,1|2)$. Therefore, the fields propagating therein must furnish themselves representations of the superconformal algebra. When the particles are in addition BPS, the representations become `short', since only half of the available supecharges act non-trivially on those. If translated to the above language, this means that the set of possible $(h, j)$ must organize into different combinations of chiral multiplets, which for us will take the form \cite{Keeler:2014bra}
\begin{eqn}\label{eq:chiralmultiplet}
    (j,j) \oplus 2 \times \left( j+\frac12,j-\frac12\right) \oplus (j+1, j-1)\,.
\end{eqn}
Since a BPS hypermultiplet contains four bosonic as well as fermionic degrees of freedom, with $SU(2)$ numbers given by \eqref{eq:SU2NumbersHyper}, one quickly realizes that there is a unique way to arrange those in terms of (towers of) chiral multiplets. Namely, one finds two copies of the set
\begin{eqn}\label{eq:chiralprimariesAdS2xS2}
    \left(k+|q_m|+\frac12, k+|q_m|+\frac12\right) \oplus 2\times \left(k+|q_m|+1, k+|q_m|\right) \oplus \left(k+|q_m|+\frac32, k+|q_m|-\frac12\right)\,,
\end{eqn}
with $k\in \mathbb{Z}_{\geq 0}$. Crucially, the reorganization of the different modes within the hypermultiplet in terms of chiral states already diagonalizes all interactions required by $\mathcal{N}=2$ supergravity. Hence, we can read directly from \eqref{eq:chiralprimariesAdS2xS2} the contribution of each of these pieces to the 1-loop path integral. For instance, the scalar sector (middle term) gives
\begin{eqn}
    \mathcal{K}^{(0)}_{\text{AdS}_2\times \mathbf{S}^2}(\tau) = \Tr \left[e^{-\tau \mathcal{D}_{\text{AdS}_2}^2} \right] 2\, e^{ \frac{\tau}{R^2} q_m^2}\left(2\sum_{k \geq 0} \left( k + |q_m| + \frac12\right) e^{ -\frac{\tau}{R^2} \left( k + |q_m| \right)(k+|q_m|+1)}\right)\,,
\end{eqn}
where we used that the contribution due to on-shell states of the form $(h,j)$ and $(h=1, j=0)$ is related as follows
\begin{eqn}\label{eq:TraceContributions(h,j)}
    \mathcal{K}^{(0)}_{\text{AdS}_2\times \mathbf{S}^2}(h, j;\tau) =  \mathcal{K}^{(0)}_{\text{AdS}_2\times \mathbf{S}^2}(h=1, j=0;\tau) e^{-\frac{\tau}{R^2} \left(h(h-1)-q_m^2\right)} (2j+1)\,,
\end{eqn}
with $\mathcal{K}^{(0)}_{\text{AdS}_2\times \mathbf{S}^2}(h=1, j=0;\tau)$ denoting the heat kernel trace of the operator \eqref{eq:helec}. Notice that this exactly matches twice the result obtained for a charged scalar shown in eq.~\eqref{eq:fullintegralspin0}, as expected since we already observed in the previous section that the hypermultiplet scalars are indeed minimally coupled to the AdS$_2 \times \mathbf{S}^2$ background (see discussion around \eqref{eq:(Anti)SelfDualW}).

\medskip

\noindent The fermionic contribution, on the other hand, arises from the first and third terms in \eqref{eq:chiralprimariesAdS2xS2}, and it yields\footnote{Here again one needs to use the spin-$\frac12$ version of \eqref{eq:TraceContributions(h,j)}, which reads

\begin{eqn}\label{eq:Spin12TraceContributions(h,j)}
    \mathcal{K}^{(1/2)}_{\text{AdS}_2\times \mathbf{S}^2}(h, j;\tau) =  \mathcal{K}^{(1/2)}_{\text{AdS}_2\times \mathbf{S}^2}(h=1, j=0;\tau) e^{-\frac{\tau}{R^2} \left(h(h-1)-q_m^2+\frac14\right)} (2j+1)\,. \notag
\end{eqn}
}
\begin{eqn}\label{eq:tracewithmixing}
\begin{aligned}
   \mathcal{K}^{(1/2)}_{\text{AdS}_2\times \mathbf{S}^2}(\tau) = &\Tr \left[e^{-\tau \slashed{D}_{\text{AdS}_2}^2} \right] 2\, e^{ \frac{\tau}{R^2} q_m^2} \bigg(2\sum_{k \geq 0} \left( k + |q_m|+1\right) e^{ -\frac{\tau}{R^2} \left( k + |q_m| \right)^2} \\
   &+ 2(k+|q_m|)\,e^{ -\frac{\tau}{R^2} \left( k + |q_m| +1 \right)^2}\bigg)\,,
\end{aligned}
\end{eqn}
which reduces after a simple relabeling to
\begin{eqn}\label{eq:HeatKernelTraceHyperFermions}
   \mathcal{K}^{(1/2)}_{\text{AdS}_2\times \mathbf{S}^2}(\tau) =\Tr \left[e^{-\tau \slashed{D}_{\text{AdS}_2}^2} \right] 2\, e^{ \frac{\tau}{R^2} q_m^2} \left(2\sum_{n \geq 1} \left( n + |q_m|\right) e^{ -\frac{\tau}{R^2} \left( n + |q_m| \right)^2} + (|q_m|+1)\,e^{ -\frac{\tau}{R^2} q_m^2}\right)\,.
\end{eqn}
Contrary to the scalar case, the Dirac fermion does not reproduce the 1-loop determinant derived for minimally coupled spin-$\frac12$ fields in \eqref{eq:fullintegralspin1/2}. However, it almost does so, with the only difference being captured by the additional `$+1$' in the second term of eq.~\eqref{eq:HeatKernelTraceHyperFermions} above. The latter gives a contribution that is formally equivalent to the presence of two additional zero modes of the Dirac operator on the sphere (cf. eq.~\eqref{eq:traceD2ferm}). Keeping track of these, we can write the complete functional trace associated with a massive BPS hypermultiplet as
\begin{eqn}\label{eq:finalhyperformulaPrequel}
\begin{aligned}
    \log{\mathcal{Z}_{\rm hm}} =&\, -\frac{V_{\text{AdS}}}{\pi R^2}\, q_e q_m \tan^{-1}\left(\frac{q_e}{q_m}\right)  +  \frac{V_{\text{AdS}}}{4\pi R^2}\int_{0^+}^{\infty} \frac{dt}{t} e^{- |q_m| t} \frac{ \cos(q_e t) }{ \sinh^2\left(\frac{t}{2}\right)}\\
    &+\frac{V_{\text{AdS}}}{\pi R^2}\int_{0^+}^{\infty} \frac{dt}{t} e^{- |q_m| t} \frac{d}{d t} \left(\frac{ \cos(q_e t) }{ \tanh\left(\frac{t}{2}\right)}\right)\,,
\end{aligned}
\end{eqn}
where the last piece arises from the non-minimal interactions in \eqref{eq:N=2HypermultipletLagrangian} and it may be recognized to be equal to the 1-loop effective action induced by two charged and massive fermions constrained to live within AdS$_2$ (cf. \eqref{eq:fermitraceAdS2}). Expanding the last term results in
\begin{eqn}\label{eq:finalhyperformula}
    \log{\mathcal{Z}_{\rm hm}} =\, -\frac{V_{\text{AdS}}}{\pi R^2} \text{Re}\left[ q_e q_m \tan^{-1}\left(\frac{q_e}{q_m}\right)  +  \frac14\int_{0^+}^{\infty} \frac{dt}{t}  \frac{e^{i(q_e + i|q_m|) t} }{ \sinh^2\left(\frac{t}{2}\right)} - i q_e\int_{0^+}^{\infty} \frac{dt}{t} \frac{ e^{i(q_e + i|q_m|) t}  }{ \tanh\left(\frac{t}{2}\right)}\right]\,.
\end{eqn}
Note that the sign in the $\csch^2(t/2)$ term has been flipped with respect to that appearing in \eqref{eq:AdS2xS2effaction}. The third term, on the other hand, is entirely new. It can be seen to cancel identically for purely magnetic backgrounds and in principle it modifies both the perturbative and non-perturbative structure of the 1-loop effective action.

\subsection{Explicit diagonalization of the $\mathcal{N}=2$ fermionic kinetic operator}\label{ss:TwistedDiracDiagonalization}

In this section, we present the explicit calculation of the 1-loop effective action induced by massive spin-$\frac12$ fields belonging to a BPS hypermultiplet in supersymmetric AdS$_2\times \mathbf{S}^2$. To do so, we first rotate the two-derivative fermionic action \eqref{eq:N=2HypermultipletLagrangianRotated} to Euclidean space, then diagonalize the relevant kinetic operator, compute the functional trace, and finally analytically continue the result back to Lorentzian signature.

\medskip

\noindent Therefore, upon Wick rotating to Euclidean time, $t \to -i \tau$, the operator $\slashed{D}=-i \gamma^\mu (\nabla_\mu-i V_\mu)$ becomes a well-defined Hermitian operator acting on sections of the spinor bundle over $\mathbb{H}^2 \times \mathbf{S}^2$. Our convention for the local frame (Euclidean) Dirac matrices in $\mathbb{H}^2\times \mathbf{S}^2$ is
\begin{eqn}
    \hat{\gamma}^0= \tau^1 \otimes \sigma^3\, , \quad \hat{\gamma}^1= \tau^2 \otimes \sigma^3\, , \quad \hat{\gamma}^2= \mathds{1} \otimes \sigma^1\, , \quad \hat{\gamma}^3= \mathds{1} \otimes \sigma^2\, ,
\end{eqn}
with $\lbrace\tau^i, \sigma^j\rbrace$ two independent sets of Pauli matrices. Similarly, the vierbeins read 
\begin{eqn}
    e^0= \frac{R}{\rho} dt\, ,\quad e^1= \frac{R}{\rho} d\rho\, ,\quad e^2= R d\theta\, ,\quad e^3= R \sin \theta d\phi\, ,
\end{eqn}
from where we readily see that the Pauli coupling appearing in \eqref{eq:N=2HypermultipletLagrangianRotated} reduces to
\begin{eqn}\label{eq:PauliTermAdS2}
    \frac14 \gamma^{\mu \nu}W_{\mu \nu}= -\frac{i}{R}\, \tau^3\otimes\mathds{1}\, ,\qquad \text{with}\quad \hat{\gamma}^a= e^a_{\ \mu} \gamma^\mu\, .
\end{eqn}
Regarding the mass term, and using the defining charge relations \eqref{eq:chargeDefSugra}, we find that
\begin{eqn}
    \mathcal{L}_{\rm mass} = \frac{1}{R} \bar{\Psi} \left(Z \bar{Z}_{\rm BH}P_+ +\bar{Z} Z_{\rm BH} P_-\right) \Psi = \frac{1}{R} \bar{\Psi} \left(q_e + q_m \gamma^5\right) \Psi\, ,
\end{eqn}
where we substituted the form of the chirality projectors $P_\pm =\frac12 (1\pm \gamma^5)$, with $\gamma^5=\gamma^0 \gamma^1 \gamma^2 \gamma^3$. Hence, putting everything together, we conclude that the relevant kinetic operator we need to diagonalize is
\begin{eqn}\label{eq:kineticoperator}
\begin{aligned}
    \mathds{D}= i\slashed{D}_{\mathbb{H}^2}\otimes \sigma^3 +\mathds{1} \otimes(i\slashed{D}_{\mathbf{S}^2}) - \frac{i}{R}\, \tau^3 \otimes \mathds{1} +\frac{q_e}{R}\, \mathds{1}\otimes\mathds{1}-\frac{i q_m}{R}\,\tau^3\otimes \sigma^3\, .
\end{aligned}
\end{eqn}
To proceed further, it is convenient to first expand the 4d spinors in a basis of direct products of eigenfunctions of $\slashed{D}_{\mathbf{S}^2}$ and $\slashed{D}_{\mathbb{H}^2}$, which verify
\begin{eqn}\label{eq:EigenFunctionsDiracAdS2S2}
    \slashed{D}_{\mathbf{S}^2} \eta_n = \zeta_n\, \eta_n\, ,\qquad \slashed{D}_{\mathbb{H}^2} \psi_{\lambda, k} = \zeta_{\lambda, k}\, \psi_{\lambda, k}\, ,
\end{eqn}
and where the spectrum of the corresponding two-dimensional Dirac operators was obtained in Section \ref{sss:fermionS2} and Section \ref{sss:fermionH2}, respectively. Recall that the latter operator features a discrete set of eigenmodes when defined on the 2-sphere, with eigenvalues given by
\begin{eqn}\label{eq:spectrucmDS2}
    \zeta_n = \frac{1}{R} \sqrt{(n+|q_m|)^2-q_m^2}\, ,\qquad n \in \mathbb{Z}_{\geq 0}\, ,
\end{eqn}
whereas in $\mathbb{H}^2$ one finds both discrete and continuous eigenstates---hence the notation $\psi_{\lambda, k}$, whose associated eigenvalues can be read directly from those shown in eqs.~\eqref{eq:eigenvaluesdiscreteDiracAdS2} and \eqref{eq:eigenvaluescontinuousDiracAdS2}.

\smallskip

\noindent Consequently, using the complete set $\lbrace \psi_{\lambda, k}\otimes \eta_n,\psi_{\lambda, k}\otimes \sigma^3\eta_n,\tau^3\psi_{\lambda, k}\otimes \eta_n,\tau^3\psi_{\lambda, k}\otimes \sigma^3\eta_n\rbrace$ as a basis of 4d spinors, one shall express the operator $\mathds{D}$ in terms of a block-diagonal matrix, with
\begin{eqn}\label{eq:MatrixFormGeneralD}
    \mathds{D}\big\rvert_{\lambda, k; n}=\begin{pmatrix}
  \frac{q_e}{R} + i\,\zeta_n & i\,\zeta_{\lambda, k}                 & -\frac{i}{R}  & -\frac{i q_m}{R}\\
  i\,\zeta_{\lambda, k}  & \frac{q_e}{R} - i\,\zeta_n & -\frac{i q_m}{R} & -\frac{i}{R}             \\
  -\frac{i}{R} & -\frac{i q_m}{R} & \frac{q_e}{R} + i\,\zeta_n & -\,i\,\zeta_{\lambda, k}\\
  -\frac{i q_m}{R} & -\frac{i}{R} & -\,i\,\zeta_{\lambda, k} & \frac{q_e}{R} - i\,\zeta_n
\end{pmatrix}\, ,
\end{eqn}
its restriction to a given sector with fixed eigenvalue of $\slashed{D}_{\mathbb{H}^2 \times \mathbf{S}^2}$.\footnote{Note that its precise eigenspectrum can be deduced by combining eqs.~\eqref{eq:SumOfSquares} and \eqref{eq:EigenFunctionsDiracAdS2S2}.}  Thus, a straightforward calculation reveals that the new eigenvalues are given by
\begin{eqn}\label{eq:eignvaluesDgeneral}
    \zeta^{(1)}_{\pm} (\lambda, k; n) &= \frac{1}{R} \left(q_e \pm \sqrt{-1-q_m^2+2 \sqrt{q_m^2+R^2 \zeta_n^2}-R^2(\zeta_{\lambda, k}^2+\zeta_n^2)}\right)\, ,\\
    \zeta^{(2)}_{\pm} (\lambda, k; n) &= \frac{1}{R} \left(q_e \pm \sqrt{-1-q_m^2-2 \sqrt{q_m^2+R^2 \zeta_n^2}-R^2(\zeta_{\lambda, k}^2+\zeta_n^2)}\right)\, .
\end{eqn}
From here, one may already determine the (Euclidean version of) the 1-loop partition function obtained by integrating out a massive Dirac fermion with kinetic operator \eqref{eq:kineticoperator} in $\mathbb{H}^2 \times \mathbf{S}^2$. However, care must be taken due to the generic existence of (chiral) zero modes of $\slashed{D}_{\mathbf{S}^2}$. Thus, in what follows we will analyze the contribution of the zero and non-zero sectors separately.

\subsubsection*{The non-zero mode sector}

Let us consider first the non-zero modes of the Dirac operator on the sphere. These correspond to the bispinors $\eta_n$ with $n\geq 1$ in \eqref{eq:EigenFunctionsDiracAdS2S2}, whose eigenvalues can be read from eq.~\eqref{eq:spectrucmDS2}. Hence, their contribution to the Euclidean path integral can be obtained by repeating steps similar to those presented in Section \ref{sss:1loopbasics}. Upon doing so, one finds that the latter is computed via the following functional trace (equivalently determinant)
\begin{equation}\label{eq:Fermion1Loop}
\log \mathcal{Z}_\Psi \supset -\log \text{det}'\, \mathds{D}=-\text{Tr}' \log \mathds{D}\,, 
\end{equation}
where the superscript indicates that we exclude the $n=0$ modes from the trace (determinant). Notice that the logarithm allows us to take the product over eigenvalue pairs with opposite sign labels in \eqref{eq:eignvaluesDgeneral}, yielding
\begin{eqn}
    \zeta^{(1)}_{+} \zeta^{(1)}_{-} &= \frac{1}{R^2} \left( q_e^2 +\zeta_{\lambda, k}^2+(n+|q_m|-1)^2\right)\, ,\\
    \zeta^{(2)}_{+} \zeta^{(2)}_{-} &= \frac{1}{R^2} \left( q_e^2 +\zeta_{\lambda, k}^2+(n+|q_m|+1)^2\right)\, ,
\end{eqn}
which results in two separate towers of states whose associated spectrum on the hyperbolic plane remains unchanged with respect to the minimally coupled case (see Section \ref{sss:spectrumH2}), whereas the Landau levels on $\mathbf{S}^2$ are shifted by one unit in opposite directions, cf. eq.~\eqref{eq:eigenvaluesDiracS2}. This implies, in turn, that the corresponding set of Landau energies remains invariant but the degeneracies no longer match with the appropriate $SU(2)$ quantum number $j=k+\frac12$. However, when combining the two towers together one restores the correct degeneracy at each level, except for a couple of additional zero modes that descend from the former $n=1$ states. Note that this also matches the behaviour exhibited by the analogous massless states in the theory \cite{Sen:2012kpz}, as reviewed in Appendix \ref{ss:masslesshyper}. We depicted this schematically in Figure \ref{fig:EigenvalueShift} below.

\smallskip

Consequently, in order to compute the functional trace, and given the positive definiteness of the operator, one may use the Euclidean version of Schwinger proper time reparametrization
\begin{eqnarray}\label{eq:EuclideanSwchwinger}
{\cal O}^{-1} = \int^{\infty}_\epsilon d\tau\, e^{-\tau\, {\cal O}} \,,
\end{eqnarray}
together with the block-diagonal structure of $\mathds{D}$, to write the 1-loop partition function in the non-zero mode sector as
\begin{eqn}\label{eq:fermionicHyperTraceH2xS2}
    \log \mathcal{Z}_\Psi \supset \int_{\epsilon}^{\infty} \frac{d\tau }{\tau} \Tr \left[e^{-\tau \left(\slashed{D}_{\mathbb{H}^2}^2 + \frac{q_e^2}{R^2} \right)}\right]\bigg\lbrace \sum_{n \geq 1} \left( n + |q_m|\right) \left(e^{ -\frac{\tau}{R^2} \left( n + |q_m| -1\right)^2} + e^{ -\frac{\tau}{R^2} \left( n + |q_m| +1\right)^2}\right)\bigg\rbrace\, .
\end{eqn}
Next, we perform an analytic continuation of the heat kernel trace, following the analysis of Section \ref{sss:fermionAdS2}. Crucially, by proceeding this way we avoid having to deal explicitly with the more complicated spectrum of the Dirac operator on $\mathbb{H}^2$.  This then transforms \eqref{eq:fermionicHyperTraceH2xS2} into 
\begin{eqn}\label{eq:fermionicHyperTraceAdS2xS2}
    \log \mathcal{Z}_\Psi \supset \int_{\epsilon}^{\infty} \frac{d\tau }{\tau} \Tr \left[e^{-\tau \left(\slashed{D}_{\text{AdS}_2}^2 + \frac{q_e^2}{R^2} \right)}\right]\bigg\lbrace\sum_{n \geq 1} \left( n + |q_m|\right) \left(e^{ -\frac{\tau}{R^2} \left( n + |q_m| -1\right)^2} + e^{ -\frac{\tau}{R^2} \left( n + |q_m| +1\right)^2}\right)\bigg\rbrace\, ,
\end{eqn}
where the first factor inside the $\tau$-integral was already computed in closed form in eq.~\eqref{eq:TraceSpin12Fourier}.

\begin{figure}[t]
    \centering
    \includegraphics[scale=0.45]{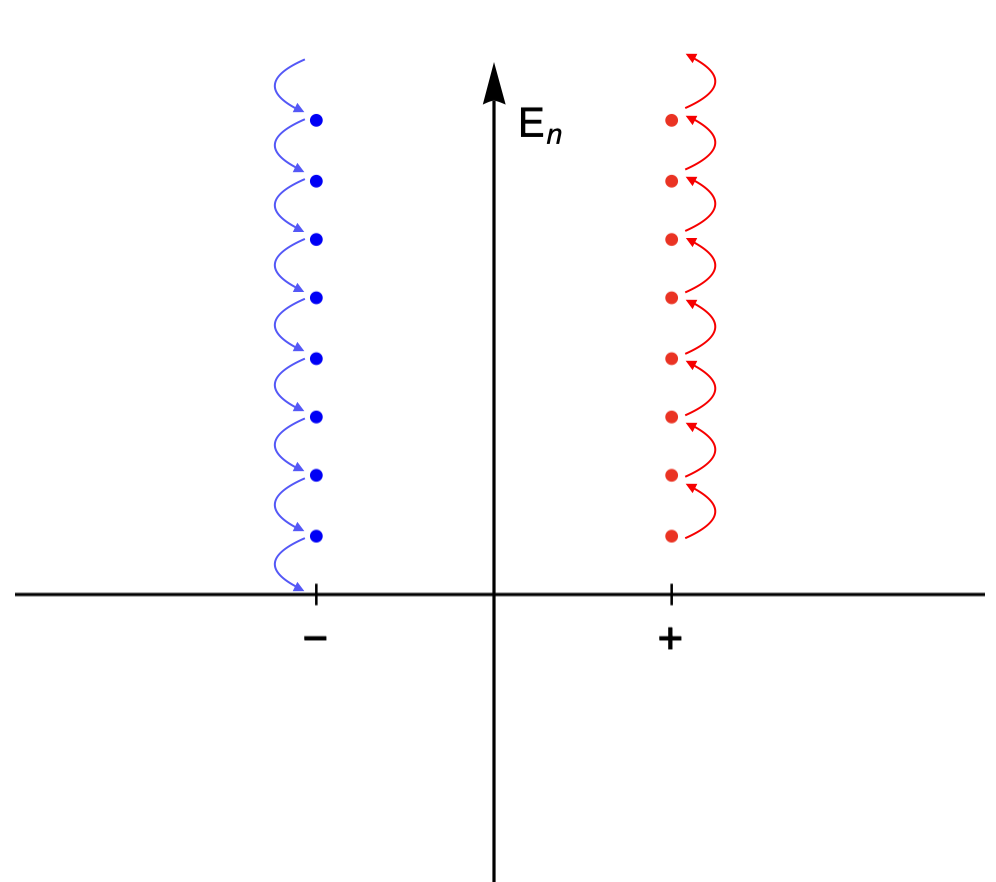}
    \caption{\small The effect of the Pauli term \eqref{eq:PauliTermAdS2} on the spectrum of the fermionic kinetic operator $\mathds{D}$ in AdS$_2\times\mathbf{S}^2$ is to shift the eigenvalues associated with the spherical sector by one unit, in opposite directions depending on the chirality of the 4d spinors. As a consequence, new zero modes can appear.}
    \label{fig:EigenvalueShift}
\end{figure}

\subsubsection*{The zero mode sector}

\begin{figure}[t]
    \centering 
    \includegraphics[scale=0.45]{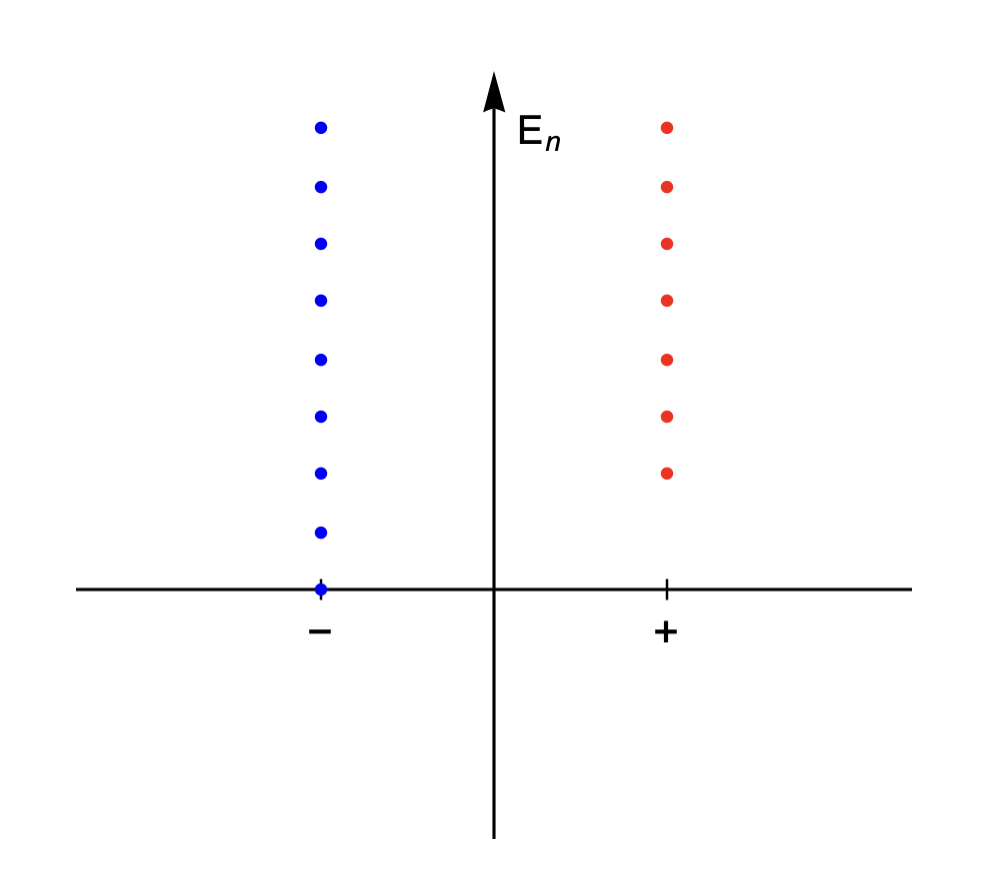}
    \caption{\small In the presence of a non-vanishing magnetic field $B$ on the sphere (cf. eq.~\eqref{eq:MetricMaxwellSphere}), the spectrum of the Dirac operator becomes chiral, see Section \ref{sss:fermionS2} for details. This results in the existence of $2R^2 |B|$ zero modes which are annihilated by $\slashed{D}_{\mathbf{S}^2}$. For concreteness, we have assumed $\text{sgn}(B)=-1$.}
    \label{fig:EigenvalueSpectrumNonZeroQm}
\end{figure}

As mentioned before, in the most general situation where $q_m \neq0$, there are additional zero modes associated with $\slashed{D}_{\mathbf{S}^2}$ (see Figure \ref{fig:EigenvalueSpectrumNonZeroQm}), which have not yet been included in the functional determinant \eqref{eq:fermionicHyperTraceAdS2xS2} above. The main difficulty in accounting for those rests on the fact that, due to their definite chirality,\footnote{See Appendix \ref{ss:ZeroModesS2} for a careful exposition of these matters.} some of the elements of our basis of four-dimensional spinors will now be linearly dependent (cf. discussion around \eqref{eq:MatrixFormGeneralD}). Thus, one might worry about whether the diagonalization performed herein remains valid when restricting to this sector.
However, by studying a simpler example, namely the minimally coupled case considered in previous sections translated to the present language (see Appendix \ref{ss:freemassivehyper} for details), one can get easily convinced that the same procedure still works, provided we take into account the reduction in the actual number of eigenmodes. Therefore, to complete our analysis we just need to study carefully what happens with the zero modes (when they exist). In particular, one sees that the eigenvalues \eqref{eq:eignvaluesDgeneral} of the kinetic operator $\mathds{D}$ simplify for $n=0$ to 
\begin{eqn}\label{eq:eignvaluesDgeneraln=0}
    \zeta^{(1)}_{\pm} (\lambda, k; n=0) &= \frac{1}{R} \left(q_e \pm i\sqrt{R^2\zeta_{\lambda, k}^2 +(|q_m| - 1)^2}\right)\, ,\\
    \zeta^{(2)}_{\pm} (\lambda, k; n=0) &= \frac{1}{R} \left(q_e \pm i\sqrt{R^2\zeta_{\lambda, k}^2 +(|q_m| + 1)^2}\right)\, .
\end{eqn}
Their associated eigenspinors read
\begin{eqn}
    \Psi^{(1)}_\pm (\lambda,k; n=0)&= \begin{pmatrix}
    R\zeta_n \mp \varepsilon \sqrt{R^2\zeta_\lambda^2 + (|q_m|-1)^2}\\
  -\left(\varepsilon R\zeta_n \mp \sqrt{R^2\zeta_\lambda^2 + (|q_m|-1)^2}\right) \\
  -\varepsilon \left(|q_m|-1\right)\\
  |q_m|-1 
\end{pmatrix}\, ,\\ \Psi^{(2)}_\pm (\lambda,k; n=0)&= \begin{pmatrix}
    -\left(R\zeta_n \pm \varepsilon \sqrt{R^2\zeta_\lambda^2 + (|q_m|+1)^2}\right)\\
  -\left(\varepsilon R\zeta_n \pm \sqrt{R^2\zeta_\lambda^2 + (|q_m|+1)^2}\right) \\
  \varepsilon \left(|q_m|+1\right)\\
  |q_m|+1 
\end{pmatrix}\, ,
\end{eqn}
where we denote $\varepsilon=\text{sgn}\, (q_m)$. Hence, since $\sigma^3 \eta_{n=0} = \varepsilon \eta_{n=0}$ (see discussion around \eqref{eq:ChiralityZeroModes}) we immediately conclude that $\Psi^{(1)}_\pm (n=0)$ vanish identically, irrespective of the sign of $q_m$. This implies that, within the $n=0$ sector, only the spinors associated with the eigenvalues $\zeta^{(2)}_\pm (n=0)$ are actually populated. Consequently, upon taking the spherical trace one obtains
\begin{eqn}
\begin{aligned}
     \Tr_{\mathbf{S}^2} &\left[e^{\tau \mathds{D}^2} \right] =2\sum_{k =0}^\infty \left( k +|q_m|+ 1\right) \left[e^{ -\frac{s}{R^2} (k+|q_m|)^2} + e^{ -\frac{s}{R^2} \left( k +|q_m| +2 \right)^2}\right]\\
     &=2 \sum_{k =0}^\infty\left[\left(k +|q_m|+ 1\right) e^{ -\frac{s}{R^2} (k+|q_m|)^2} + (k+|q_m|) e^{ -\frac{s}{R^2} \left( k+|q_m| +1 \right)^2}\right]- 2|q_m| e^{-\frac{s}{R^2} (|q_m|+1)^2}\, ,
\end{aligned}
\end{eqn}
for the $n\geq 1$ states, as well as a contribution
\begin{eqn}
     2|q_m| e^{-\frac{s}{R^2} (|q_m|+1)^2}\, ,
\end{eqn}
due to the remaining $n=0$ eigenmodes. Therefore, we find that the complete 1-loop partition function for the hypermultiplet fermions is (cf. eq.~\eqref{eq:fermionicHyperTraceAdS2xS2})
\begin{eqn}\label{eq:fermionicHyperTraceAdS2xS2Complete}
    \log \mathcal{Z}_\Psi = \int_{\epsilon}^{\infty} \frac{d\tau }{\tau} \Tr \left[e^{-\tau \left(\slashed{D}_{\text{AdS}_2}^2 + \frac{q_e^2}{R^2} \right)}\right]\left(2\sum_{n \geq 1} \left( n + |q_m|\right) e^{ -\frac{\tau}{R^2} \left( n + |q_m| \right)^2} + (|q_m|+1)\,e^{ -\frac{\tau}{R^2} q_m^2}\right)\, ,
\end{eqn}
which reproduces the result of Section \ref{ss:ExactHyperDeterminant} derived by other indirect methods.

\section{Conclusions and Outlook}
\label{s:conclusions}

In this work, we have carried out an exact analytic computation of 1-loop determinants for fields propagating in AdS$_2 \times \mathbf{S}^2$ geometries threaded by constant electric and magnetic fluxes. Our analysis is performed  within the constant background field approximation and employs the Schwinger proper-time formalism \cite{Schwinger:1951nm}. The resulting expressions are fully non-perturbative at one loop level and retain exact dependence on mass, charge, and background field parameters. 

\smallskip

A first outcome of our investigation is the calculation of the 1-loop path integral associated with massive spin-0 and spin-$\frac{1}{2}$ particles minimally coupled to the aforementioned gauge and gravitational backgrounds. We achieved this by exploiting the separability of the kinetic operators to reduce the overall functional determinant to the product of AdS$_2$ and $\mathbf{S}^2$ heat kernel traces. Making use of the Hubbard–Stratonovich transformation \cite{Stratonovich1957OnAM,Hubbard:1959ub}, we were able to express the result as a compact double line-integral. The latter allowed us, in turn, to address the non-perturbative stability of the background. In particular, we explicitly showed that only super-extremal particles---in the sense of Section \ref{ss:Instability&Schwinger}---can trigger an instability. Given that AdS$_2 \times \mathbf{S}^2$ describes the near-horizon region of asymptotically flat, static and extremal black hole solutions, this matches the classical expectations that super-extremal states are necessary for black holes to decay \cite{Arkanihamed:2006dz}. A detailed analysis of the precise bound that this mechanism enforces on the theory spectrum will be presented elsewhere \cite{WGCandSPP}.

\medskip

We then specialized these results to BPS particles in $\mathcal{N}=2$ supersymmetric AdS$_2 \times \mathbf{S}^2$ spacetimes. Supersymmetry imposes specific constraints relating the electric-magnetic fluxes experienced by charged probe particles, as well as their mass and the AdS curvature radius \cite{Billo:1999ip, Simons:2004nm, Castellano:2025rvn}. Therefore, enforcing the appropriate BPS mass-charge relations, we found out that the double integral collapses to a single Schwinger-like integral, thereby obtaining new exact expressions for the 1-loop effective actions of minimally coupled scalars and spin-$\frac12$ fields. Exploiting this simple representation, we showed that the non-perturbative structure becomes manifest upon a proper change of contour in the complex Schwinger plane. In particular, the coupling constant can be interpreted as the ratio of the Compton wavelength of the particle, $\ell_{\rm c}=m^{-1}$, to the AdS$_2$ curvature radius $R$. As a byproduct, we verified explicitly that 4d $\mathcal{N}=2$ BPS black hole solutions, which lead to supersymmetric AdS$_2\times \mathbf{S}^2$ near-horizon geometries, are non-perturbatively stable under Schwinger pair production.

\smallskip

Finally, we computed the 1-loop determinant of a $\mathcal{N}=2$ hypermultiplet. The main difference with the previous cases is the presence of non-minimal couplings of the fermions to the graviphoton $U(1)$ field strength. We performed the calculation following two orthogonal approaches (cf. sections \ref{ss:ExactHyperDeterminant} and \ref{ss:TwistedDiracDiagonalization}) that nevertheless led to the same answer. Interestingly, compared to the simpler case with minimal couplings, one finds additional contributions that mimic those of two extra massive spin-$\frac12$ fields propagating solely in AdS$_2$. This provides the full hypermultiplet 1-loop effective action, representing the main result in this work. 

\medskip

More precisely, our findings reveal three distinct contributions to the 1-loop partition function. The first one takes the form of a quantum-induced topological term with an effective theta angle depending on the particle and black hole central charges via $\theta=\pi/2+ \text{arg}(Z \bar{Z}_{\rm BH})$. The second term, on the other hand, resembles the structure of the original Gopakumar--Vafa integral evaluated at the attractor geometry \cite{Gopakumar:1998ii, Gopakumar:1998jq, Dedushenko:2014nya} (see also \cite{Hattab:2024ssg} for a recent discussion). Despite the fact that both our calculation and the Gopakumar--Vafa construction arise in 4d $\mathcal{N}=2$ theories and involve integrating out BPS hypermultiplets in maximally supersymmetric backgrounds with constant graviphoton field strength, the two setups are crucially different. However, this observation still suggests that the 1-loop determinants associated with different supermultiplets in AdS$_2 \times \mathbf{S}^2$ may encode in a non-trivial way structures closely related to the non-perturbative amplitudes of an auxiliary topological string theory \cite{Witten:1988xj,Witten:1991zz, Labastida:1991qq}. Lastly, we find a third contribution stemming from the presence of non-minimal couplings of the fermions to the graviphoton. As explained in Section \ref{ss:ExactHyperDeterminant}, including the latter has two separate effects. Firstly, it modifies the previous Gopakumar-Vafa type term by flipping its sign, bringing it into a form that agrees with the $\mathbb{R}^{1,3}$ Euler–Heisenberg Lagrangian of scalar QED in (anti-) self-dual backgrounds. Notably, for massless states this precisely reverses the sign of the logarithmic corrections to the entropy of extremal black holes in the hypermultiplet sector, providing an exact match with the literature in the massless and uncharged sector \cite{Sen:2012kpz, Keeler:2014bra}. The remaining contribution appears to be new and is proportional to the electric charge $q_e$. Therefore, it vanishes identically in purely magnetic backgrounds or for neutral particles. 

\medskip

Our results may open up several promising avenues for future research. Indeed, one of the main motivations for pursuing this analysis is the broader objective of understanding the non-perturbative structure of black holes in string theory or, more generally, quantum gravity. Given that BPS solutions with AdS$_2 \times \mathbf{S}^2$ near-horizon geometries provide an ideal setting for explicit computations to be addressed, they serve as a natural starting point for our investigation. In this context, the techniques developed here aim to provide a systematic characterization of the effective action induced by massive fields around such backgrounds, with direct implications for black hole physics and quantum entropy functions \cite{Sen:2008yk,Sen:2008vm}. This can be compared with the predictions made via other methods in string theory \cite{LopesCardoso:1998tkj,LopesCardoso:1999cv,LopesCardoso:1999fsj}.

\noindent Relatedly, and given the similarity with the original Gopakumar-Vafa integral in flat space, it would be interesting to perform a precise comparison of the 1-loop partition function obtained herein with the one given in that work, clarifying how the latter encodes the AdS$_2 \times \mathbf{S}^2$ result and its implications for the $\mathcal{N}=2$ (single-centered) black hole partition functions \cite{Ooguri:2004zv}.
Another possible direction would be to extend the present analysis to other supermultiplets appearing in 4d $\mathcal{N}=2$ string theory, including higher-spin D-brane wrapped states, or perhaps to adapt the computation so as to accommodate other similar near-horizon geometries in higher dimensions, such as AdS$_3 \times \mathbf{S}^2$ or AdS$_2 \times \mathbf{S}^3/\mathbb{Z}_k$. Finally, it would also be interesting to connect our results with other approaches in the literature, including microscopic counting or supersymmetric localization procedures (see e.g., \cite{Sen:2007qy,Cassani:2025sim} and references therein). 

\medskip

We hope that the results presented in this work may offer new valuable insights into the quantum description of extremal black holes and serve as a useful basis for future explorations.

\section*{Acknowledgements}
	
We are indebted to Ivano Basile, Ralph Blumenhagen, José Calderón-Infante, Jinwei Chu, Niccoló Cribiori, Bernardo Fraiman, Aleksandar Gligovic, Manvir Grewal, Damian van de Heisteeg, Álvaro Herráez, Elias Kiritsis, Dieter Lüst, Puxin Lin, Emil Martinec, Miguel Montero, Hirosi Ooguri, Tomás Ortín, Klaas Parmentier, Savdeep Sethi, Gary Shiu, Timo Weigand, Max Wiesner, and Cumrun Vafa for illuminating discussions and very useful comments on the manuscript. We also grateful to Dieter Lüst for collaboration on related topics. A.C. and C.M. would like to thank Harvard University and its Swampland Initiative for hosting and providing a stimulating environment where parts of this work were completed. A.C. also thanks U.W. Madison, Caltech, MPP Münich, and the Aspen Center for Physics, funded by the NSF grant PHY-2210452, for hospitality during the different stages of this work. C.M. also thanks DESY and Universit\"at Hamburg for the kind hospitality during the final stages of this work. The work of A.C. is supported by a Kadanoff and an Associate KICP fellowships, as well as through the NSF grants PHY-2014195 and PHY-2412985. The work of M.Z. is supported by the Alexander von Humboldt Foundation. A.C. and M.Z. are also grateful to Teresa Lobo and Miriam Gori for their continuous encouragement and support.
	

\appendix

\section{Density of States in AdS$_2$ and Digamma Functions}\label{ap:densityAdS2&polygamma}

In this section, we show explicitly that the densities appearing in the heat kernel operators \eqref{eq:heatkernelAdS2spin0complete} and \eqref{eq:heatkernelAdS2spin12complete} reduce to \eqref{eq:densityspin0AdS2} and \eqref{eq:densityspin12AdS2}. 

\medskip

The digamma function, $\psi(z)$, is defined as the derivative of the logarithm of the $\Gamma$-function
\begin{equation}
    \psi(z) = \partial_z \log{\Gamma(z)}\,.
\end{equation}
The starting point to relate the digamma and trigonometric functions is the reflection formula (see e.g., \cite{Abramowitz})
\begin{equation} \label{eq:reflection}
    \psi(z) - \psi(1-z) = -\pi \cot{\left( \pi z \right)}\, .
\end{equation}
Let us consider then the combination 
\begin{equation}
    \mathcal{A} \equiv  \text{Im}\left[ \psi\left( \frac{1}{2} + g + i\lambda \right) \right] + \text{Im}\left[ \psi\left( \frac{1}{2} - g + i\lambda \right) \right]\,. \label{eq:combinationA}
\end{equation}
In order to massage \eqref{eq:combinationA}, we observe that replacing $z = \frac{1}{2} + \omega$ in \eqref{eq:reflection} with $\omega \in \mathbb{C}$, one gets
\begin{equation}
    \psi\left( \frac{1}{2} + \omega \right) - \psi\left( \frac{1}{2} - \omega \right) = \pi \tan{\left( \pi \omega \right)}\,,
\end{equation}
and taking $\omega = \pm g + i\lambda$, we obtain
\begin{equation}
         \psi\left( \frac{1}{2} \pm g + i\lambda \right) - \psi\left( \frac{1}{2} \mp g - i\lambda \right) = \pi \tan \left( \pi(\pm g + i\lambda)\right)\,.  \label{eq:gamma1st} 
\end{equation}
Using the holomorphic properties of the digamma function and eq.~\eqref{eq:gamma1st}, $\mathcal{A}$ simplifies to
\begin{equation}
    \mathcal{A} = \frac{\pi}{2} \bigg[ \tanh\left( \pi(\lambda - ig) \right) + \tanh{\left( \pi(\lambda + ig) \right)} \bigg] \,,
\end{equation}
whereas upon inserting the (hyperbolic) trigonometric identity
\begin{equation}
    \tanh(x+iy)= \frac{\sinh(2x)+i\sin(2y)}{\cosh(2x)+\cos(2y)}\, ,
\end{equation}
we arrive at 
\begin{equation}
    \mathcal{A} = \pi\, \frac{\sinh{\left( 2\pi\lambda \right)}}{\cosh{\left( 2\pi\lambda \right)} + \cos{\left( 2\pi g \right)}}\,.
\end{equation}
Finally, performing the analytic continuation $g \rightarrow i e$, we obtain the desired relation 
\begin{equation}
         \text{Im}\left[ \psi\left( \frac{1}{2} + i e + i\lambda \right) \right] + \text{Im}\left[ \psi\left( \frac{1}{2} - i e + i\lambda \right) \right]=  \frac{\pi \sinh{\left( 2\pi\lambda \right)}}{\cosh{\left( 2\pi\lambda \right)} + \cosh{\left( 2\pi e \right)}}\, .
\end{equation}

\medskip

For fermions, we consider instead the quantity
\begin{equation}
 \mathcal{B}  \equiv \text{Im} \left[ \psi\left(i\lambda-g\right)+
    \psi\left(i\lambda+g\right) + \psi\left(i\lambda-g+1\right)+ \psi\left(i\lambda+g+1\right)\right] \,, \label{eq:combinationB}
\end{equation}
which can be written as
\begin{equation}
\begin{split}
    2i\mathcal{B}= &-\pi \big[ \cot \left(\pi \left( i\lambda +g\right)\right) +  \cot \left(\pi \left( i\lambda -g\right)\right) \\
    & + \cot \left(\pi \left( i\lambda -g +1\right)\right) + \cot \left(\pi \left( i\lambda +g+1\right)\right)\big]\, .
\end{split}    
\end{equation}
From here, one deduces that
\begin{eqn}\label{DensityfermiAdS}
\begin{aligned}
    \mathcal{B} = \pi \left( \coth\left( \pi(\lambda - ig) \right) + \coth{\left( \pi(\lambda + ig) \right)} \right)\, ,
\end{aligned}
\end{eqn}
such that using the (hyperbolic) trigonometric identity
\begin{equation}
    \coth(x+iy)= \frac{\sinh(2x)-i\sin(2y)}{\cosh(2x)-\cos(2y)}\, ,
\end{equation}
we arrive at
\begin{equation}\label{eq:densityspin12H2&AdS2}
 \mathcal{B} = 2 \pi \frac{\sinh{\left( 2\pi\lambda \right)}}{\cosh{\left( 2\pi\lambda \right)} - \cos{\left( 2\pi g \right)}}\, .
\end{equation}
Finally, performing the analytic continuation $g \rightarrow i e$, we obtain
\begin{equation}
 \text{Im} \left[ \psi\left(i\lambda-ie\right)+
    \psi\left(i\lambda+i e\right) + \psi\left(i\lambda-ie+1\right)+ \psi\left(i\lambda+ie+1\right)\right] = \frac{ 2 \pi \sinh{\left( 2\pi\lambda \right)}}{\cosh{\left( 2\pi\lambda \right)} - \cosh{\left( 2\pi e \right)}} \,.
\end{equation}    

\section{Details on the Integration Out Procedure in AdS$_2 \times \mathbf{S}^2$}\label{ap:DetailsonAdS2xS2Traces}

\subsection{Derivation of the AdS$_2$ trace formulae}\label{ss:TraceDetailsAdS2}

In this subsection, we clarify some mathematical aspects of the formulae and manipulations presented in Section \ref{sss:HalfAdS2}. In particular, recall that for the scalar case and in order to perform the spectral density transformation in \eqref{eq:FourierAdS2}, we made use of the formal identity
\begin{equation}\label{eq:BernardoHaRottoGliOcchiali}
    \mathcal{F}^{-1}[\mathcal{F}[\rho_B]] = \rho_B \,,
\end{equation}
where\footnote{See Appendix \ref{ap:densityAdS2&polygamma} for more details on the densities of states and their representations.}
\begin{subequations}
    \begin{align}
        \rho_B(\lambda) & = \frac{V_{\text{AdS}}}{4\pi R^2} \lambda 
        \left[\tanh(\pi \lambda + \pi q_e)+\tanh(\pi \lambda - \pi q_e)\right] \,, \\ 
        \mathcal{F}\left[\rho_B\right](t) 
        & = \frac{V_{\text{AdS}}}{2 \pi R^2}\frac{1}{\sqrt{2 \pi}} 
        \frac{d}{dt} \left[\frac{\cos (q_e t)}{\sinh{\left(\frac{t}{2}\right)}}\right] 
        = \frac{V_{\text{AdS}}}{2 \pi R^2}\frac{1}{\sqrt{2 \pi}} W_B(t) \,.
        \label{eq:PandaConGliOcchiali}
    \end{align}
\end{subequations}
However, the inverse Fourier transform of \eqref{eq:PandaConGliOcchiali} is not well defined. 
This issue arises because the integrand is singular along the real axis (it has a double pole at $t=0$). To deal with this, we proposed to shift the contour of integration and rewrite the left-hand-side of \eqref{eq:BernardoHaRottoGliOcchiali} as
\begin{equation} 
\label{eq:GrilloTalpaConGliOcchiali}
    \frac{V_{\text{AdS}}}{(2 \pi R)^2} 
    \int_{\mathbb{R}+ i \delta_t} dt \,  e^{- i \lambda t} \, W_B(t) \,.
\end{equation}
Strictly speaking, this prescription is not entirely correct, since one can verify that it does not reproduce the original integral. 
This subtlety was first pointed out in \cite{Sun:2020ame}, where it is shown how to construct a well-defined Fourier transform in some simple cases.  The same method was later used in \cite{Grewal:2021bsu} to obtain the regularization encoded in equation \eqref{eq:PathIntegralBosonAdS2}. 
The main idea is to evaluate explicitly the integral over $\mathbb{R} + i \delta_t$ using the residue theorem and then verify whether the resulting series reproduces the expansion of $\rho_B(\lambda)$. 
To this end, we introduce
\begin{equation}
    W^\pm_B (t) = \frac{d}{dt}\left( \frac{e^{\pm i q_e t}}{2 \sinh\left(\frac{t}{2}\right)} \right)\,,
\end{equation}
and define
\begin{equation}
    I^\pm (\lambda) =  \frac{V_{\text{AdS}}}{(2 \pi R)^2} 
    \int_{\mathbb{R}+ i \delta_t} dt \,  e^{- i \lambda t} \, W_B^{\pm}(t) \,.
\end{equation}
Let us first consider $I^+$. For large but finite $q_e$ (so that the factor $\sinh(t/2)$ is subleading), the asymptotic behaviour of the integrand is dominated by the exponential term $e^{i t (q_e - \lambda)}$.

\smallskip

If $q_e - \lambda \ge 0$, we can close the contour by adding an arc in the upper half of the complex plane, thereby enclosing the poles at $t = 2\pi i k$ with $k \in \mathbb{Z}_{>0}$. Conversely, if $q_e - \lambda < 0$, we may close the contour by adding an arc in the lower half-plane, which encloses the poles at $t = 2\pi i k$ with $k \in \mathbb{Z}_{<0}$, as well as the pole at $t = 0$. The explicit result of this procedure is
\begin{equation}
    I^+(\lambda) = \frac{V_{\text{AdS}}}{4 \pi R^2} \lambda
    \begin{cases}
         -2 \sum_{k\ge 1} (-1)^k e^{-2\pi k q_e} e^{2 \pi k \lambda} 
         & \text{if } q_e - \lambda \ge 0 \,, \\
         2 + 2\sum_{k\ge 1} (-1)^k e^{2\pi k q_e} e^{-2 \pi k \lambda} 
         & \text{if } q_e - \lambda < 0 \,.
    \end{cases}
\end{equation}
Comparing this result with the following expansion of $\tanh(x)$,
\begin{equation}
    \tanh(x) =
    \begin{cases}
         1 + 2 \sum_{k\ge 1} (-1)^k e^{-2 k x}
         & \text{if } \mathrm{Re}(x) \ge 0 \,, \\
         -1 - 2 \sum_{k\ge 1} (-1)^k e^{2 k x}
         & \text{if } \mathrm{Re}(x) < 0 \,,
    \end{cases}
\end{equation}
we immediately obtain
\begin{equation}\label{eq:I+}
    I^+(\lambda) = \frac{V_{\text{AdS}}}{4 \pi R^2} \lambda
    \left[1 + \tanh(\pi \lambda - \pi q_e)\right] \,.
\end{equation}
Repeating the same steps for $I^-$, we find
\begin{equation}\label{eq:I-}
    I^- (\lambda) = \frac{V_{\text{AdS}}}{4 \pi R^2} \lambda
    \left[1 + \tanh(\pi \lambda + \pi q_e)\right] \,,
\end{equation}
which, together with \eqref{eq:I+}, implies that
\begin{equation}
    I^+ + I^- \ne \rho_B \,.
\end{equation}
For this reason, we introduce the related integrals
\begin{equation}
    J^\pm (\lambda) = \frac{V_{\text{AdS}}}{(2 \pi R)^2}
    \int_{\mathbb{R}- i \delta_t} dt \, e^{- i \lambda t} \, W_B^{\pm}(t) \,,
\end{equation}
differing from $I^\pm$ in the position of the pole at $t=0$. In this case, the pole contributes only to the contour integrals whose closing arc lies in the upper half-plane. Proceeding as before, one can show that
\begin{equation}
    J^\pm (\lambda) = \frac{V_{\text{AdS}}}{4 \pi R^2} \lambda \left[-1 + \tanh(\pi \lambda \mp \pi q_e)\right] \,.
\end{equation}
One can now construct several combinations of $I^\pm$ and $J^\pm$ that reproduce $\rho_B$. Among these, we may choose the prescription used in \cite{Sun:2020ame, Grewal:2021bsu}, namely
\begin{equation}\label{eq:correctintegraldensity}
    \rho_B(\lambda) = \frac{1}{2}\left[\int_{\mathbb{R}+ i \delta_t} + \int_{\mathbb{R}- i \delta_t}\right] dt \,  e^{- i \lambda t} \, W_B(t) \,.
\end{equation}
Crucially, it is straightforward to verify that this combination is equivalent to taking the \textit{principal value} of the original integral (cf. eq.~\eqref{eq:tracefermiAdS2ppalvalue})
\begin{eqn}\label{eq:principalvalue}
\left[\text{P.V}\left(\frac{1}{x}\right) \right]\left(u\right) = \lim_{\epsilon \rightarrow 0^+} \int_{\mathbb{R}/\left[-\epsilon, \epsilon\right]} \frac{u(x)}{x}dx\,.
\end{eqn}
Indeed, the contributions from the small semicircles around the pole at $t=0$ in \eqref{eq:correctintegraldensity} have opposite orientations in the two contours and therefore cancel each other.

\medskip
 
One might then wonder whether using this more appropriate regularization would change the results obtained in Section \ref{ss:ExactComputation}. In fact, the answer is negative, since the two regularizations differ only by the residue of the pole at the origin, which ultimately vanishes. For concreteness, we consider eq.~\eqref{eq:LogZintermediate}. According to the above discussion, the latter should read
\begin{eqnarray}
     \log \mathcal{Z}_\phi = {-}\frac{V_{\text{AdS}}}{4\pi R^2}
     \frac{1}{2}\left[\int_{\mathbb{R}+ i \delta_t} + \int_{\mathbb{R}- i \delta_t}\right]
     \frac{dt}{\sqrt{t^2 + \epsilon^2}} \, W_B(t)\,
     f_B(i\sqrt{t^2 + \epsilon^2})\,.
\end{eqnarray}
Because of $\epsilon > \delta_t$, the only singularity of the integrand between the lines $\mathbb{R}\pm i \delta_t$ is due to $W_{B}(t)$ at $t=0$. However, as one can easily compute, the corresponding residue is zero. Therefore, we may safely replace the two integrals above with just one
\begin{equation}
     \frac{1}{2}\left[\int_{\mathbb{R}+ i \delta_t} + \int_{\mathbb{R}- i \delta_t}\right]
     \;\rightarrow\;
     \int_{\mathbb{R}+ i \delta_t}\,.
\end{equation}
As argued in the main text, we can also remove the $\delta_t$-regulator and deform the contour toward the real axis, thereby taking the principal value of the resulting integral
\begin{eqn}\label{eq:CorrectRegBoson}
    \log \mathcal{Z}_\phi = -\frac{V_{\text{AdS}}}{4\pi R^2}  \lim_{\tilde{\epsilon} \rightarrow 0^+} \int_{\mathbb{R}/\left[-\tilde{\epsilon}, \tilde{\epsilon}\right]}\frac{dt}{\sqrt{t^2 + \epsilon^2}} \, W_B(t) f_B(i\sqrt{t^2 + \epsilon^2})\,,
\end{eqn}
where $\tilde{\epsilon} < \epsilon$. Exploiting the parity of the integrand and using the notation $\tilde{\epsilon}=0^+$, we can finally express \eqref{eq:CorrectRegBoson} as
\begin{eqn}
     \log \mathcal{Z}_\phi = -\frac{V_{\text{AdS}}}{2\pi R^2}  \int_{0^+}^{\infty} \frac{dt}{\sqrt{t^2 + \epsilon^2}} \, W_B(t) f_B(i\sqrt{t^2 + \epsilon^2})\,,
\end{eqn}
cf. eq.~\eqref{SintegralBoson0}. Notice that the same argument can be readily applied to the fermionic case.

\subsection{A comment on the domain of integration in Schwinger proper time}\label{ss:domainoftauintegration}

In this subsection, we address a small subtlety that arises when recalling that the actual computation of the 1-loop determinant of interest must be performed in Lorentzian signature. As explained in Section \ref{sss:1loopbasics}, this implies that the integral over Schwinger proper time $\tau$ takes the following form (e.g., for a complex scalar field)
\begin{eqn}
    \log \mathcal{Z}_\phi = -\int_{i 0}^{i\infty} \frac{d\tau }{\tau}\, e^{-\frac{\epsilon^2}{4\tau}}\, \Tr \left[e^{-\tau \left(\mathcal{D}^2 + m^2 \right)}\right]\,.
\end{eqn}
On the other hand, when computing the AdS$_2\times \mathbf{S}^2$ partition function in Section \ref{ss:AdS2xS2Trace}, our strategy was to first express the traces in integral form using the Hubbard-Stratonovich trick, and subsequently perform the integration over proper time. The corresponding integral is shown in eq.~\eqref{FirstSintegral}. Instead, in here we would like to evaluate
\begin{eqn}\label{eq:I_SLorentzian}
    I_S = \frac{1}{4\pi} \int_{i 0}^{i\infty} \frac{d\tau }{\tau^2}\, e^{-\frac{\epsilon^2+t^2 +u^2}{4\tau}}\, ,
\end{eqn}
where we have already restricted ourselves to the BPS case and thus we set $\Delta =0$. Ideally, one would like to deform the integration contour towards the positive real axis and proceed as explained in Section \ref{ss:AdS2xS2Trace}. However, care must be taken regarding the singularity structure of the integrand in \eqref{eq:I_SLorentzian}, which appears to have an essential singularity at $\tau=0$. To remedy this, we first perform a change of variables $\tau=1/x$, yielding (cf. eq.~\eqref{eq:intBes}) 
\begin{eqn}
    I_S = \frac{1}{4\pi} \int_{-i 0}^{-i\infty}  dx\, e^{-x\,\frac{\epsilon^2+t^2 +u^2}{4}}\, .
\end{eqn}
The advantage of this expression is that the new integrand now exhibits no singularity at all, and therefore we are free to deform the contour towards the positive real axis. Then, using Cauchy's residue theorem and the fact that the arc at infinity does not contribute when $\text{Re}\, x >0$, we conclude that 
\begin{eqn}
    I_S = \frac{1}{4\pi} \int_{-i 0}^{-i\infty}  dx\, e^{-x\,\frac{\epsilon^2+t^2 +u^2}{4}} = \frac{1}{4\pi} \int_{0}^{\infty}  dx\, e^{-x\,\frac{\epsilon^2+t^2 +u^2}{4}} = \frac{1}{\pi} \frac{1}{\epsilon^2 + t^2 + u^2}\, ,
\end{eqn}
in agreement with eq.~\eqref{BPSLimit} from the main text.

\subsection{The flat-spacetime limits}\label{ss:flatspacelimit}

In this section, we consider the flat space limit to illustrate how our results from Section \ref{s:integrationAdS2xS2} relate to the original works on exact 1-loop determinants in flat space(-time) backgrounds. We start in Section \ref{ss:summaryLandauR2} by reviewing the known material on the subject \cite{Heisenberg:1936nmg,Weisskopf:1936hya,Schwinger:1951nm}, and then proceed to evaluate the aforementioned limits in Section \ref{ss:AdS2xS2toMinkowski4}.

\subsubsection{Review of the Landau Problem in $\mathbb{R}^2$ and $\mathbb{R}^{1,1}$}\label{ss:summaryLandauR2}

We briefly summarize the spectral Landau problem on $\mathbb{R}^2$ for massive spin-$0$ and spin-$\frac{1}{2}$ fields minimally coupled to background gauge and gravitational fields \cite{Heisenberg:1936nmg,Weisskopf:1936hya,Schwinger:1951nm, Dunne:1991cs, Tong:2016kpv}. To facilitate the comparison with the flat-space limits of the Landau problems on $\mathbb{H}_2$ and $\mathbf{S}^2$ (cf. sections \ref{ss:LandauAdS2H2} and \ref{ss:spectralAdS2}), we introduce the (anti-)holomorphic coordinates 
\begin{eqn}
    z = \sqrt{\frac{2}{B}} (x + i y)\,, \qquad \bar{z} = \sqrt{\frac{2}{B}} (x - i y)\,,
\end{eqn} 
where $(x,y)$ denote the usual Cartesian variables and we assume $B>0$. Using this coordinate system, the metric and gauge connection take the form
\begin{equation}\label{eq:antiholomorphicR2}
ds^2 = \frac{2}{B} \, dz\, d\bar{z}\,, 
\qquad 
A = -\frac{i}{2}\left(z d\bar{z} - \bar{z} dz \right)\,,
\end{equation}
whereas the Hamiltonian for a charged scalar particle (cf. eq.~\eqref{eq:2dHamiltonian}) becomes
\begin{equation}\label{eq:HamiltonianR2}
H = B \left( \frac{|z|^2}{4} - \partial \bar{\partial}
- \frac{1}{2}\left( z \partial - \bar{z} \bar{\partial} \right) \right)\, .
\end{equation}
To solve the associated spectral problem, we adopt an algebraic approach and introduce the following ladder operators \cite{Dunne:1991cs}
\begin{align}
J_+ = -\sqrt{2B}\, \partial + \sqrt{\frac{B}{2}}\, \bar{z}
\equiv \sqrt{2B}\, a_+\,, \qquad 
J_- = \sqrt{2B}\, \bar{\partial} + \sqrt{\frac{B}{2}}\, z
\equiv \sqrt{2B}\, a_-\, ,
\end{align}
together with the angular momentum operator
\begin{equation}
L_0 = z \partial - \bar{z} \bar{\partial}\, .
\end{equation}
These operators satisfy the algebra
\begin{equation}
[a_-, a_+] = 1\,,
\qquad
[H, a_\pm] = \pm a_\pm\,,
\qquad 
[H, L_0] = 0\, .
\end{equation}
The Hamiltonian \eqref{eq:HamiltonianR2} can then be rewritten as
\begin{equation}
H = \frac{1}{4}\left(J_+ J_- + J_- J_+ \right)
= B \left( a_+ a_- + \frac{1}{2} \right)\,,
\end{equation}
which coincides with that of a simple harmonic oscillator. Consequently, one can find simultaneous eigenstates of $H$ and $L_0$ using standard methods of quantum mechanics
\begin{align}\label{eq:HOR2}
H \ket{\psi_{\ell,n}} &= B\left(n + \frac{1}{2}\right) \ket{\psi_{\ell,n}}\,, 
\qquad n \geq 0 \,, \\
L_0 \ket{\psi_{m,n}} &= \ell \ket{\psi_{m,n}}\,, 
\qquad \ell \geq -n\, ,
\end{align}
where $n$ is the Landau level (radial quantum number) and $\ell$ the magnetic quantum number. The minimum value of the angular momentum, $\ell_{\text{min}}$, is obtained by solving
\begin{equation}
    J_- \ket{\psi_{\ell_{\text{min}},n}} = 0\,,
\end{equation}
together with the requirement that the energy spectrum be non-negative or, equivalently, that the wavefunctions be well-defined and normalizable \cite{Dunne:1991cs, Tong:2016kpv}.
This leads to a spectrum that is bounded from below, and such that for each Landau level $n$ the degeneracy is infinite due to the allowed range $\ell \geq -n$.

The lowest Landau level wavefunctions are those annihilated by $L_-$ for arbitrary magnetic quantum number $\ell$. The corresponding equation can be explicitly solved, yielding
\begin{equation}
    \psi_{\ell,0}(z,\bar{z}) = f_\ell(z)\, e^{-\frac{|z|^2}{4}}\,,
\end{equation}
where $f_\ell(z)$ is an arbitrary holomorphic monomial of order $\ell$.
The density of states, namely the number of states per unit area in $\mathbb{R}^2$, can be extracted from the normalization of the lowest Landau level wavefunctions and is given by \cite{Tong:2016kpv}
\begin{equation}
    \rho(B) = \frac{B}{2\pi}\, .
\end{equation}

\medskip

The Landau problem for a spin-$\frac{1}{2}$ particle differs from the scalar case due its magnetic moment coupling. The relevant Lagrangian reads (cf. eq.~\eqref{eq:fermionicAction})
\begin{equation*}
\mathcal{L}_{\Psi} = \bar{\Psi}\left[ (\slashed{\nabla} -i \slashed{A})+m\right] \Psi \equiv
\bar{\Psi}\left[ i\slashed{D} + m \right]\Psi\,.
\end{equation*}
One can easily show that
\begin{equation}
\slashed{D}^2 = -D^2 - B\, \sigma_z\, ,\qquad \text{with}\quad \sigma_z =
\begin{pmatrix}
1 & 0 \\
0 & -1
\end{pmatrix}\, ,
\end{equation}
where $D^2$ is the scalar Laplacian.
Thus, the spin-$\frac{1}{2}$ spectrum is analogous to the scalar case, with the crucial difference that the energy levels are shifted oppositely for positive and negative chirality modes (with $\gamma^3 = \sigma_z$). For $B>0$, the energy eigenvalues are
\begin{equation}
E_n^+ = Bn\,,
\qquad 
E_n^- = B(n+1)\,,
\qquad n \geq 0\, .
\end{equation}
As in the bosonic case, there exist infinitely many zero modes of definite chirality, with density $B/2\pi$, in agreement with the Atiyah–Singer index theorem \cite{Atiyah:1963zz}.

\subsubsection*{From $\mathbb{R}^2$ to $\mathbb{R}^{1,1}$}

The electric Landau problem on $\mathcal{M}_2=\mathbb{R}^{1,1}$ can be understood as the Lorentzian counterpart of the standard (magnetic) Landau problem on $\mathbb{R}^2$, namely a charged particle propagating in two-dimensional Minkowski spacetime in the presence of a constant electric field $E$. 

\smallskip

To relate the two systems, we closely follow the procedure of \cite{Pioline:2005pf} and perform a double analytic continuation,
\begin{equation}\label{eq:DoubleWickR11}
   y = i t\,, 
    \qquad 
   B = - i E\, .
\end{equation}
As in the AdS$_2$ case (cf. Section \ref{sss:analyticcontAdS2}), it is convenient to work in $(t,x)$ coordinates.
The Hamiltonian for a charged spin-$0$ particle (with the convention $p_\mu \rightarrow i\partial_\mu$) then reads
\begin{equation}\label{eq:HamiltonianR11}
    H = \frac{1}{2}\left(p_x^2-(p_t + E x)^2\right)\, .
\end{equation}
Since $p_t$ is conserved, the eigenstates can be written in terms of continuous momentum eigenvalues as
\begin{equation}
    \phi_{n,p_t}(t,x) 
    = \chi_n\!\left(x + \frac{p_t}{E}\right) 
      e^{i p_t t}\, .
\end{equation}
Upon inserting this ansatz into \eqref{eq:HamiltonianR11}, the problem reduces to an inverted one-dimensional harmonic oscillator, i.e.\ the Schwinger Hamiltonian \cite{Schwinger:1951nm, Dunne:2004nc}. Under this analytic continuation, the discrete modes in \eqref{eq:HOR2} become states with imaginary energy, and therefore do not belong to the normalizable spectrum, in analogy with the AdS$_2$ example \cite{Pioline:2005pf}.

\bigskip

\noindent For later comparison, we compute the traces of $\mathcal{D}^2$ and $\slashed{D}^2$ in the electric and magnetic Landau problems by summing over their corresponding quantum-mechanical energy eigenvalues.

\smallskip

For the scalar Laplacian in $\mathbb{R}^{1,1}$, the 1-loop determinant is thus computed by integrating over the continuous modes \cite{Heisenberg:1936nmg,Schwinger:1951nm, Dunne:2004nc}:
\begin{equation}\label{eq:flatBosonLandauLorentz}
    \Tr\!\left[e^{-i\tau \mathcal{D}^2}\right] 
    = \int d^2x \int \frac{d^2\lambda}{(2\pi)^2 } 
      |\phi_\lambda(x)|^2 e^{-i \tau E_\lambda} 
    = \int d^2x \,\frac{E}{4\pi} \frac{1}{\sin(E\tau)}\,,
\end{equation}
where $\phi_{\lambda}(x) = \langle x | \lambda \rangle$ denotes the position-space wavefunction associated with the eigenvalue $E_\lambda$.
Similar to the scalar case, the spin-$\frac{1}{2}$ system exhibits the same structure, up to the energy splitting induced by the spin degree of freedom, exactly as in the Euclidean theory \cite{Heisenberg:1936nmg,Schwinger:1951nm, Dunne:2004nc}:
\begin{equation}\label{eq:flatFermionLandauLorentz}
    \Tr\!\left[e^{-i\tau \slashed{D}^2}\right] 
    = \sum_{s =  \pm \frac{1}{2}} 
      \int d^2x \int \frac{d^2\lambda}{(2\pi)^2 } 
      |\phi^\lambda_s (x)|^2 e^{-i \tau E_\lambda^s} 
    = \int d^2x \,\frac{E}{2\pi} \frac{1}{\tan(E\tau)}\,,
\end{equation}
where $s=\pm\frac{1}{2}$ labels the two spin projections.

\smallskip

For the scalar Laplacian in $\mathbb{R}^2$, we find instead
\begin{equation}\label{eq:flatBosonLandau}
\Tr\!\left[e^{-\tau D^2}\right]
=
\frac{B}{2\pi}
\int d^2x
\sum_{n\geq 0}
e^{-2B\tau \left(n + \frac{1}{2}\right)}
=
\int d^2x\, \frac{B}{4\pi}\, \frac{1}{\sinh{B \tau}}\, ,
\end{equation}
whereas the Dirac operator is such that
\begin{equation}\label{eq:flatFermionLandau}
\Tr\!\left[e^{-\tau \slashed{D}^2}\right]
=
\frac{B}{2\pi}
\int d^2x
\left(
\sum_{n\geq 0} e^{-2B\tau n}
+
\sum_{n\geq 1} e^{-2B\tau n}
\right)
=
\int d^2x\, \frac{B}{2\pi}\, \frac{1}{\tanh{(B\tau)}}\, .
\end{equation}

\subsubsection{Relating the Landau problems in AdS$_2 \times \mathbf{S}^2$ and $\mathbb{R}^{1,3}$}\label{ss:AdS2xS2toMinkowski4}


In this section, we study the flat-spacetime limit of the 1-loop determinant for a massive spin-$0$ and spin-$\frac{1}{2}$ field on AdS$_2 \times \mathbf{S}^2$. As an intermediate step, we first analyze the corresponding limits on AdS$_2$ and $\mathbf{S}^2$ separately, and then combine them to obtain the result for the full product space. Finally, we consider the flat-space limit of the supersymmetric 1-loop determinant for a minimally coupled four-dimensional $\mathcal{N}=2$ hypermultiplet (cf. \eqref{eq:AdS2xS2effaction}).

\subsubsection*{From $\mathbf{S}^2$ to $\mathbb{R}^2$}
\label{sss:fromS2toR2}

We begin by analyzing how the Landau problem on $\mathbf{S}^2$ reduces to its flat-space counterpart on $\mathbb{R}^2$, and how the corresponding trace formulas match in the appropriate limit. 
To this end, we must define a well-controlled flat-space limit. This is achieved by sending both the radius of the sphere and the total magnetic charge to infinity while keeping the magnetic field strength fixed:
\begin{equation}\label{eq:flatS2limit}
R,\, q_m \to \infty\,,
\qquad 
\text{with}
\qquad 
B = \frac{q_m}{R^2}
\quad \text{fixed}\,.
\end{equation}

\paragraph{Bosonic case.} Starting from the trace formula \eqref{eq:traceLaplacianS2}, we obtain
\begin{equation}\label{eq:traceboseLaplacianS2flatlimit}
\begin{aligned}
\Tr \left[e^{-\tau \mathcal{D}^2} \right] 
&= \frac{V_{\mathbf{S}^2}}{4\pi R^2}
\sum_{n \geq 0} 
2\left( n + q_m + \frac{1}{2} \right)
e^{ -\frac{2q_m\tau}{R^2}
\left( n + \frac{1}{2} + \frac{n(n+1)}{2q_m} \right)} 
\\
&\stackrel{\eqref{eq:flatS2limit}}{\longrightarrow}\,
V_{\mathbb{R}^2}\,
\frac{B}{2\pi}
\sum_{n \geq 0}
e^{ -2B\tau \left( n + \frac{1}{2}\right)}
=
V_{\mathbb{R}^2}\,
\frac{B}{4\pi}\,
\frac{1}{\sinh(B\tau)}\,,
\end{aligned}
\end{equation}
in perfect agreement with the classic results
\cite{Heisenberg:1936nmg,Weisskopf:1936hya,Schwinger:1951nm}
(see also \eqref{eq:flatBosonLandau}).\footnote{
Here we use the notation $V_{\mathbb{R}^n}=\int d^nx$.
}

\medskip

Interestingly, the same result can be recovered by first performing the HS transformation following eq.~\eqref{eq:HStransfbosonsS2}, and subsequently taking the flat space limit. Indeed, starting from \eqref{eq:HStransfbosonsS2}, and extracting the leading contribution in the large-$R$ expansion, we find
\begin{equation}
\Tr \left[e^{-\tau \mathcal{D}^2} \right]
=
\frac{V_{\mathbf{S}^2}}{4\pi}
\int_{\mathbb{R}} du\,
\frac{R}{\sqrt{4 \pi \tau }}
e^{-\frac{R^2}{4\tau}(u - 2 i \tau B)^2}
\frac{i B}{\sin(u/2)}
\left[
1 + \mathcal{O}\!\left(\frac{1}{R^2}\right)
\right]\, .
\end{equation}
The crucial observation is that, as $R \to \infty$, the Gaussian factor becomes increasingly localized. 
Interpreted as a distribution, it converges to a Dirac delta:
\begin{equation*}
\frac{R}{\sqrt{4 \pi \tau }}
e^{-\frac{R^2}{4\tau}(u - 2 i \tau B)^2}
\;\longrightarrow\;
\delta(u - 2 i \tau B)\,.
\end{equation*}
Therefore,
\begin{equation}
\Tr \left[e^{-\tau \mathcal{D}^2} \right]
\stackrel{\eqref{eq:flatS2limit}}{\longrightarrow}
\frac{V_{\mathbb{R}^2}}{4\pi}
\int_{\mathbb{R}} du\,
\delta(u - 2 i \tau B)
\frac{i B}{\sin(u/2)}
=
V_{\mathbb{R}^2}\,
\frac{B}{4\pi}\,
\frac{1}{\sinh(B\tau)}\,,
\end{equation}
reproducing again \eqref{eq:traceboseLaplacianS2flatlimit}.

\paragraph{Fermionic case.}

Let us now consider the fermionic trace. Using the same limit \eqref{eq:flatS2limit}, the expression \eqref{eq:traceD2ferm} becomes
\begin{equation}\label{eq:tracefermiLaplacianS2flatlimit}
\begin{aligned}
\Tr \left[e^{-\tau \slashed{D}^2} \right]
&=
\frac{V_{\mathbf{S}^2}}{2\pi R^2}
\left[
\sum_{n \geq 1}
2\left( n + q_m\right)
e^{ -\frac{2q_m\tau}{R^2}
\left( n + \frac{n^2}{2q_m} \right)}
+ q_m
\right]
\\
&\longrightarrow\,
V_{\mathbb{R}^2}\,
\frac{B}{\pi}
\left[
\sum_{n \geq 1}
e^{ -2B\tau n}
+\frac{1}{2}
\right]
=
V_{\mathbb{R}^2}\,
\frac{B}{2\pi}\,
\coth(B\tau)\,,
\end{aligned}
\end{equation}
again in agreement with the original flat-space results
\cite{Heisenberg:1936nmg,Weisskopf:1936hya,Schwinger:1951nm}.

\medskip

Proceeding instead from the HS representation and keeping the leading large-$R$ terms,
\begin{equation}
\Tr \left[e^{-\tau \slashed{D}^2} \right]
=
2\,
\frac{V_{\mathbf{S}^2}}{4\pi R^2}
e^{\tau R^2 B^2}
\int_{\mathbb{R}} du\,
\frac{R}{\sqrt{4\pi \tau}}
e^{ -\frac{R^2u^2}{4\tau} }
\frac{e^{i R^2 B u}}{\tan\left(\frac{u}{2}\right)}
\, i R^2 B
+ \ldots \,,
\end{equation}
the same delta-function argument yields
\begin{equation}
\Tr \left[e^{-\tau \slashed{D}^2} \right]
\longrightarrow
\frac{V_{\mathbb{R}^2}}{2\pi}
\int_{\mathbb{R}} du\,
\delta(u - 2 i \tau B)
\frac{i B}{\tan\left(\frac{u}{2}\right)}
=
V_{\mathbb{R}^2}\,
\frac{B}{2\pi}\,
\coth(B\tau)\,.
\end{equation}
This confirms the consistency of the HS representation with the spectral derivation in the flat spacetime case \eqref{eq:flatFermionLandau}.


\subsubsection*{From AdS$_2$ to $\mathbb{R}^{1,1}$}
\label{sss:fromA2toR2}

We now study the flat-space limit along the AdS$_2$ sector. 
As before, we take
\begin{equation}
R,\, q_e \to \infty,
\qquad 
\text{with}
\qquad 
E = \frac{q_e}{R^2}
\quad \text{fixed}\, .
\end{equation}
This limit requires additional care. 
At first sight, one might attempt to evaluate the heat kernel \eqref{KE2} by closing the contour in the complex plane and applying the residue theorem. 
However, this procedure fails because the quadratic dependence of the energy eigenvalues on the spectral parameter $\lambda$ restricts the allowed contour: the infinite arc must lie within the wedge $-\pi/4 \leq \theta \leq \pi/4$, preventing a standard contour closure.

For this reason, it is technically simpler to take the flat-space limit \emph{after} performing the Hubbard–Stratonovich transformation.\footnote{\label{fnote:flatH2->AdS2}We point out that if we first take the flat space limit in $\mathbb{H}^2$ and then Wick rotate to AdS$_2$, one also retrieves the correct result. This follows from the fact that if $q_m=g \to \infty$, the continuous spectrum gets infinitely heavy and it does not contribute to the heat kernel, whilst the discrete states give \begin{equation*}\label{eq:traceboseLaplacianAdS2flatlimit} \begin{aligned} &\Tr \left[e^{-\tau \mathcal{D}^2} \right] =\frac{V_{\mathbb{H}^2}}{2\pi R^2} \sum_{n=0}^{\lfloor g-\frac12\rfloor} \left( g -n - \frac{1}{2} \right) e^{ -\frac{2g\tau}{R^2} \left( n + \frac{1}{2} - \frac{n(n+1)}{2g} \right)} \stackrel{R, g \to \infty}{\DOTSB\relbar\joinrel\relbar\joinrel \rightarrow} V_{\mathbb{R}^2}\, \frac{B}{2\pi} \sum_{n = 0}^{\infty} e^{ -2B\tau \left( n + \frac{1}{2}\right)}\ \stackrel{g \to-iq_e}{=}\ V_{\mathbb{R}^2}\, \frac{E}{4\pi}\, \frac{1}{\sin(E\tau)}\,, \end{aligned} \end{equation*} in agreement with the original works \cite{Heisenberg:1936nmg,Weisskopf:1936hya,Schwinger:1951nm}.}

\paragraph{Bosonic case.}

Keeping the leading term in the large-$R$ expansion of \eqref{eq:PathIntegralBosonAdS2}, we obtain
\begin{equation}
\Tr \left[e^{-\tau \mathcal{D}^2} \right]
=
\frac{V_{\text{AdS}}}{4\pi R^2}
e^{\tau R^2 E^2}
\int_{\mathbb{R}} dt \,
\frac{R}{\sqrt{4 \pi \tau}}
e^{- \frac{R^2 t^2}{4 \tau}}
\left(
- E R^2 \frac{\sin(R^2 E t)}{\sinh(t/2)}
+ \dots
\right)\, .
\end{equation}
Using the parity properties of the integrand, the sine function can be traded for an exponential representation, which allows us to complete the square in the Gaussian. 
We then obtain
\begin{equation}
\label{eq:FlatLimitBosA2}
\Tr \left[e^{-\tau \mathcal{D}^2} \right]
=
-\frac{V_{\text{AdS}}}{4\pi}
\int_{\mathbb{R}} dt\,
\frac{R}{\sqrt{4 \pi \tau}}
e^{- \frac{R^2}{4 \tau}\left(t - 2 i E \tau \right)^2}
\frac{E}{i \sinh(t/2)}
+ \dots\, .
\end{equation}
As $R \to \infty$, the Gaussian becomes sharply peaked and again converges, in the sense of distributions, to a Dirac delta, yielding
\begin{equation}
\Tr \left[e^{-\tau \mathcal{D}^2} \right]
\longrightarrow
V_{\mathbb{R}^2}\,
\frac{E}{4\pi}\,
\frac{1}{\sin(E\tau)}\,,
\end{equation}
in agreement with the flat-space Landau result \eqref{eq:flatBosonLandauLorentz}.

\paragraph{Fermionic case.}

The same strategy applies to the fermionic trace. 
At leading order in $1/R$, we have
\begin{equation}
\Tr \left[e^{-\tau \slashed{D}^2} \right]
=
\frac{V_{\text{AdS}}}{2\pi R^2}
e^{\tau R^2 E^2}
\int_{\mathbb{R}} dt \,
\frac{R}{\sqrt{4 \pi \tau}}
e^{- \frac{R^2 t^2}{4 \tau}}
\left(
- E R^2 \frac{\sin(R^2 E t)}{\tanh(t/2)}
+ \dots
\right)\, .
\end{equation}
Proceeding exactly as in the bosonic case and exploiting the parity of the integrand, we obtain
\begin{equation}
\Tr \left[e^{-\tau \slashed{D}^2} \right]
=
-\frac{V_{\text{AdS}}}{2\pi}
\int_{\mathbb{R}} dt\,
\frac{R}{\sqrt{4 \pi \tau}}
e^{- \frac{R^2}{4 \tau}\left(t - 2 i E \tau \right)^2}
\frac{E}{i \tanh(t/2)}
+ \dots\, .
\end{equation}
Taking again the large-$R$ limit and identifying the Gaussian with a delta distribution gives
\begin{equation}
\Tr \left[e^{-\tau \slashed{D}^2} \right]
\longrightarrow
V_{\mathbb{R}^2}\,
\frac{E}{2\pi}\,
\frac{1}{\tan(E\tau)}\,,
\end{equation}
which coincides with the standard flat-space result (cf. \eqref{eq:flatFermionLandauLorentz}).

\subsubsection*{From AdS$_2 \times \mathbf{S}^2$ to $\mathbb{R}^{1,3}$}

We are now ready to take the flat-spacetime limit of the 1-loop determinant on the full product space AdS$_2 \times \mathbf{S}^2$. 
More precisely, we implement the simultaneous limit
\begin{equation}\label{eq:flatSTfromA2S2}
    R,\, q_e,\, q_m \rightarrow \infty\,,
    \qquad 
    \text{with}
    \qquad 
    E = \frac{q_e}{R^2}\,, 
    \quad 
    B = \frac{q_m}{R^2}
    \quad 
    \text{fixed}\,,
\end{equation}
so that the background electric-magnetic fields remain finite.

\smallskip

As shown in Section \ref{sss:1loopbasics}, the trace of the four-dimensional kinetic operator $\mathcal{D}_{\text{AdS}_2 \times \mathbf{S}^2}^2$ factorizes into two commuting operators acting on the AdS$_2$ and $\mathbf{S}^2$ subspaces. Consequently, the flat limit can be taken independently in each sector using the results from previous sections.

\paragraph{Spin-$0$ case.}

Starting from \eqref{eq:fullintegralspin0} and performing the limit \eqref{eq:flatSTfromA2S2}, we obtain
\begin{align}\label{eq:BosonR4}
\log \mathcal{Z}_\phi 
&= -\lim_{\epsilon\rightarrow 0}
\int_\epsilon^{\infty} \frac{d\tau}{\tau}\,
e^{-\tau m^2}\,
\mathcal{K}^{(0)}_{\rm AdS_2}(\tau)\,
\mathcal{K}^{(0)}_{\mathbf{S}^2}(\tau)
\nonumber\\
&\longrightarrow{}
-\frac{V_{\mathbb{R}^4}}{(4\pi)^2}
\lim_{\epsilon\rightarrow 0}
\int_\epsilon^{\infty} \frac{d\tau}{\tau}\,
e^{-\tau m^2}\,
\frac{EB}{\sin(E\tau)\sinh(B\tau)}\,,
\end{align}
where $\mathcal{K}_X(\tau)$ denotes the heat kernel trace associated with kinetic operator $\mathcal{D}_X^2$.
\paragraph{Spin-$\frac{1}{2}$ case.}

Proceeding analogously from \eqref{eq:fullintegralspin1/2} and applying the same limit \eqref{eq:flatSTfromA2S2}, we find
\begin{align}
\log \mathcal{Z}_\Psi 
&= \frac12\lim_{\epsilon\rightarrow 0}
\int_\epsilon^{\infty} \frac{d\tau}{\tau}\,
e^{-\tau m^2}\,
\mathcal{K}^{\left(\frac{1}{2}\right)}_{\rm AdS_2}(\tau)\,
\mathcal{K}^{\left(\frac{1}{2}\right)}_{\mathbf{S}^2}(\tau)
\nonumber\\
&\longrightarrow 
2\,\frac{V_{\mathbb{R}^4}}{(4\pi)^2}
\lim_{\epsilon\rightarrow 0}
\int_\epsilon^{\infty} \frac{d\tau}{\tau}\,
e^{-\tau m^2}\,
\frac{EB}{\tan(E\tau)\tanh(B\tau)}\,.
\end{align}
These expressions are in agreement with the classic flat-space computations of the 1-loop effective action for a charged complex scalar and a Dirac fermion in constant electromagnetic fields (see e.g., \cite{Dunne:2004nc}).\footnote{
Strictly speaking, \cite{Dunne:2004nc} differs by an overall sign compared to our result. 
This discrepancy originates from their convention for the 1-loop effective action, inherited from \cite{Schubert:2001he}, where $\log \mathcal{Z}_\phi$ and $\log \mathcal{Z}_\Psi$ are defined with an opposite sign relative to ours.
}

\paragraph{Anti-self-dual background.}

Several interesting features emerge in the special case of an anti-self-dual background satisfying $E = \pm i B$.

\smallskip

For a complex scalar and a Dirac fermion one obtains
\begin{equation}
\log \mathcal{Z}(E=\pm iB)
=
\log \mathcal{Z}_\phi
+
\log \mathcal{Z}_\Psi
=
\frac{V_{\mathbb{R}^4}}{(4\pi)^2}
\lim_{\epsilon \rightarrow 0}
\int_{\epsilon}^{\infty}
\frac{d\tau}{\tau}\,
e^{-m^2 \tau}\,
\frac{B^2}{\sinh^2(B\tau)}
\cosh(2B\tau)\, .
\end{equation}
In contrast, for two complex scalars and one Dirac fermion (cf. Section \ref{ss:Susic1LoopMinimal}), one finds
\begin{equation}\label{eq:AffectiveActionASD}
\log \mathcal{Z}(E=\pm iB)
=
2\log \mathcal{Z}_\phi
+\log \mathcal{Z}_\Psi
=
2\frac{V_{\mathbb{R}^4}}{(4\pi)^2}
\lim_{\epsilon \rightarrow 0}
\int_{\epsilon}^{\infty}
\frac{d\tau}{\tau}\,
e^{-m^2 \tau}\,
B^2\, .
\end{equation}
Notice how in the second case, all perturbative corrections cancel except for the logarithmic running of the gauge coupling, and the non-perturbative poles in the proper-time integrand disappear (see discussion in Section \ref{sss:NonPerturbative}). This reflects the hidden supersymmetry of the anti-self-dual background combined with the chosen matter content, rendering the vacuum both perturbatively and non-perturbatively stable. Indeed, the imaginary part of the four-dimensional effective Lagrangian vanishes identically in this limit. Using the results of \cite{Kim:2003qp},
\begin{align}
2\,\mathrm{Im}\,\mathcal{L}_{\rm ferm}
&=
\frac{EB}{8\pi^2}
\sum_{n=1}^{\infty}
\frac{1}{n}
\coth\!\left(\frac{n\pi B}{E}\right)
e^{-\frac{n\pi m^2}{E}}\,,
\\
2\,\mathrm{Im}\,\mathcal{L}_{\rm bos}
&=
\frac{EB}{8\pi^2}
\sum_{n=1}^{\infty}
\frac{(-1)^{n+1}}{n}
\csch\!\left(\frac{n\pi B}{E}\right)
e^{-\frac{n\pi m^2}{E}}\,,
\end{align}
one verifies that their sum vanishes upon taking $B \to \mp iE$.

\medskip

At this point one may wonder how the original Gopakumar--Vafa computation in flat space \cite{Gopakumar:1998ii,Gopakumar:1998jq}, which involves precisely an anti-self-dual graviphoton background, can generate a non-trivial (non-)perturbative result beyond the familiar logarithmic running derived in \eqref{eq:AffectiveActionASD}. 
This issue was explained in detail in \cite{Dedushenko:2014nya}, where it was shown that the non-minimal Pauli couplings studied in Section \ref{s:SusyLoopDeterminant} are responsible for collapsing the 1-loop determinant of the $\mathcal{N}=2$ hypermultiplet essentially to the scalar contribution.

\section{Details on the Fermionic Operator Diagonalization}\label{ap:DetailsFermionOps}

This appendix provides supplementary material to the discussion of the diagonalization procedure for the fermionic operator employed in Section \ref{ss:TwistedDiracDiagonalization} of the main text. We first consider two simpler examples of kinetic operators analogous to \eqref{eq:kineticoperator}, corresponding to the cases of massless hypermultiplets and BPS, albeit minimally coupled, hypermultiplets. This is the content of sections \ref{ss:masslesshyper} and \ref{ss:freemassivehyper}, respectively. Finally, in Section \ref{ss:ZeroModesS2} we derive the chirality condition of the 2d spinors in the zero mode sector of the Dirac operator on the sphere.

\subsection{Uncharged hypermultiplets with kinetic mixing}\label{ss:masslesshyper}

Considering massless states in 4d $\mathcal{N}=2$ supergravity is tantamount to turning off the charges under the graviphoton field. Thus, the relevant kinetic operator becomes now (cf. eq.~\eqref{eq:kineticoperator})
\begin{eqn}\label{eq:kineticoperatormassless}
    \mathds{D}_{m=0}=\slashed{\nabla}_{\mathbb{H}^2}\otimes \sigma^3 +\mathds{1} \otimes\slashed{\nabla}_{\mathbf{S}^2} - \frac{i}{R}\, \tau^3 \otimes \mathds{1}\, ,
\end{eqn}
which, in matrix notation, reads
\begin{eqn}\label{eq:MatrixFormMasslessD}
    \mathds{D}\big\rvert_{\lambda, n}=\begin{pmatrix}
   i\,\zeta_n & i\,\zeta_\lambda  & -\frac{i}{R}  & 0\\
  i\,\zeta_\lambda  & - i\,\zeta_n & 0 & -\frac{i}{R}             \\
  -\frac{i}{R} & 0 & i\,\zeta_n & -\,i\,\zeta_\lambda\\
  0 & -\frac{i}{R} & -\,i\,\zeta_\lambda & - i\,\zeta_n
\end{pmatrix}\, ,
\end{eqn}
and we have restricted to a fixed eigensector of $\slashed{D}_{\mathbb{H}^2 \times \mathbf{S}^2}$, using the same basis as in \eqref{eq:MatrixFormGeneralD}. The eigenvalues are given by\footnote{One may arrive at the same result by noting that \eqref{eq:MatrixFormMasslessD} can be written as a sum of two anti-commuting operators \cite{Sen:2012kpz}, namely $\mathds{D}_{m=0} = \mathcal{D}_1 + \mathcal{D}_2$, with $\mathcal{D}_1 = \slashed{\nabla}_{\mathbb{H}^2}\otimes \sigma^3$ and $\mathcal{D}_2=\mathds{1} \otimes\slashed{\nabla}_{\mathbf{S}^2} - \frac{i}{R}\, \tau^3 \otimes \mathds{1}$. The latter have eigenvalues $i \zeta_\lambda$ and $i (\zeta_n \pm 1)$, respectively. This, together with the identity $\mathds{D}^2_{m=0}=\mathcal{D}_1^2+\mathcal{D}_2^2$, implies \eqref{eq:eignvaluesDmassless}.}
\begin{eqn}\label{eq:eignvaluesDmassless}
    \zeta^{(1)}_{\pm} (\lambda, n) &= \pm\frac{i}{R} \sqrt{1+2 R \zeta_n+R^2(\zeta_\lambda^2+\zeta_n^2)} =  \pm \frac{i}{R} \sqrt{\lambda^2+(n + 1)^2}\, ,\\
    \zeta^{(2)}_{\pm} (\lambda, n) &= \pm\frac{i}{R} \sqrt{1-2 R \zeta_n+R^2(\zeta_\lambda^2+\zeta_n^2)} =  \pm \frac{i}{R} \sqrt{\lambda^2+(n - 1)^2}\, ,
\end{eqn}
where in the last step we substituted $R^2\zeta_\lambda^2= \lambda^2$ for $\lambda \in \mathbb{R}_{\geq 0}$, and $R^2\zeta_n^2= n^2$ with $n \in \mathbb{N}$. Notice that in this case there is no discrete part in the fermionic spectrum on $\mathbb{H}^2$, nor are there any normalizable zero modes associated with the Dirac operator on the compact 2-sphere. The eigenspinors corresponding to the eigenvalues shown in \eqref{eq:eignvaluesDmassless} are found to be
\begin{eqn}\label{eq:eigenspinorsDMassless}
    \Psi^{(1)}_\pm (\lambda, n) &=\begin{pmatrix}
   1 + \left( R\zeta_n \pm \sqrt{1+2R\zeta_n +R^2 (\zeta_n^2+\zeta_\lambda^2)}\right)\\
  R\zeta_\lambda \\
  -1 - \left( R\zeta_n \pm \sqrt{1+2R\zeta_n +R^2 (\zeta_n^2+\zeta_\lambda^2)}\right)\\
  R\zeta_\lambda 
\end{pmatrix}\, ,\\
\Psi^{(2)}_\pm (\lambda, n) &=\begin{pmatrix}
   1 - \left( R\zeta_n \pm \sqrt{1-2R\zeta_n +R^2 (\zeta_n^2+\zeta_\lambda^2)}\right)\\
  -R\zeta_\lambda \\
  1 - \left( R\zeta_n \pm \sqrt{1-2R\zeta_n +R^2 (\zeta_n^2+\zeta_\lambda^2)}\right)\\
  R\zeta_\lambda 
\end{pmatrix}\, .
\end{eqn}
We thus see that the net effect of the kinetic mixing is to shift the (non-chiral) spectrum of the Dirac operator on $\mathbf{S}^2$, leaving that on $\mathbb{H}^2$ untouched. Consequently, the 1-loop partition function can be readily computed in the present case as follows
\begin{equation} 
      \log \mathcal{Z}_\Psi = \frac{1}{2} \int_{\epsilon}^\infty \frac{d\tau}{\tau}\,  \Tr \left[e^{\tau \slashed{\nabla}_{\mathbb{H}^2}^2} \right]\,\Tr \left[e^{\tau \left(\mathds{1} \otimes\slashed{\nabla}_{\mathbf{S}^2} - \frac{i}{R}\, \tau^3 \otimes \mathds{1}\right)^2} \right]\,,
\end{equation}
with the only difference being captured by the trace over the spherical modes, which yields
\begin{eqn}\label{eq:tracewithmixingmassless}
\begin{aligned}
     \Tr_{\mathbf{S}^2} \left[e^{\tau \mathds{D}_{m=0}^2} \right] &=2\sum_{k =0}^\infty \left( k + 1\right) \left[e^{ -\frac{s}{R^2} k^2} + e^{ -\frac{s}{R^2} \left( k +2 \right)^2}\right]= 2\sum_{k =0}^\infty \left[\left( k + 1\right) e^{ -\frac{s}{R^2} k^2} + k e^{ -\frac{s}{R^2} \left( k +1 \right)^2}\right]\\
     &=2 \left[1+2\sum_{k =0}^\infty \left( k + 1\right) e^{ -\frac{s}{R^2} (k+1)^2}\right]\, .
\end{aligned}
\end{eqn}
Upon analytic continuation, one obtains the analogous partition function in AdS$_2 \times \mathbf{S}^2$
\begin{eqn}
    \log \mathcal{Z}_\Psi = \int_{\epsilon}^{\infty} \frac{d\tau }{\tau} \Tr \left[e^{\tau \slashed{\nabla}^2_{\text{AdS}_2}}\right]\bigg(1+2\sum_{k =0}^\infty \left( k + 1\right) e^{ -\frac{s}{R^2} (k+1)^2}\bigg)\, ,
\end{eqn}
in perfect agreement with previous works \cite{Sen:2012kpz, Keeler:2014bra}. Intuitively, what happens then is that the $\slashed{D}_{\mathbf{S}^2}$ spectrum gets shifted oppositely depending on the chirality of the accompanying AdS$_2$ bispinors. Notice that this gives the same pattern observed in the more complicated case studied in Section \ref{ss:TwistedDiracDiagonalization}, where we saw that the set of Landau levels in $\mathbf{S}^2$ remains invariant but the degeneracies no longer match with the corresponding $SU(2)$ quantum number $j=k+\frac12$. Nevertheless, when combining the two towers together one reproduces the correct degeneracy for each $n\in \mathbb{Z}_{\geq 0}$, modulo two additional chiral zero modes that descend from the former $n=1$ states, see Figure \ref{fig:EigenvalueShift}.

\subsection{Charged hypermultiplets without kinetic mixing}\label{ss:freemassivehyper}

As a second simple example, we consider the case in which the fields in the supermultiplet are minimally coupled to the gauge and gravitational backgrounds. In this situation no kinetic mixing occurs, due to the absence of Pauli couplings. The relevant fermionic operator reads
\begin{eqn}\label{eq:kineticoperatorBPSnoMixing}
    \mathds{D}=i\slashed{D}_{\mathbb{H}^2}\otimes \sigma^3 +\mathds{1} \otimes(i\slashed{D}_{\mathbf{S}^2}) +\frac{q_e}{R}\, \mathds{1}\otimes\mathds{1}-\frac{i q_m}{R}\,\tau^3\otimes \sigma^3\, ,
\end{eqn}
or, in matrix notation,
\begin{eqn}
    \mathds{D}\big\rvert_{\lambda,k; n}=\begin{pmatrix}
  \frac{q_e}{R} + i\,\zeta_n & i\,\zeta_{\lambda, k}                 & 0  & -\frac{i q_m}{R}\\
  i\,\zeta_{\lambda, k}  & \frac{q_e}{R} - i\,\zeta_n & -\frac{i q_m}{R} & 0             \\
  0 & -\frac{i q_m}{R} & \frac{q_e}{R} + i\,\zeta_n & -\,i\,\zeta_{\lambda, k}\\
  -\frac{i q_m}{R} & 0 & -\,i\,\zeta_{\lambda, k} & \frac{q_e}{R} - i\,\zeta_n
\end{pmatrix}\, .
\end{eqn}
The eigenvalues in this case are given by
\begin{eqn}\label{eq:eignvaluesDchargedNoMixing}
    \zeta_\pm (\lambda,k; n) = \frac{1}{R} \left(q_e\pm i\sqrt{q_m^2+R^2(\zeta_{\lambda, k}^2+\zeta_n^2)}\right)\, ,
\end{eqn}
each of them having double degeneracy. Correspondingly, the eigenspinors are found to be
\begin{eqn}\label{eq:eigenspinorsDchargedNoMixing}
    \Psi^{(1)}_\pm (\lambda,k; n) &=\begin{pmatrix}
    R\zeta_n \pm \sqrt{q_m^2 +R^2 (\zeta_n^2+\zeta_{\lambda, k}^2)}\\
  R\zeta_{\lambda, k} \\
  0\\
  -q_m 
\end{pmatrix}\, ,\\
\Psi^{(2)}_\pm (\lambda,k; n) &=\begin{pmatrix}
    R\zeta_{\lambda, k}\\
   -R\zeta_n \pm \sqrt{q_m^2 +R^2 (\zeta_n^2+\zeta_{\lambda, k}^2)}\\
   -q_m\\
  0 
\end{pmatrix}\, .
\end{eqn}
This example, however, provides a good opportunity to understand how the diagonalization procedure deals with the zero modes of the Dirac operator acting on the spherical submanifold. Indeed, from the Atiyah-Singer theorem \cite{Atiyah:1963zz}, we know that there should be exactly $2|q_m|$ zero modes of $\slashed{D}_{\mathbf{S}^2}$, corresponding to the $SU(2)$ quantum number $j=|q_m|-\frac12$ (see discussion around eq.~\eqref{eq:algebraiceigenfunctionsS2spin1/2}). Furthermore, since these states have definite chirality, i.e., they satisfy
\begin{eqn}\label{eq:ChiralityZeroModes}
    \sigma^3 \eta_{n=0} = \varepsilon\, \eta_{n=0}\, ,\qquad \text{with} \quad \varepsilon = \text{sgn}\,(q_m)\, ,
\end{eqn}
one deduces that the eigenspinors \eqref{eq:eigenspinorsDchargedNoMixing} get simplified in the $n=0$ sector to
\begin{eqn}\label{eq:SpinorsZeroModeSectorS2}
\begin{aligned}
    \Psi^{(1)}_\pm (\lambda,k; n=0) &= \left( \varepsilon\,  R\zeta_{\lambda, k}\pm \sqrt{q_m^2+R^2 \zeta_{\lambda, k}^2}\right)\psi_{\lambda, k}\otimes \eta_{n=0} - \varepsilon\,  q_m \tau^3\psi_{\lambda, k}\otimes \eta_{n=0}\, ,\\
    \Psi^{(2)}_\pm (\lambda,k; n=0) &= \left( R\zeta_{\lambda, k} \pm \varepsilon \sqrt{q_m^2+R^2 \zeta_{\lambda, k}^2}\right)\psi_{\lambda, k}\otimes \eta_{n=0} -  q_m\tau^3\psi_{\lambda, k}\otimes \eta_{n=0}\, .
\end{aligned}
\end{eqn}
Hence, we conclude that only two of the a priori four independent eigenspinors are actually linearly independent, each of them being associated with a different eigenvalue $\zeta_\pm(\lambda,k;n=0)$, such that their product yields
\begin{eqn}\label{eq:EigenvalueProductNoMixing}
    \zeta_{+} \zeta_{-} = \frac{1}{R^2} \left( q_e^2 +q_m^2+R^2\zeta_{\lambda, k}^2+R^2 \zeta_n^2\right)\, ,
\end{eqn}
in agreement with the analysis of Section \ref{s:integrationAdS2xS2}. Notice that \eqref{eq:EigenvalueProductNoMixing} suggests that the functional determinant of \eqref{eq:kineticoperatorBPSnoMixing} does not depend on the phase of the hypermultiplet central charge appearing in the mass term, which may in fact be reabsorbed via some chiral redefinition of the spin-$\frac12$ fermion. However, one should recall that the zero-mode sector of the \emph{full} Dirac operator is actually sensitive to this phase, as can be readily seen from eqs.~\eqref{eq:eignvaluesDchargedNoMixing} and \eqref{eq:SpinorsZeroModeSectorS2}. Indeed, when $k=0$, one finds that only the eigenvalue $\zeta_\delta(k=n=0) =\frac{1}{R} (q_e+i \delta q_m)$ survives, with $(\gamma^5-\delta \mathds{1}) \psi_{k=0} \otimes \eta_{n=0}=0$, and $\delta=\pm1$. This will, in turn, affect the $\theta$-term in \eqref{eq:AdS2xS2effaction}.

\subsection{Chirality of the zero modes of the Dirac operator on $\mathbf{S}^2$}\label{ss:ZeroModesS2}

In this subsection, we show how, in the presence of a constant magnetic field on $\mathbf{S}^2$, the chirality of the Dirac zero-modes is correlated with the sign of the magnetic flux $\frac{1}{2\pi}\int_{\mathbf{S}^2} F$. This may also be seen as a direct consequence of the Atiyah-Singer theorem \cite{Atiyah:1963zz}.

\medskip

To make this relation explicit it is more convenient to work in spherical polar coordinates. In this basis, the $SU(2)$ ladder operators together with $J_0$ (cf. eq.~\eqref{eq:OperatorSU2fermi}) take the form \cite{Grewal:2021bsu}
\begin{eqn}\label{eq:ladderoperatorSpherical}
J_\pm
&= e^{\pm i\phi}\left(
\pm \partial_\theta
+ i\cot\theta\,\partial_\phi
+ \frac{\sigma_z}{2\sin\theta}
- g\frac{1-\cos\theta}{\sin\theta}
\right)\,,\\
J_0 &= -i\partial_\phi - g\,.
\end{eqn}
We now determine the explicit form of the spinor eigenfunctions with lowest weight $m=-j$ for each $SU(2)$ irrep, characterized by the quantum number\footnote{We use the convention $g =|g|$ as discussed in Section \ref{ss:LandauAdS2H2}.}
\begin{eqn}
j = n + g - \frac12 = \ell - g\, .
\end{eqn}
To this end, we introduce the following ansatz for the positive/negative chirality modes
\begin{eqn}
\psi^{\pm}_{j,-j}
=
f^{\pm}_j(\theta)e^{-i(j-g)\phi}\,,
\qquad \text{with}\quad \sigma_z \psi^{\pm}_{j, -j}=\pm \psi^{\pm}_{j, -j}\, ,\quad \text{and}\quad 
J_0 \psi^{\pm}_{j,-j} = -j\,\psi^{\pm}_{j,-j}\, .
\end{eqn}
Therefore, the condition that selects the lowest-weight state is
\begin{eqn}
J_- \psi^{\pm}_{j,-j}=0\, ,
\end{eqn}
which, using the explicit form of the operators \eqref{eq:ladderoperatorSpherical}, reduces to the
following ordinary differential equation for $f_j^{\pm}(\theta)$
\begin{eqn}
-\partial_\theta f^{\pm}_j(\theta)
+ (j-g)\cot\theta\, f^{\pm}_j(\theta)
\pm \frac{1}{2\sin\theta}f^{\pm}_j(\theta)
- g\left(\frac{1-\cos\theta}{\sin\theta}\right)
f^{\pm}_j(\theta)
=0 \,.
\end{eqn}
The latter can be conveniently rewritten as
\begin{eqn}
\frac{d}{d\theta}\left(\ln f^{\pm}_j(\theta)\right)
=
\csc\theta
\left[
j\cos\theta-(g\pm\frac12)
\right]\, ,
\end{eqn}
whose solutions may be found via direct integration, yielding
\begin{eqn}
f^{\pm}_j(\theta)
\propto
\left(\sin\frac{\theta}{2}\right)^{\gamma_1}
\left(\cos\frac{\theta}{2}\right)^{\gamma_2}\,.
\end{eqn}
The exponents above are given by
\begin{eqn}
\gamma_1 = n + |g| - \frac12(1\mp1)\,,\qquad
\gamma_2 = n + |g| - \frac12(1\pm1)\,,
\end{eqn}
for each $n=0,1,\ldots, \infty$. 

\smallskip

\noindent Finally, imposing normalizability with respect to the integration measure on $\mathbf{S}^2$---which includes the factor $\sin\theta\,d\theta$, implies that
when $|g|=\pm g$, only $f^{\pm}_{n=0}(\theta)$ becomes square-integrable.
	
\bibliography{ref.bib}
\bibliographystyle{JHEP}

\end{document}